\documentclass[traditabstract]{aa} 
\pdfoutput=1
\usepackage{graphicx,natbib,longtable,txfonts,lscape}
\newcommand{\FR}{FRI{\sl{CAT}}}

\newcommand{\ergs}{\>{\rm erg}\,{\rm s}^{-1}}

\usepackage[switch]{lineno}
\begin{document}
\title{COMP2{\sl{CAT}}: hunting compact double radio sources in the local Universe}

  \author{A. Jimenez-Gallardo\inst{1,2}
          \and
          F. Massaro\inst{1,2,3,4}
          \and
          A. Capetti\inst{2}
          \and
          M. A. Prieto\inst{5,6}
          \and
          A. Paggi\inst{1,2,3}
          \and
          R.D. Baldi\inst{7}
          \and 
          R. Grossova\inst{1,8}
          \and
          L. Ostorero\inst{1,3}
          \and
          A. Siemiginowska\inst{9}
          \and
          S. Viada\inst{1}
          }
          
  \institute{Dipartimento di Fisica, Universit\`a degli
     Studi di Torino, via Pietro Giuria 1, 10125 Torino, Italy,
    \and
    INAF-Osservatorio Astrofisico di Torino, via Osservatorio 20, 10025 Pino Torinese, Italy,
    \and
    Istituto Nazionale di Fisica Nucleare, Sezione di Torino, I-10125 Torino, Italy,
    \and
    Consorzio Interuniversitario per la Fisica Spaziale (CIFS), via Pietro Giuria 1, I-10125, Torino, Italy,
    \and
    Departamento de Astrof\'isica, Universidad de La Laguna, E-38206 La Laguna, Tenerife, Spain,
    \and
    Instituto de Astrof\'isica de Canarias (IAC), E-38200 La Laguna, Tenerife, Spain,
    \and
    Department of Physics and Astronomy, University of Southampton, Highfield, SO17 1BJ, UK,
    \and
    Department of Theoretical Physics and Astrophysics, Faculty of Science, Masaryk University, Kotl\'a\v{r}sk\'a 2, Brno, 611 37, Czech Republic,
    \and
    Smithsonian Astrophysical Center, 60 Garden Street, Cambridge, MA 02138, USA.
    }
   \date{\today}

   \abstract {We present a catalog of compact double radio galaxies (hereafter COMP2$CAT$) listing 43 edge-brightened radio sources whose projected linear size does not exceed 60 kpc, the typical size of their host galaxies. This is the fifth in a series of radio source catalogs recently created, namely: FRI$CAT$, FRII$CAT$, FR0$CAT$ and WAT$CAT$, each focused on a different class of radio galaxies. The main aim of our analysis is to attain a better understanding of sources with intermediate morphologies between FR\,IIs and FR\,0s. COMP2$CAT$ sources were selected from an existing catalog of radio sources based on NVSS, FIRST and SDSS observations for having, mainly, i) edge-brightened morphologies, typical of FR\,IIs, ii) redshifts $z < 0.15$ and iii) projected linear sizes smaller than 60 kpc. With radio luminosities at 1.4 GHz $10^{38} \lesssim L_{1.4} \lesssim 10^{41}$ erg s$^{-1}$, COMP2$CAT$ sources appear as the low radio luminosity tail of FR\,IIs. However, their host galaxies are indistinguishable from those of large-scale radio sources: they are luminous ($-21 \gtrsim M_{r} \gtrsim -24$), red, early-type galaxies with black hole masses in the range of $10^{7.5} \lesssim M_{\rm BH} \lesssim 10^{9.5}\, \rm{M}_\odot$. Moreover, all but one of the COMP2$CAT$ sources are optically classifiable as low excitation radio galaxies, in agreement with being the low radio-power tail of FR\,Is and FR\,IIs. This catalog of compact double sources, which is $\sim 47 \%$ complete at $z<0.15$, can potentially be used to clarify the role of compact double sources in the general evolutionary scheme of radio galaxies.}


\titlerunning{COMP2{\sl{CAT}}: hunting compact double radio sources}
\authorrunning{Jimenez-Gallardo et al.}

\maketitle
\section{Introduction}
\label{sec:intro}
In 1974 Fanaroff \& Riley proposed to classify extragalactic radio sources on the basis of the morphology of their extended structures at 178 MHz. These authors distinguished two main classes of radio sources: edge-darkened, known as FR\,Is, and edge-brightened, known as FR\,IIs. \citet{fanaroff74} also discovered a link between this morphological classification and the total radio power, where FR\,Is tend to be less luminous at 178 MHz than FR\,IIs. Afterward, \citet{ledlow96} found that this classification was even sharper when comparing the optical luminosity of their host galaxies with their total radio power. However, several authors such as \citet{Best2009}, \citet{Lin2010}, \citet{Wing2011} and \citet{Capetti2017II} showed that the dichotomy in the optical-radio diagram disappears when considering samples selected at lower radio power, thus such distinction is probably due to high flux thresholds adopted in previous sample selections.

Another population, represented by ``compact'' radio galaxies, confined within a region of a few kpc and lacking large-scale jets, the formation and propagation of which is determined by plasma instabilities (see, e.g., \citealt{Bodo2013}, for a theoretical analysis), was later identified by \citet{baldi15}. These sources, known as FR\,0s, share almost all the characteristics of FR\,Is, but are lacking extended radio emission \citep[see also][]{Ghisellini2011}.

Motivated by the necessity of having homogeneous and complete samples of radio galaxies to investigate their properties and those of their large-scale environments, some of us recently built different catalogs of FR\,Is  \citep{Capetti2017I}, FR\,IIs \citep{Capetti2017II} and FR\,0s \citep{Baldi2018}. 

The \FR\ lists 219 sources, all hosted in red early-type galaxies, spectroscopically classified as low excitation radio galaxies (LERGs), at redshift $z < 0.15$ and with radio luminosity at 1.4 GHz, $L_{1.4}$, in the range $\sim 10^{39.5} - 10^{41.3} \ergs$. On the other hand, the FRII$CAT$ is composed of 122 edge-brightened radio sources within the same redshift range of the previous catalog and with $L_{1.4} \sim 10^{39.5} - 10^{42.5} \ergs$. A large fraction (i.e., $\sim$90\%) of the FR\,IIs listed therein are LERGs with the same type of host galaxies as FR\,Is. The remaining $\sim$10\% show optical spectra typical of high excitation radio galaxies (HERGs) and their hosts are bluer in the optical band and redder in the mid-IR than FR\,II LERGs.

Here, we aim at performing a study similar to that carried out for the \FR\ and the FRII$CAT$, by searching for radio galaxies with a classical FR\,II morphology at 1.4 GHz but showing a projected linear size smaller than 60 kpc.

From this study, we expect to find young radio sources classified as Gigahertz Peaked-Spectrum (GPS) as well as Compact Steep-Spectrum (CSS) sources. GPSs have typical sizes smaller than $\sim1$ kpc and their radio spectra peak between 500 MHz and 10 GHz, in the observer's frame, while CSSs are larger (i.e., between 1 and 20 kpc) and with radio spectra peaking at lower frequencies ($< 500$ MHz). Both classes include extremely powerful radio galaxies at 1.4 GHz, reaching $L_{1.4} \gtrsim 10^{41}$ erg s$^{-1}$ \citep[see e.g.,][for a review]{Odea1998}. 

In contrast to FR\,0s, GPSs and CSSs are resolved in the radio band showing a typical double-lobed structure. This is the reason underlying the morphological subclassification proposed by \citet{Readhead1995}, distinguishing between Compact Symmetric Objects (CSO; $<$ 1 kpc), Medium-sized Symmetric Objects (MSO; 1-15 kpc) and Large Symmetric Objects (LSO; $>$ 15 kpc).

Several hypotheses have been proposed in the literature (see, e.g., \citealt{Fanti1995} and \citealt{Odea1998}) to explain the relation between GPS/CSSs and large-scale radio galaxies. According to the most accepted scenario, GOS/CSSs are ``young" versions of large-scale radio sources; therefore, GPS may evolve into CSSs, and CSSs may then evolve either into FR Is or into FR IIs. 

Another popular hypothesis interprets the small sizes of GPSs and CSSs by assuming that they have the same ages as large-scale radio sources, but that they have been confined by interactions with dense gas in their environment. However, observations do not support this scenario, since the gas surrounding GPS/CSSs seems to be similar to that of FR II sources (\citealt{Orienti2014}; see, however, \citealt{Sobolewska2018}).

The creation of a catalog of ``small-size'' FR\,IIs will enable us to attain a better understanding of sources with morphologies between FR\,0s and FR\,IIs. Furthermore, such catalog could eventually let us distinguish whether these sources represent a completely different population of radio galaxies or whether they are ``young'' stages of the evolution of FR\,IIs.

The paper is organized as follows. In \S~\ref{sec:sample} we present the sample and the selection criteria of the catalog. The radio, optical and infrared properties of the selected sources are described in \S~\ref{sec:properties}. In \S~\ref{sec:discussion}, we present a comparison between the optical and radio properties of the sources, having then \S~\ref{sec:conclusions} devoted to our discussion and conclusions. The tables with the properties of the selected sources and their images are collected in the Appendixes \ref{ap:VLA}, \ref{ap:NVSS}, \ref{ap:properties}, \ref{ap:compcat} and \ref{ap:contours}.

We adopt cgs units for numerical results and we also assume a flat cosmology with $H_0=69.6$ km s$^{-1}$ Mpc$^{-1}$, $\Omega_\mathrm{M}=0.286$ and $\Omega_\mathrm{\Lambda}=0.714$ \citep{bennett14}, unless otherwise stated.\footnote{Thus, 1\arcsec\ corresponds to 2.634 kpc at $z=0.15$.}
Spectral indices are based on the definition of the flux density as $S_{\nu} \propto \nu^{-\alpha}$.

In general, the uncertainty on the radio luminosity is $< 5\%$, while for the linear size it is $< 0.25$ kpc. Therefore, we will not include error bars in our plots.

\section{Sample selection}
\label{sec:sample}

The sources in the catalog were selected from the sample of 18286 radio sources presented by \citet{best12}, hereafter BH12, but limited to the 3357 sources with redshift $z < 0.15$ and classified as AGN according to the criteria reported therein. The BH12 sample was built by cross-matching different catalogs: the optical spectroscopic catalogs based on data from the 7$^{th}$ Sloan Digital Sky Survey (DR7/SDSS; \citealt{abazajian09})\footnote{Available at {\tt http://www.mpa-garching.mpg.de/SDSS/}.} and produced by the group from the Max Planck Institute for Astrophysics and The Johns Hopkins University \citep{bri04,tre04}; the National Radio Astronomy Observatory Very Large Array Sky Survey (NVSS; \citealt{condon98}); and the Faint Images of the Radio Sky at Twenty centimeters survey (FIRST; \citealt{becker95}).
In BH12 a flux density threshold of 5 mJy was chosen.

Source selection was based on visual examination of FIRST images carried out for all objects. Radio contours at 1.4 GHz were built starting at a surface brightness level of 0.45 mJy/beam, approximately three times the typical rms of the FIRST images for objects at $z = 0.15$. This minimum surface brightness level was increased by a factor $\big[(1+0.15)/(1+z)\big]^4$ for closer objects to compensate for its cosmological dimming. This level corresponds to a correction factor of $\sim$1.75 for $z=0$. We also applied a $K$ correction by assuming a spectral index of 0.7 between 178 MHz and 1.4 GHz ($\alpha_{0.178-1.4}$), as done by \citet{Schoenmakers2000} and \citet{Capetti2017I}. Overall this correction was rather small, being $\sim$10\%.

Selected sources were those displaying radio emission at the sensitivity limit of the FIRST images within a circle of 30 kpc radius centered on the optical position. This size is large enough to ensure that the radio emission is still contained within the host galaxy. We selected only radio sources with a FRII-like morphology (i.e., edge-brightened) as done in \citet{Capetti2017II}. Thus, by applying the cut of having extended radio structure with projected size less than 60 kpc and selecting sources with FRII radio morphology, we considered those that were excluded from the FRII$CAT$. Five collaborators carried out the morphological classification independently, and only sources selected by at least three of them were included in the final catalog.

We called ``compact doubles" FR II radio sources with radio emission contained in the host galaxy and having two peaks of surface brightness; therefore, our catalog was called ``COMP2$CAT$" (COMPact Doubles CATalog). The term ``compact double" was first coined by \citet{Phillips1982}, although they called ``symmetric compact doubles" sources with the same radio morphology as those we selected, but with projected linear sizes $\leq 1$ kpc. Since the selection carried out was based on morphology, COMP2$CAT$ is analogous to a CSO/MSO/LSO sample with projected linear sizes limited to 60 kpc.

\begin{figure}
    \centering
\includegraphics[width=9.5cm]{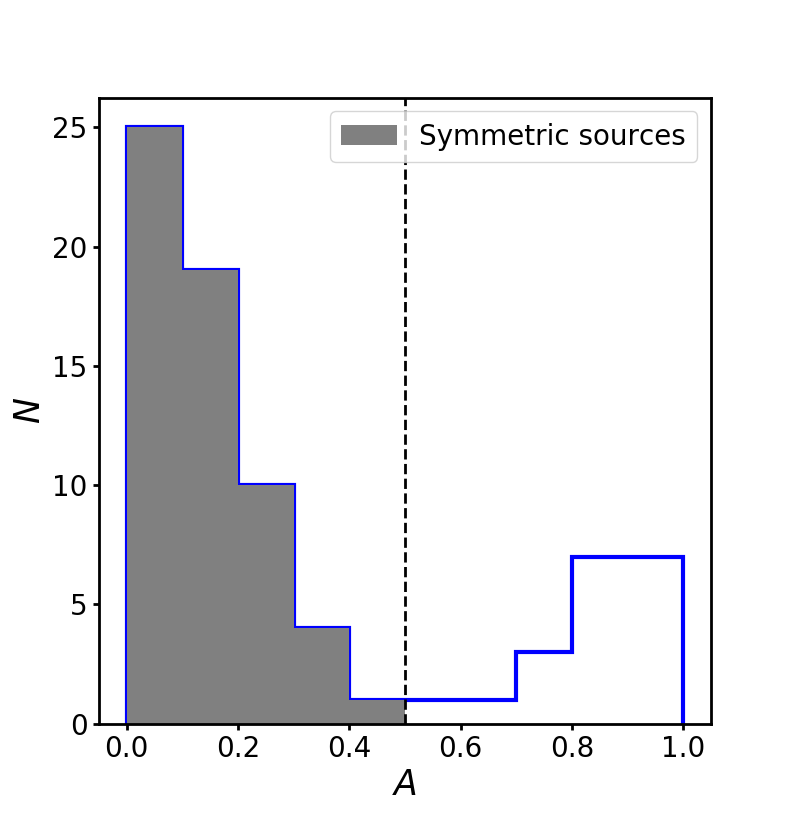}
\caption{Asymmetric index distribution of the 78 sources initially selected as defined in eq. \ref{eq:asym}. The black vertical dashed line indicates the separation between symmetric and asymmetric sources defined as $A = 0.5$. Symmetric sources are those with $A \leq 0.5$.}
\label{asym}
\end{figure}

\begin{figure*}
    \centering
    \includegraphics[width=6.cm,height=6.3cm]{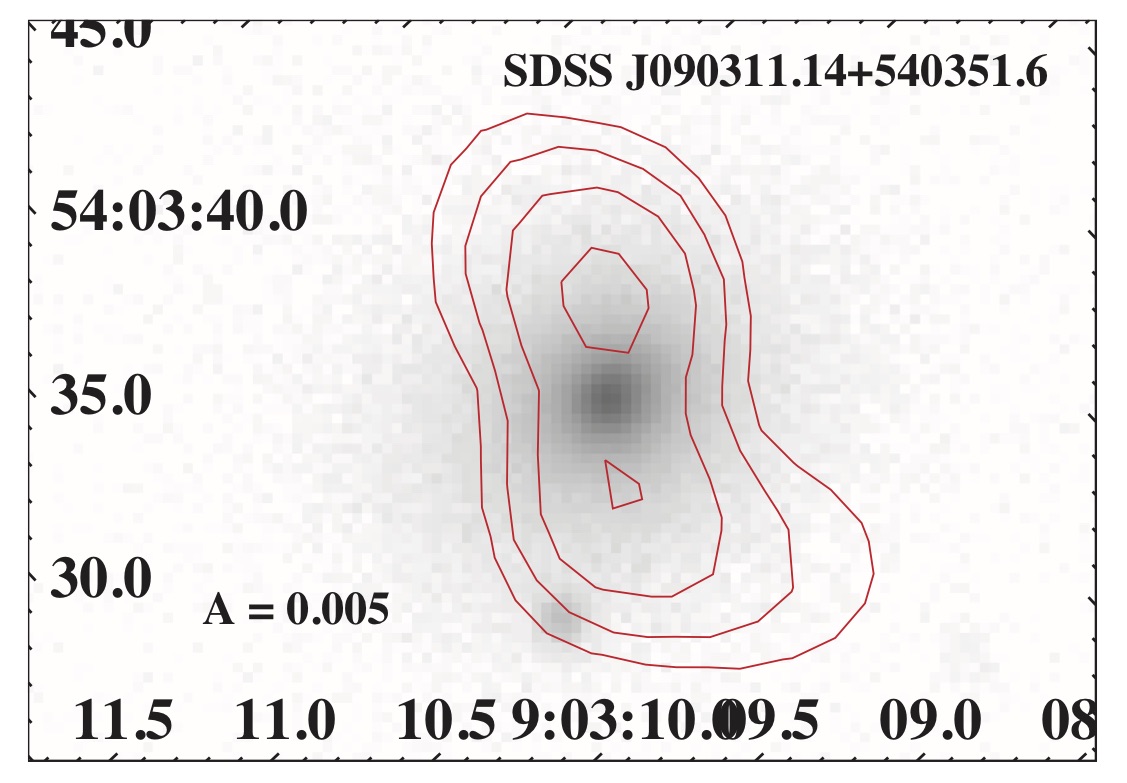}
    \includegraphics[width=6.cm,height=6.3cm]{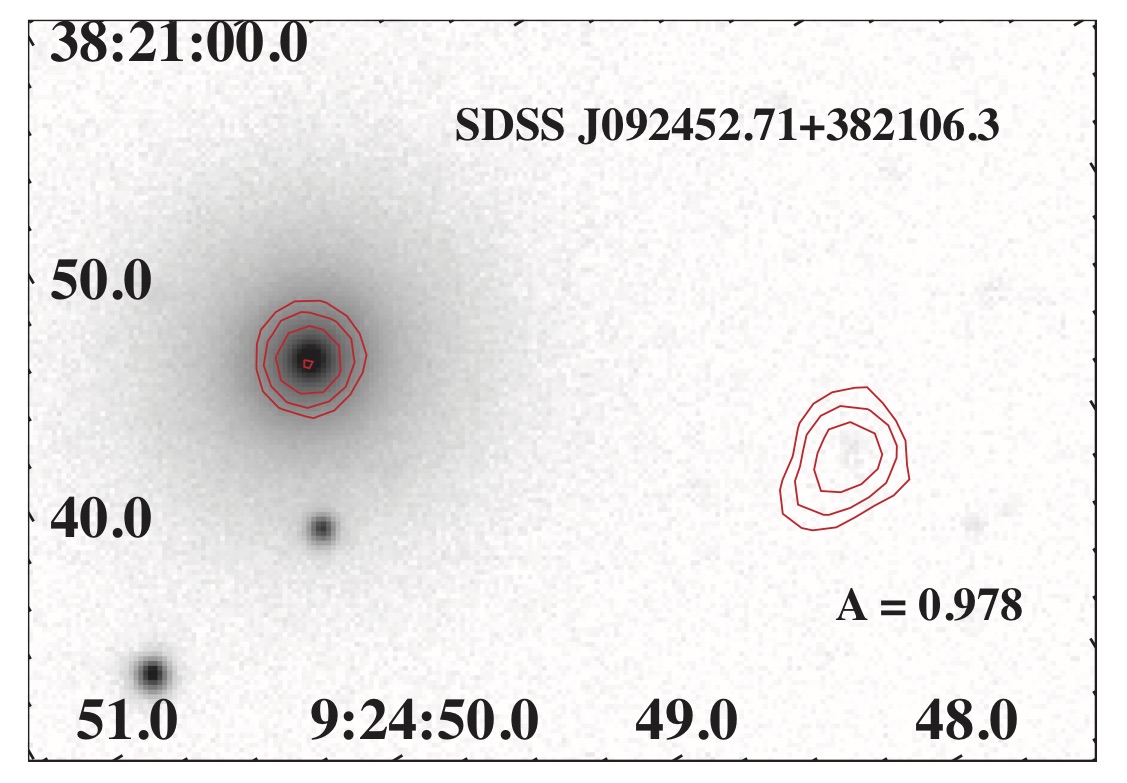}
    \includegraphics[width=6.cm,height=6.3cm]{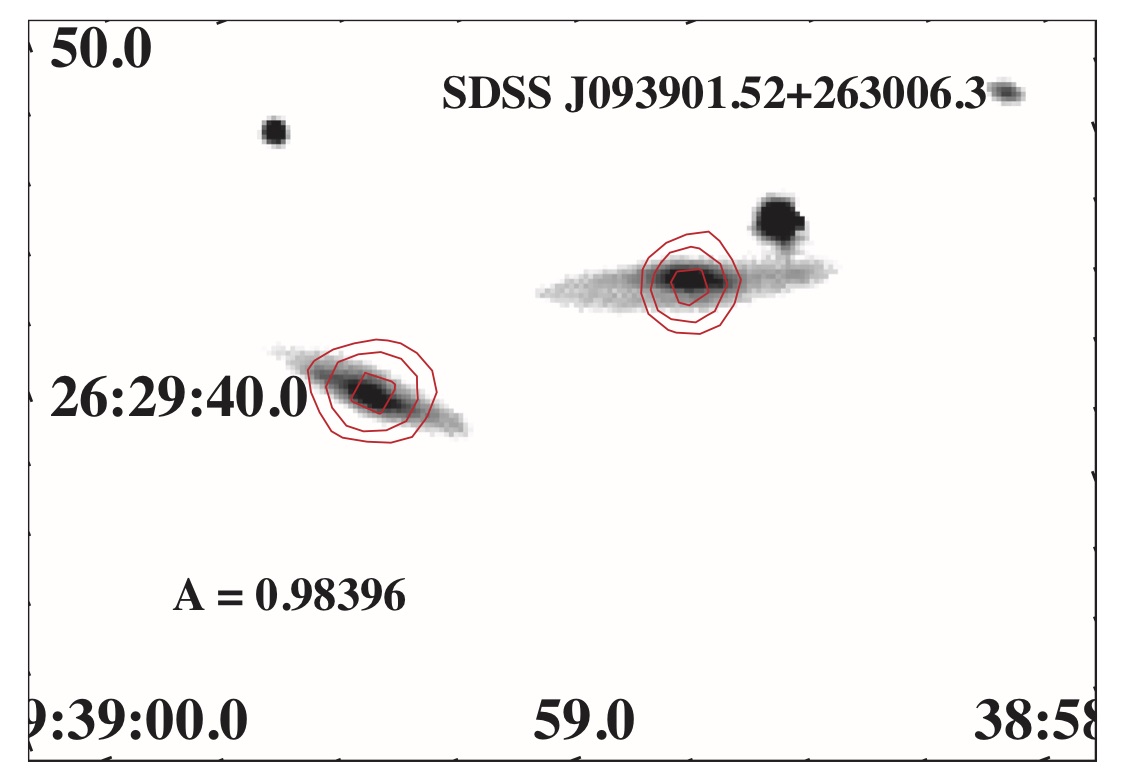}
    \caption{$r$-band SDSS images with FIRST contours superimposed for a symmetric (left) and two asymmetric (middle and right) sources, according to their FIRST contours. Left: contours start at 2 mJy beam$^{-1}$ and increase by a factor 2. Middle: contours start at 0.6 mJy beam$^{-1}$ and increase by a factor 2. Right: contours start at 0.5 mJy beam$^{-1}$ and increase by a factor 2.}
    \label{fig:sym}
\end{figure*}

Following our definition, we selected 78 sources from the original sample. However, there were cases of selected sources having one of the peaks of the radio surface brightness closer to the SDSS optical position than the other. These sources could be, for example, chance alignments of unrelated sources (i.e., background or foreground objects), close counterparts or FRI-like sources, instead of true compact doubles. To distinguish between these cases, we defined the asymmetric index, $A$, as:

\begin{center}
\begin{equation}
    A=\frac{\left|r_{2}-r_{1}\right|}{r_{2}+r_{1}}
    \label{eq:asym}
\end{equation}
\end{center}
where $r_{1}$ and $r_{2}$ are the distances of each radio surface brightness peak from the optical position. The $A$ parameter ranges between 0 and 1: symmetric sources display $A\sim0$, whereas $A=1$ corresponds to extremely asymmetric sources. Fig. \ref{asym} represents the distribution of the asymmetric indices for all the selected sources. Given that the bi-modal distribution peaks around $A=0.1$ (i.e. more symmetric) and around $A=0.9$ (i.e. very asymmetric ones) and since we wanted to focus on the most symmetric sources, we cut out those with $A \geq 0.5$, which corresponds to one of the peaks of surface brightness being at a distance from the optical position three times larger than the other one. A comparison between symmetric and asymmetric sources is shown in Fig. \ref{fig:sym}, which shows an example of a COMP2$CAT$ source identified as symmetric using our criterion (left panel), as well as two of the sources identified as asymmetric (central and right panels). As shown in Fig. \ref{fig:sym}, this definition of asymmetric index enabled us to exclude double sources with asymmetric radio morphology. 

Then, we also considered a few more exclusions based on radio and optical properties:
\begin{enumerate}
    \item Sources with unclear radio classification based on high-resolution radio maps obtained from the National Radio Astronomy Observatory (NRAO) Very Large Array (VLA) archive\footnote{Available at: \newline {\tt http://www.aoc.nrao.edu/$\sim$vlbacald/ArchIndex.shtml}}. (see App. \ref{ap:VLA}): SDSS J111025.09+032138.8, SDSS J125724.35+272952.1 and SDSS J132451.44+362242.7 (using radio maps at 1.4 GHz with resolutions of 12.40, 1.43 and 1.31 arcsec, respectively) and SDSS J125935.70+275733.3 and SDSS J161531.36+272657.3 (using radio maps at 5 GHz with resolutions of 1.51 and 1.18 arcsec).
    
    \item ``Restarted'' radio sources (or double-double radio galaxies, see \citealt{Schoenmakers2000} and App. \ref{ap:NVSS}) based on their extended 1.4 GHz and 150 MHz emission: SDSS J152804.95+054428.1, SDSS J132345.01+313356.7, SDSS J215305.08-071106.9, SDSS J115905.68+582035.5 and SDSS J083830.99+194820.4.
    
    \item Sources with a FRI-like morphology at scales of hundreds of kiloparsecs (see \citealt{Capetti2017I} and App. \ref{ap:NVSS}) based on their extended 1.4 GHz and 150 MHz emission: SDSS J091443.12+073554.9 and SDSS J083224.82+184855.4
\end{enumerate}

After finalizing the selection, COMP2$\sl{CAT}$ includes 43 sources whose properties and contours from FIRST (at 1.4 GHz with a resolution of 5 arcsec), NVSS (at 1.4 GHz with a resolution of 45 arcsec) and the Tata Institute of Fundamental Research (TIFR) Giant Metrewave Radio Telescope (GMRT) Sky Survey (TGSS; at 150 MHz with a resolution $> 25$ arcsec) are represented in App. \ref{ap:properties} and \ref{ap:compcat}, in Table \ref{tab:compcat} and in Fig. \ref{images1}. We used FIRST radio maps to carry out the morphological identification of sources while obtaining the 1.4 GHz fluxes from NVSS. The reason why we chose to use the 1.4 GHz fluxes from NVSS instead of those from FIRST is that FIRST could have missed some of the flux from large-scale structures due to its lack of short baselines.

Other catalogs of compact radio galaxies, with sources mainly selected on the basis of their radio properties, are present in the literature. In particular, \citet{Snellen2004} published the CORALZ catalog, a catalog of compact radio sources at low redshifts. This catalog was built by selecting those sources in FIRST with an optical counterpart in the APM Palomar Sky Survey (APM/POSS-I) catalog and radio flux densities at 1.4 GHz $> 100$ mJy and angular sizes $< 2$ arcsec (which translates into a projected linear size $< 5.6$ kpc using our cosmology). Following these criteria, the CORALZ catalog is made of 28 sources at $0.005<z<0.16$, of which 17 form a $95\%$ statistically complete sample. 

On the other hand, for COMP2$CAT$ we did not impose any limits on the radio flux density of the sources and we included sources with projected linear sizes up to 60 kpc; thus, we expected to find physically larger and less radio luminous sources in COMP2$CAT$ than those in the CORALZ catalog.

\section{COMP2$CAT$ host and radio properties}
\label{sec:properties}

\subsection{Radio properties}

\begin{figure}
    \centering
\includegraphics[width=9.cm]{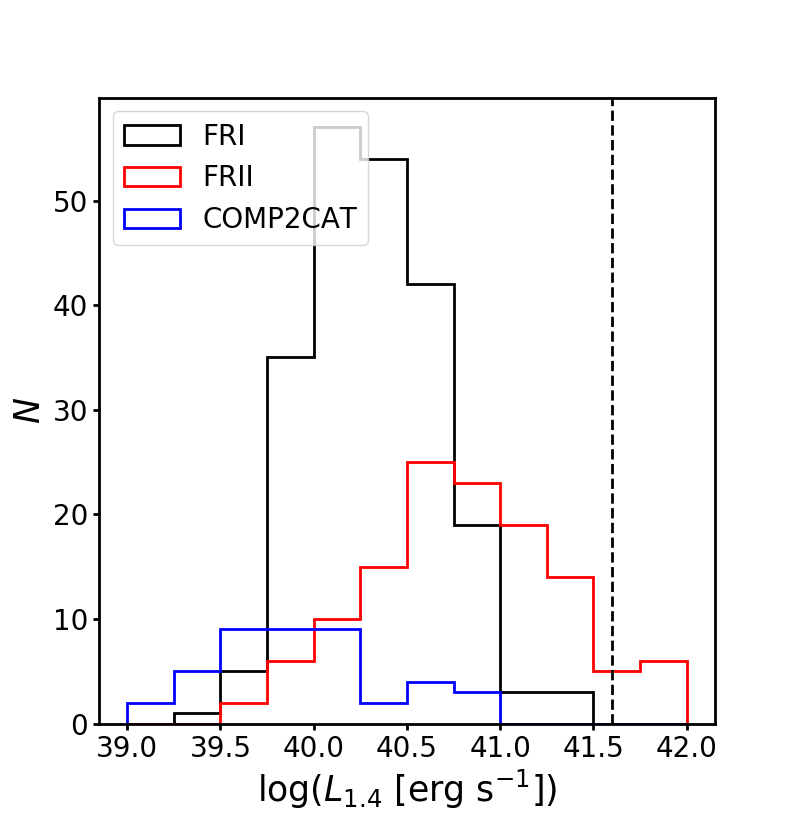}
\caption{Radio luminosity distribution of COMP2$CAT$, \FR\ and FRII$\sl{CAT}$ (blue, black and red dots respectively). The dashed vertical line indicates the transition power between FR I and FR II reported by \citet{fanaroff74}. }
\label{lhist}
\end{figure}

COMP2$CAT$ sources appear as the low radio power tail of FR Is and FR IIs, with a distribution of NVSS 1.4 GHz radio luminosities that ranges in the interval $10^{38.5} \lesssim L_{1.4} \lesssim 10^{41}$ $\ergs$ and peaks around $\sim 10^{39.8} \ergs$, as shown in Fig. \ref{lhist}. This Figure also shows the separation between FR Is and FR IIs established by \citet{fanaroff74}, $L_{0.178}\sim 2.8 \times10^{41}$ erg s$^{-1}$ sr$^{-1}$, which we adapted to the cosmology chosen here and assuming $\alpha_{0.178-1.4} = 0.7$ ($L_{1.4} \sim 10^{41.6}$ $\ergs$). Most objects included in COMP2$CAT$, \FR\ and FRII$CAT$ fall below this threshold.

\begin{figure*}
\includegraphics[width=9.cm]{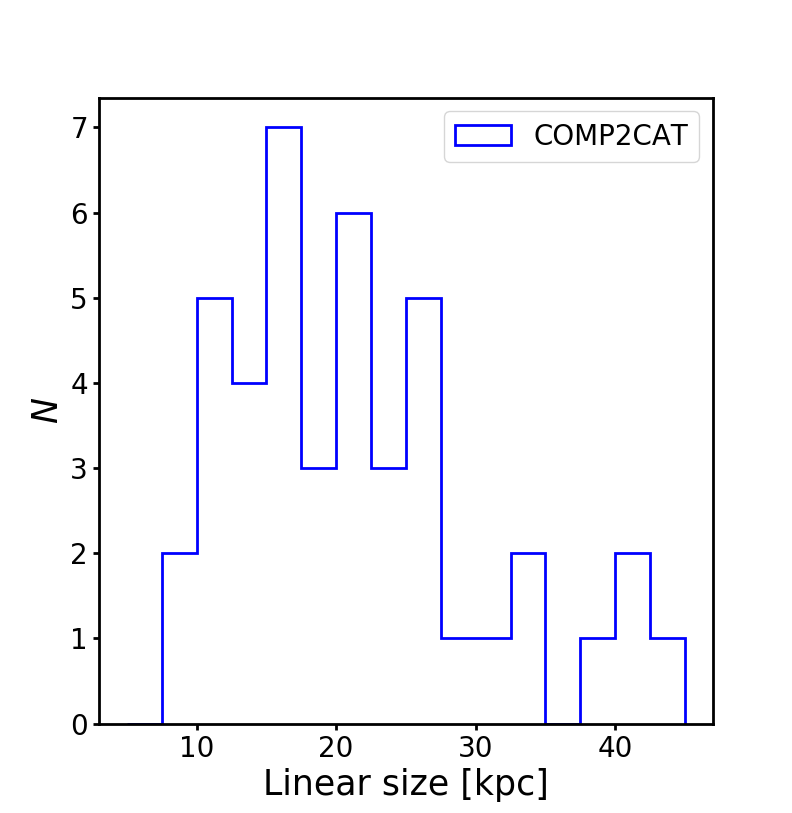}
\includegraphics[width=9.cm]{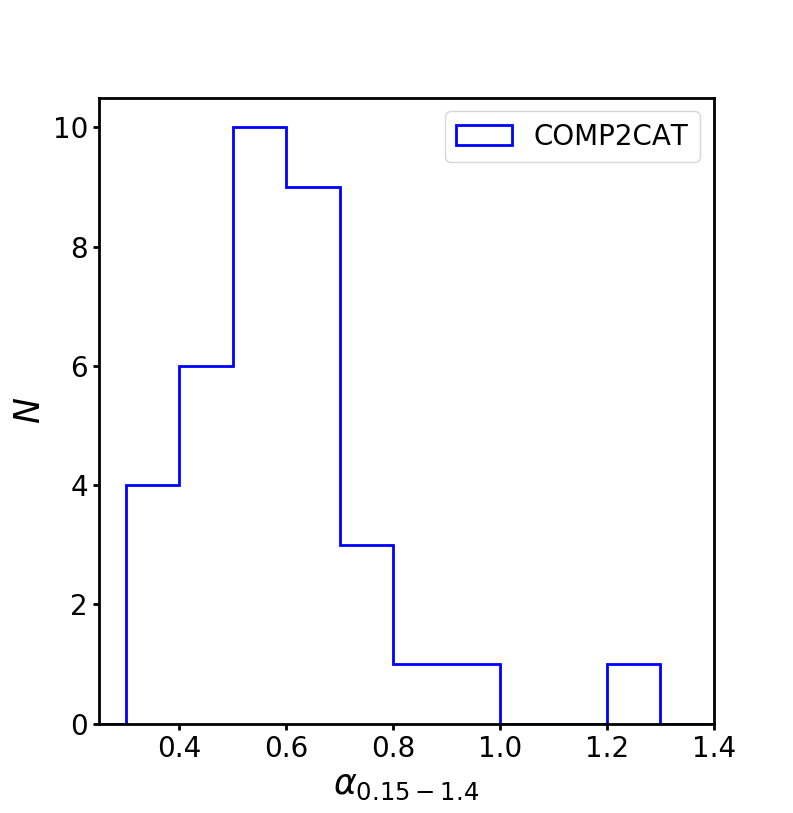}
\caption{Left: Projected linear sizes distribution of COMP2$CAT$ sources. Right: Distribution of the spectral index between 150 MHz and 1.4 GHz for sources in COMP2$CAT$. This spectral index was obtained using 1400 and 150 MHz fluxes from NVSS and TGSS as shown in eq. \ref{eq:alpha}.} 
\label{linsize}
\end{figure*}

The left panel of Fig. \ref{linsize} shows the projected radio linear size distribution of COMP2$CAT$ sources. Their projected radio linear sizes range from 5 to 45 kpc, peaking around 15 kpc and correspond to the distances between the two peaks of radio surface brightness at 1.4 GHz. Then, according to the morphological classification for compact doubles presented by \citet{Readhead1995}, COMP2$CAT$ includes 11 MSOs and 32 LSOs.

The spectral index between 150 MHz and 1.4 GHz, $\alpha_{0.15-1.4}$, was computed as follows:

\begin{center}
    \begin{equation}
       \alpha_{0.15-1.4} = -1.03\log\left(\frac{S_{1.4}}{S_{0.15}}\right)
    \end{equation}
    \label{eq:alpha}
\end{center}
where $S_{1.4}$ is the NVSS flux density and $S_{0.15}$ the TGSS one.

We show the distribution of spectral indices in the right panel of Fig. \ref{linsize}. The bulk of COMP2$CAT$ sources have $0.3 \leq \alpha_{0.15-1.4} \leq 0.9$ and their distribution peaks around $\alpha_{0.15-1.4}=0.5$, while only 8 COMP2$CAT$ sources do not have TGSS counterparts and, hence, their spectral index could not be estimated.

\begin{figure}
    \centering
    \includegraphics[width=9.5cm]{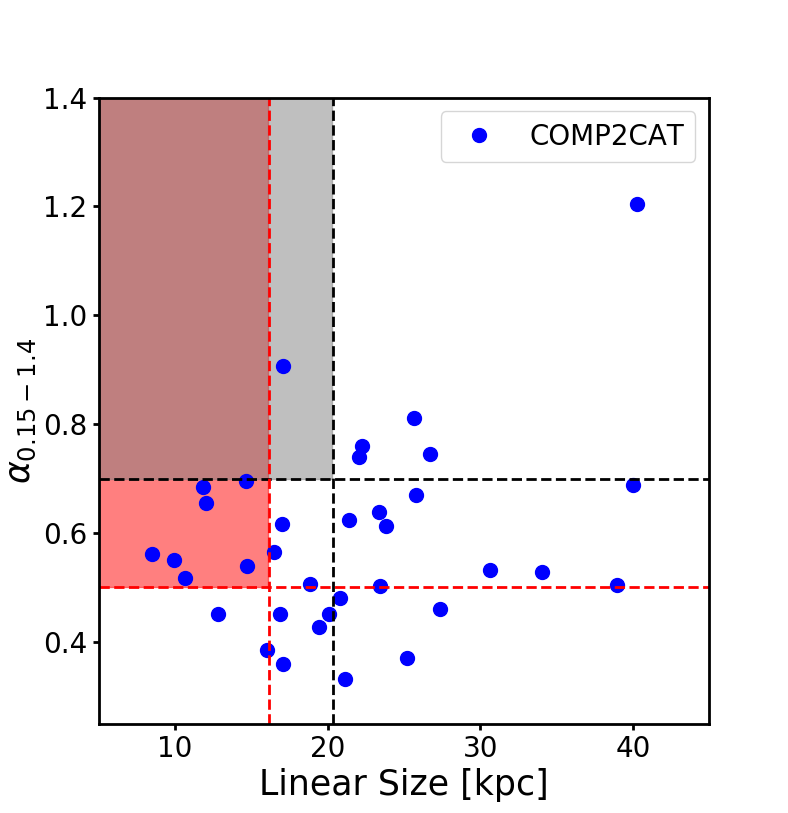}
    \caption{Spectral index between 150 MHz and 1.4 GHz vs. linear size of COMP2$CAT$ sources. The red lines mark the criteria chosen by \citet{Labiano2006} to define CSSs: projected linear sizes below 15 kpc and $\alpha_{0.15-1.4}\gtrsim 0.5$ (corrected using the cosmology adopted). The area shaded in red corresponds to sources that would be considered CSSs according to these criteria. On the other hand, the black lines correspond to criteria presented by \citet{Orienti2014}: projected linear sizes smaller than 20 kpc and $\alpha_{0.15-1.4} \gtrsim 0.7$ (also corrected with our cosmology). The area shaded in gray marks the CSS area in the diagram based on \citet{Orienti2014} selection.}
    \label{size}
\end{figure}

\citet{Labiano2006} defined CSSs as those sources with linear sizes smaller than 15 kpc and $\alpha > 0.5$, while according to a more recent analysis of \citet{Orienti2014} this definition could be extended to sources with linear sizes smaller than 20 kpc and $\alpha > 0.7$. In Fig. \ref{size}, $\alpha_{0.15-1.4}$ is represented against the projected linear size. Thus, possible CSSs in the COMP2$CAT$ are those lying in the upper left corner of Fig. \ref{size}, whereas the previous definitions are corrected for the adopted cosmological parameters (see \S\,\ref{sec:intro}). According to the criteria adopted by \citet{Labiano2006}, there are 7 possible CSSs in COMP2$CAT$ (SDSS J073600.87+273926.0, SDSS J074641.45+184405.4, SDSS J090311.14+540351.6, SDSS J111109.58+393552.0, SDSS J113643.49+545446.8, SDSS J144731.24+330606.2 and SDSS J164452.86+341251.3), while according to the definition by \citet{Orienti2014}, only one of the COMP2$CAT$ sources, SDSS J113305.52+592013.7, could be considered a CSS. Therefore, adopting \citet{Orienti2014} criteria, CSSs do not constitute an important fraction of COMP2$CAT$.

Additionally, we estimated the spectral index between 1.4 and 5 GHz using the Green Bank 6-cm (GB6) Radio Source Catalog and the NVSS flux. Only 14 COMP2$CAT$ sources have GB6 counterparts. Furthermore, the ``Full Width at Half Maximum", $FWHM$, of the primary beam of GB6 is $\sim 3.5$ arcmin; thus, the fluxes obtained at 5 GHz are only upper limits, since there are multiple FIRST sources inside the beam that could contaminate the result. We checked the sources individually and found that the only ones that could have important contamination from neighbor sources are SDSS J132649.30+164948.0, SDSS J135338.43+360802.4 and SDSS J155749.61+161836.6. In those cases, the spectral indices obtained are regarded as lower limits.

\begin{figure}
    \centering
\includegraphics[width=9.5cm]{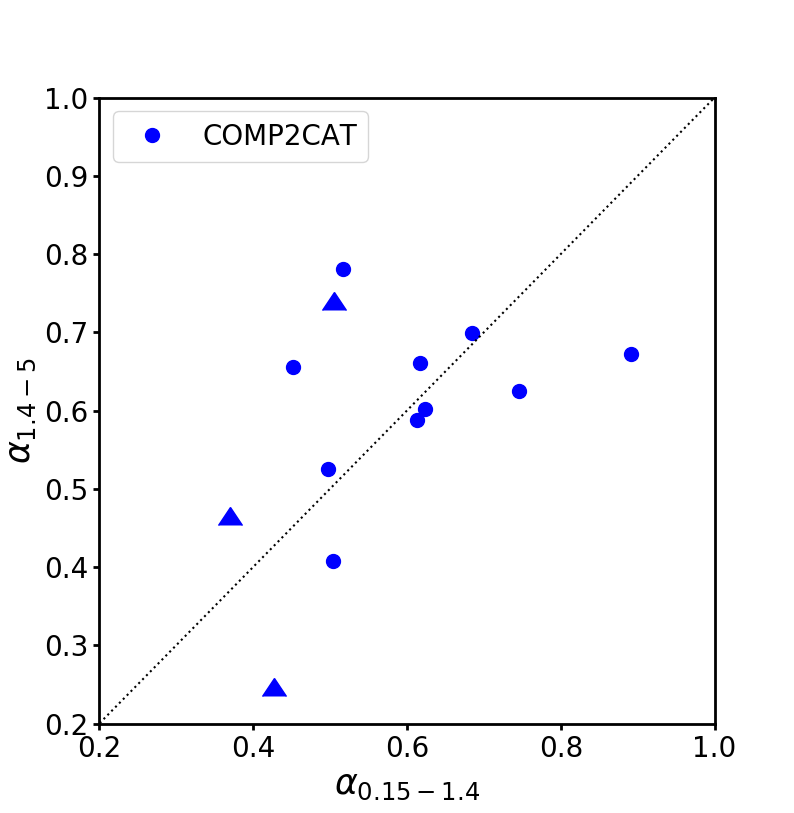}
\caption{Distribution of spectral indices between 1.4 and 5 GHz for COMP2$CAT$ sources with GB6 counterparts. Upward arrows indicate lower limits.}
\label{alphaghz}
\end{figure}

The comparison between the spectral indices at low and high frequencies is shown in Fig. \ref{alphaghz}. The distribution of $\alpha_{1.4-5}$ ranges from 0.2 to 0.9 and peaks around $\alpha_{1.4-5}=0.65$. Five COMP2$CAT$ sources are actually out of GB6 footprint, while the flux at 5 GHz of the remaining sources of the sample, assuming either a flat spectrum or the same spectral index as from 150 MHz to 1.4 GHz, is below the completeness level of GB6 (50 mJy); the only exceptions are (i) SDSS J125935.70+275733.3 and (ii) SDSS J091134.75+125538.1. For them, we could estimate lower limits for their spectral indices between 1.4 and 5 GHz: (i) $\alpha_{1.4-5} > 0.56$ and (ii) $\alpha_{1.4-5} > 0.98$. 

\begin{figure}
    \centering
    \includegraphics[width=9.5cm]{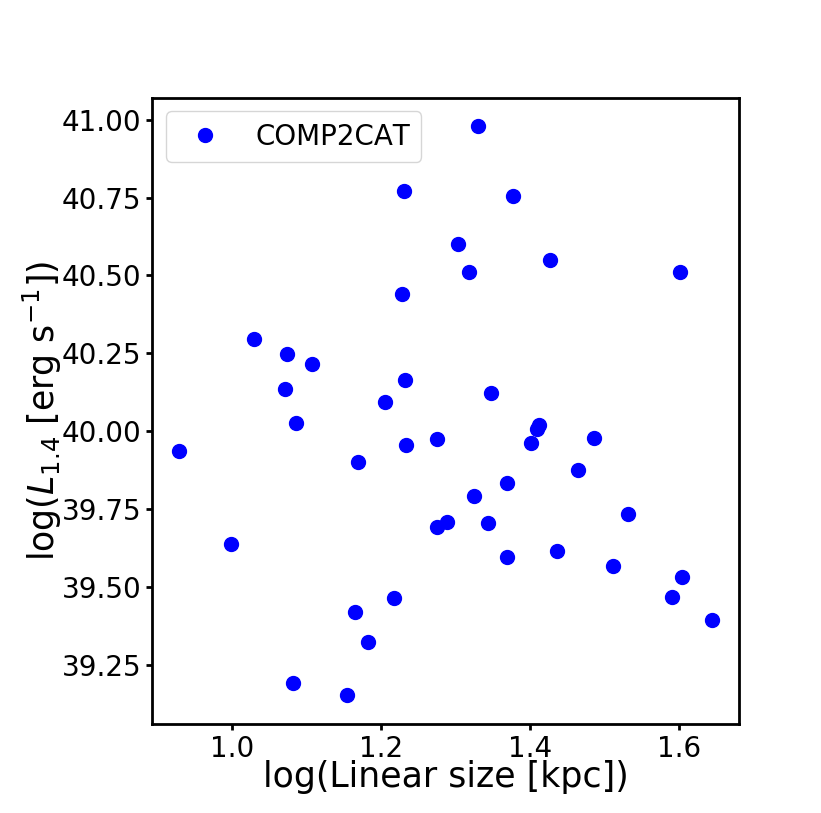}
    \caption{Radio luminosities at 1.4 GHz as a function of projected linear sizes of COMP2$CAT$ sources.}
    \label{fig:radsize}
\end{figure}

Lastly, we show a comparison between $L_{1.4}$ and the projected linear size in Fig. \ref{fig:radsize}. Should COMP2$CAT$ sources be the predecessors of FRII$CAT$ sources, we would see an increase of the luminosity with size. However, we do not see a clear trend. This could indicate either that these sources could evolve into low-luminosity FRIIs or that they represent a different population of radio sources. 

Images with the radio contours for COMP2$CAT$ sources are represented in App. \ref{ap:compcat}, in Fig. \ref{images1}. For each source, we show the FIRST\footnote{Available at ${\tt https://third.ucllnl.org/cgi-bin/firstcutout}$.} (black), the NVSS \footnote{Available at ${\tt https://www.cv.nrao.edu/nvss/postage.shtml}$.} (red) and the TGSS \footnote{Available at ${\tt https://vo.astron.nl/tgssadr/q_fits/imgs/form}$.} (blue) contours. The contours at a given frequency  are drawn by choosing a starting surface brightness level and increasing this value by a chosen factor. The starting level and the increase factor for each source and frequency are listed in Table \ref{tab:contourscomp}.

\subsection{Optical and infrared properties}

\begin{figure}
    \centering
    \includegraphics[width=9.5cm]{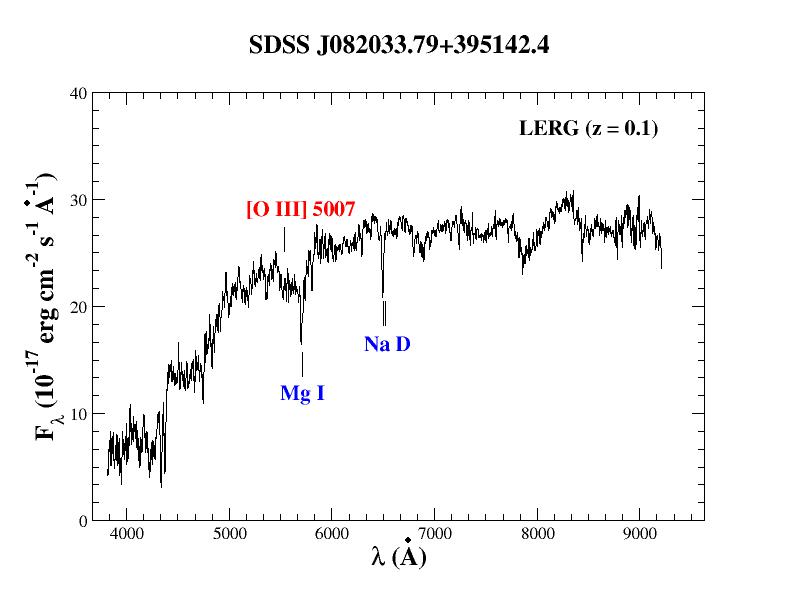}
    \caption{Optical spectrum of SDSS J082033.79+395142.4, selected in COMP2$CAT$, from 4000 \AA\ and 9000 \AA\ available in the SDSS database. This source is a clear LERG. Optical emission (red) and absorption (blue) lines identified are marked in the figure.}
    \label{fig:LERGspec}
\end{figure}

\begin{figure}
    \centering
    \includegraphics[width=9.5cm]{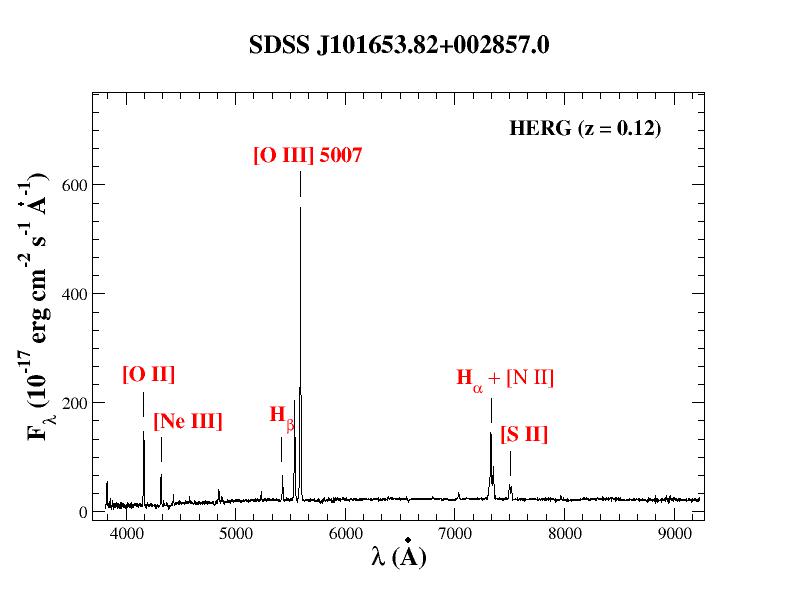}
    \caption{Same as Fig. \ref{fig:LERGspec} but for SDSS J101653.82+002857.0, the only HERG belonging to COMP2$CAT$.}
    \label{fig:HERGspec}
\end{figure}

\begin{figure}
    \centering
    \includegraphics[width=9.5cm]{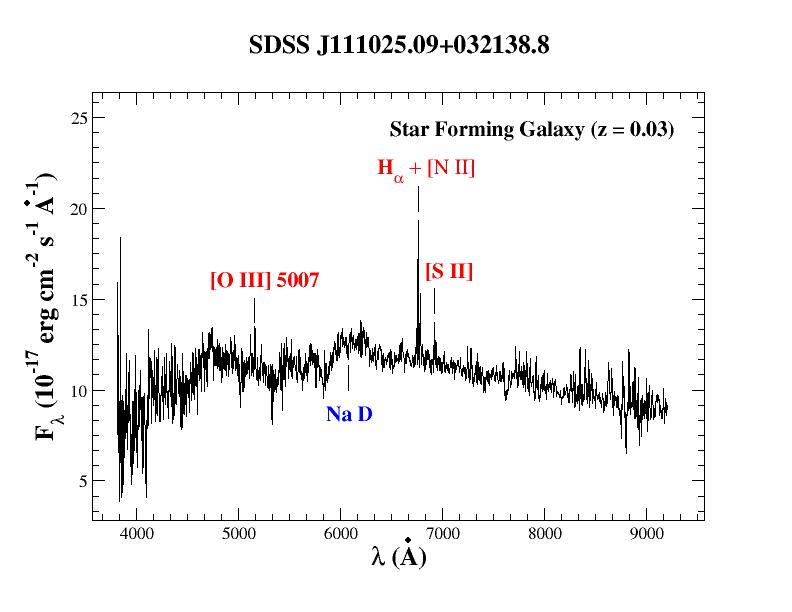}
    \caption{Same as Fig. \ref{fig:LERGspec} but for the case of SDSS J111025.09+032128.8, the starforming galaxy excluded from our final selection on the basis of its radio morphology shown in the VLA radio image.}
    \label{fig:SFspec}
\end{figure}

All COMP2$CAT$ sources are classified as Low Excitation Radio Galaxies (LERGs); the only exception is SDSS J101653.82+002857.0, which is a possible HERG. Differences in the spectra of LERGs, HERGs and star-forming galaxies are represented in Figs. \ref{fig:LERGspec}, \ref{fig:HERGspec} and \ref{fig:SFspec}. \citet{Baldi2010} claimed that there is contamination of $\sim 10 \%$ from radio-quiet AGN in the SDSS/NVSS sample. In our case, the contamination would mainly come from Seyfert galaxies, since we do not expect other radio-quiet galaxies to form double structures with sizes exceeding a few kiloparsecs. Actually, we only found one object with strong optical emission features that could be either a Seyfert galaxy or a HERG; therefore, the contamination from radio-quiet AGN is negligible in our case.

\begin{figure}
    \centering
\includegraphics[width=9.5cm]{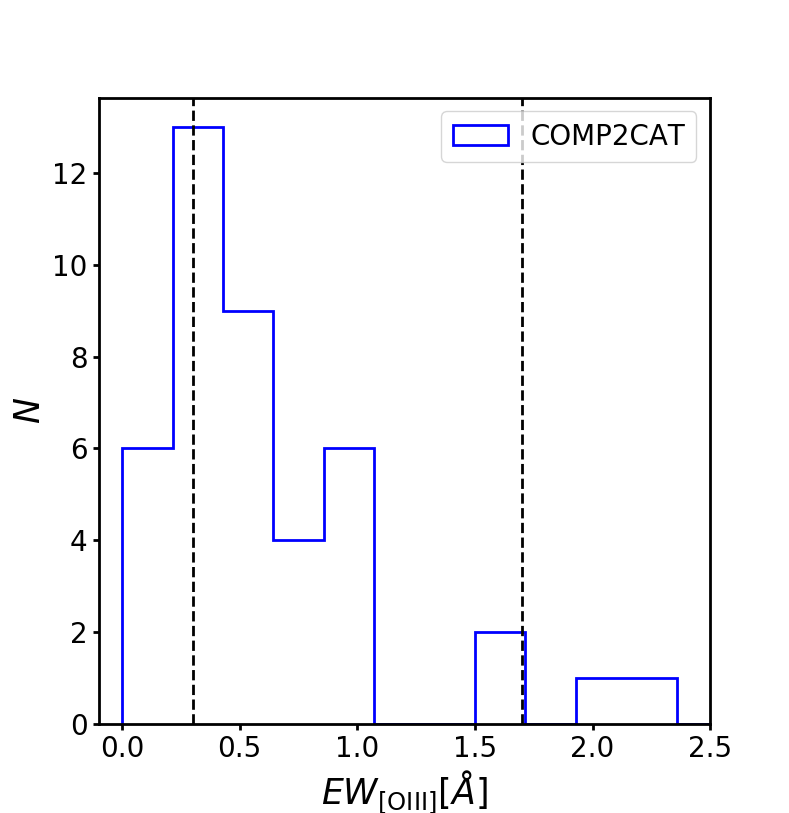}
\caption{Distribution of equivalent widths of the [O III] line for COMP2$CAT$ sources. The black vertical dashed lines limit the region where $L_{[\rm{OIII}]}$ is dominated by populations of old stars instead of by the AGN according to \citet{Capetti2011}.}
\label{EW}
\end{figure}

We obtained the equivalent width of the [O III], $EW_{[\rm{OIII}]}$, from the SDSS database, and show its distribution on Fig. \ref{EW}. The HERG source has not been included in this plot due to its high $EW_{[\rm{OIII}]}$ value\footnote{SDSS J101653.82+002857.0: $EW_{[\rm{OIII}]} \sim 190\, \AA$}. The $EW_{[\rm{OIII}]}$ values range between $0.5 \lesssim EW_{[\rm{OIII}]} \lesssim 3\, \AA$ and peak at $EW_{[\rm{OIII}]} \sim 0.5\, \AA$. According to \citet{Capetti2011} stellar processes (instead of the AGN) can dominate the [O III] line emission, especially for low radio luminosity sources and this can be distinguished on the basis of the $EW_{[\rm{OIII}]}$, in such a way that the [O III] line emission from COMP2$CAT$ sources presenting $EW_{[\rm{OIII}]}\leq 1.7\, \AA$ is mostly due to stellar processes. This is the case for $\sim 80\%$ of COMP2$CAT$ sources, as we will show in the following section. Even considering this effect, we found that COMP2$CAT$ sources have [O III] luminosities, $L_{[OIII]}$, thousands of times smaller than the ones that \citet{Labiano2009} found for GPSs and CSSs, which highlights the fact that sources in COMP2$CAT$ constitute a different population than GPS/CSSs, as already shown in Fig. \ref{size}.

\begin{figure*}
    \centering
\includegraphics[width=8.9cm]{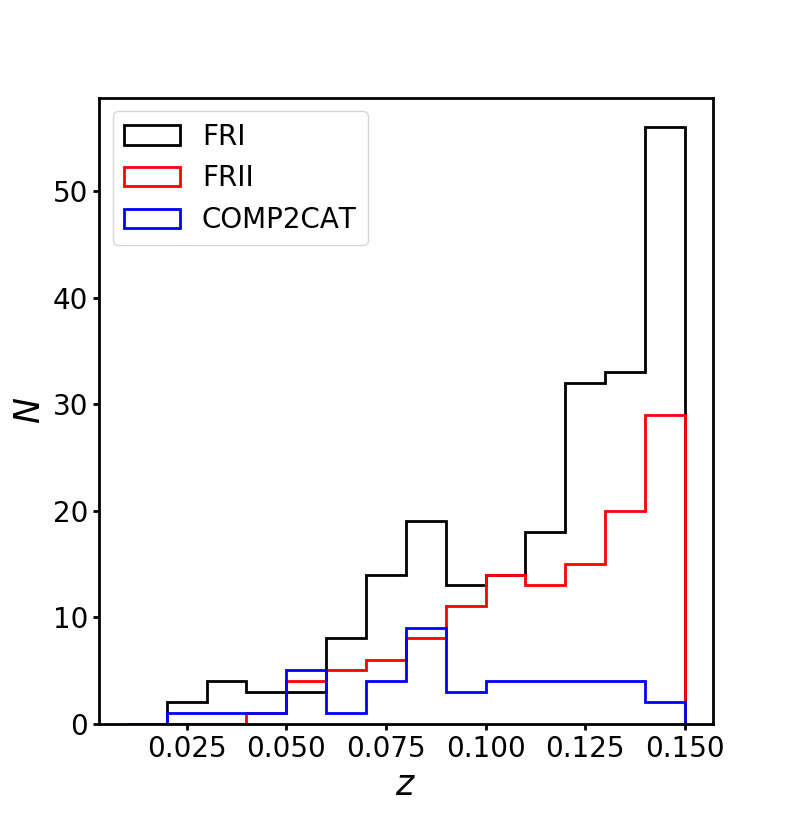}
\includegraphics[width=9.3cm]{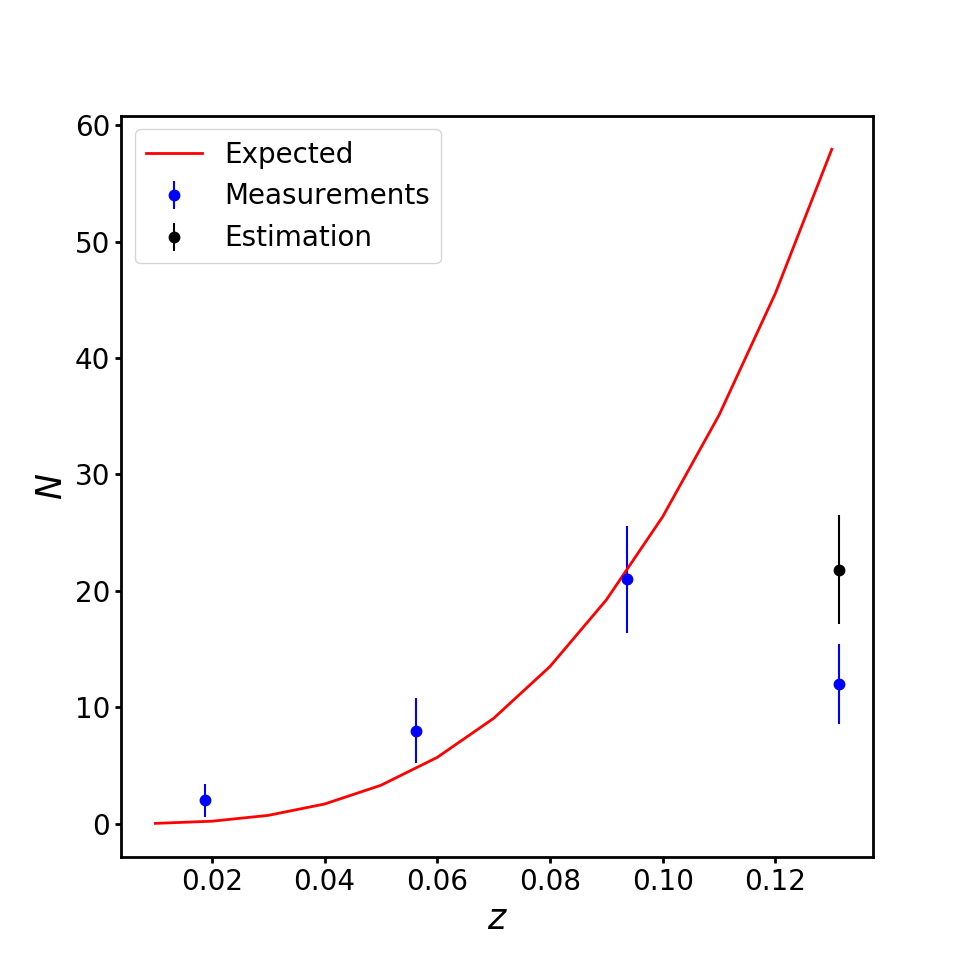}
\caption{Left: Histograms showing the redshift distribution of COMP2$CAT$ (blue), \FR\ (black) and FRII$\sl{CAT}$ (red) sources. Right: Observed redshift distribution (blue dots) and expected redshift distribution (red line). The black dot corresponds to the number of galaxies that would be present if the only effects playing a role are the low luminosities and the small sizes of the sources.}
\label{zhist}
\end{figure*}

The left panel of Fig. \ref{zhist} shows the redshift distribution of COMP2$CAT$ in comparison with the same distribution for \FR\ and FRII{\sl{CAT}}. The COMP2$CAT$ redshift distribution appears rather flat. As can be seen in the right panel of Fig. \ref{zhist}, we binned the redshift distribution and fitted it as $N \propto z^{3}$, leaving out the higher redshift tail of the distribution; i.e., the last bin. In that way, we expect $\sim 91$ sources in the whole redshift range. However, we actually observed only 47$\%$ of them, so COMP2$CAT$ is only $\sim 47 \%$ complete at $z<0.15$.

A possible, simple explanation is that the remaining sources are lost due to either their small sizes, their faint radio luminosities, or a combination of these two effects. To estimate the number of sources potentially lost, we took the radio luminosities and linear sizes of the subsample of COMP2$CAT$ sources at $0.04 \leq z \leq 0.08$ and assigned a random value of redshift in the range $0.11 \leq z \leq 0.15$ to each of them. We computed their radio fluxes and angular sizes at the new redshifts and checked how many of them fall below the sensitivity limit (5 mJy) and angular resolution (5 arcsec) of FIRST. We estimated that the low radio luminosities and the small linear sizes of the sources account for the loss of $\sim 25 \%$ and $\sim 20 \%$ of the sources between $z=0.11$ and $z=0.15$, which corresponds to a loss of $\sim 11 \%$ over the whole catalog. It is important to highlight the fact that this test is highly sensitive to the number of sources taken as the low redshift sample, due to the poor number of sources in the lower redshift range.

Out of the $42 \%$ of lost sources remaining, at least a $10 \%$ can be explained by the incompleteness of the SDSS; since, according to \citet{strauss02} and \citet{montero09}, the SDSS is complete up to $\sim$90\% for apparent magnitudes in the range $14.5 < r < 17.77$. This incompleteness is mostly due to the SDSS fiber collision, that does not allow to place the fibers closer than 55\arcsec\ apart. Indeed the apparent magnitude distribution of COMP2$CAT$ sources ranges from 12.5 to 18 magnitudes, peaking at 15 magnitudes and with only one source with $r > 17.7$ and 6 with $r \leq 14.5$. The $\sim 32 \%$ of loss left is consistent with the uncertainties of our analysis and the possible non-uniform selections performed.

In order to determine the completeness level of our catalog at lower redshift, namely at $z<0.1$, we binned the redshift distribution up to $z=0.1$ and we fitted it leaving out the last bin as previously done. Thus, our catalog is $\sim 83 \%$ complete up to $z=0.1$. Nevertheless, due to the low number of sources at $z<0.1$, this analysis is not statistically significant.  

\begin{figure*}
\includegraphics[width=9.9cm]{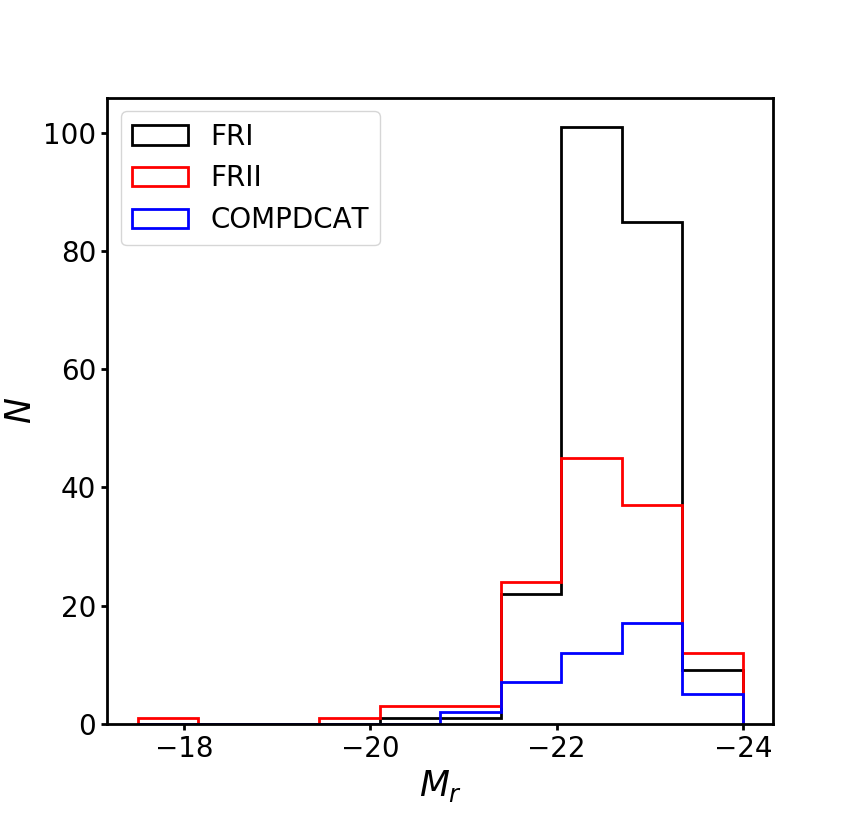}
\includegraphics[width=9.1cm]{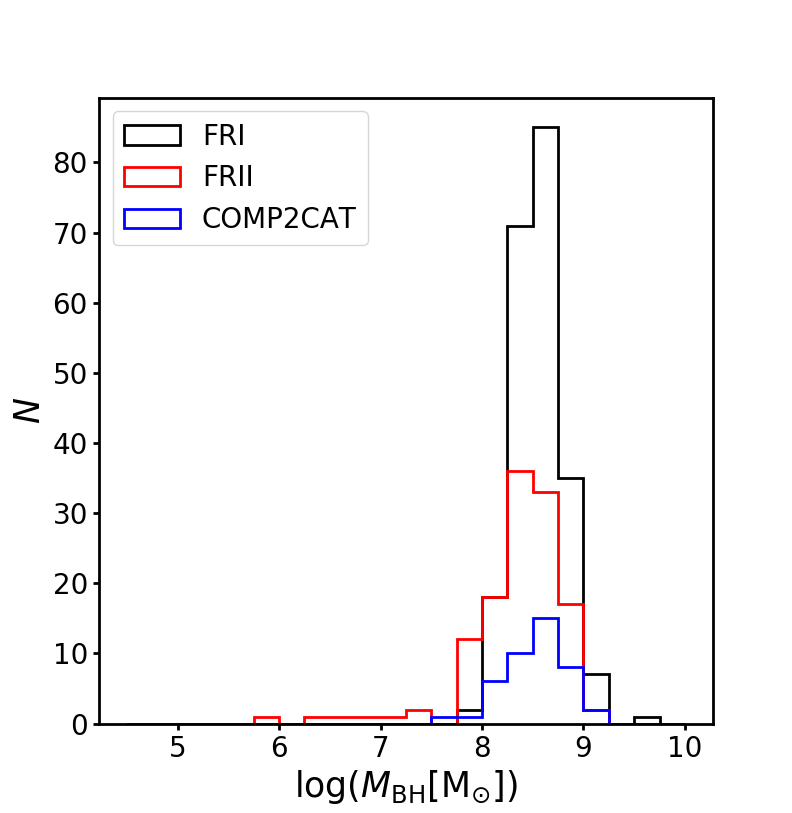}
\caption{Distributions of the $r$ band absolute magnitude (left) and black hole masses (right), for COMP2$CAT$ (blue), for \FR\, (black) and for FRII$\sl{CAT}$ (red).}
\label{mhist}
\end{figure*}

Hosts galaxies of COMP2$CAT$ sources show a distribution of absolute magnitudes in the $r$-band, $M_r$ and of black hole masses, $M_{\rm{BH}}$ (Fig. \ref{mhist}, left and right panels, respectively), similar to \FR\ and FRII{\sl{CAT}} sources. The $M_r$ ranges from $-21$ to $-24$ and peaks at $M_r \sim -22.5$, whereas the black hole mass is in the range of $7.5 \lesssim \log M_{\rm BH} \lesssim 9.5\, \rm{M}_\odot$, peaking at $\sim10^{8.5}\, \rm{M}_\odot$.

$M_{\rm{BH}}$ was computed using its correlation with the stellar velocity dispersion, $\sigma_{*}$, published by \citet{tremaine02}:
\begin{equation}
    \log\big(M_{\rm{BH}}/\rm{M}_{\odot}\big)=\alpha+\beta \log\big(\sigma_{*}/\sigma_{0}\big)
\end{equation}
with $\alpha = 8.13 \pm 0.06$, $\beta = 4.02 \pm 0.32$ and $\sigma_{0} =  200$ km s$^{-1}$. The error in $M_{\rm{BH}}$ is dominated by the spread of the relation, so the $M_{\rm{BH}}$ presented have an uncertainty of a factor $\sim2$.

As previously performed for \FR\ and FRII{\sl{CAT}}, we computed the concentration index, $C_{r}$, defined as the ratio of the radii including the 90\% and the 50\% of the light in the $r$-band, respectively. This index tends to have higher values (i.e. $\gtrsim$ 2.86, according to \citealt{nakamura03,shen03}, or $\gtrsim$ 2.6, according to \citealt{strateva01,kauffmann03b,bell03}) for Early-type galaxies (ETGs) than for Late-type galaxies.

In addition, we also estimated the Dn(4000) index, defined as the ratio of the flux density in the ``red" side of the Ca-II break (4000--4100 \AA) and in the ``blue  side" (3850--3950 \AA) \citep{balogh99}. The Dn(4000) index is lower in the presence of young stars and non-stellar emission and, according to \citet{capetti15}, red galaxies at $z < 0.15$ have $\rm{Dn(4000)} = 1.95 \pm 0.05$.

\begin{figure*}  
\includegraphics[width=9.5cm]{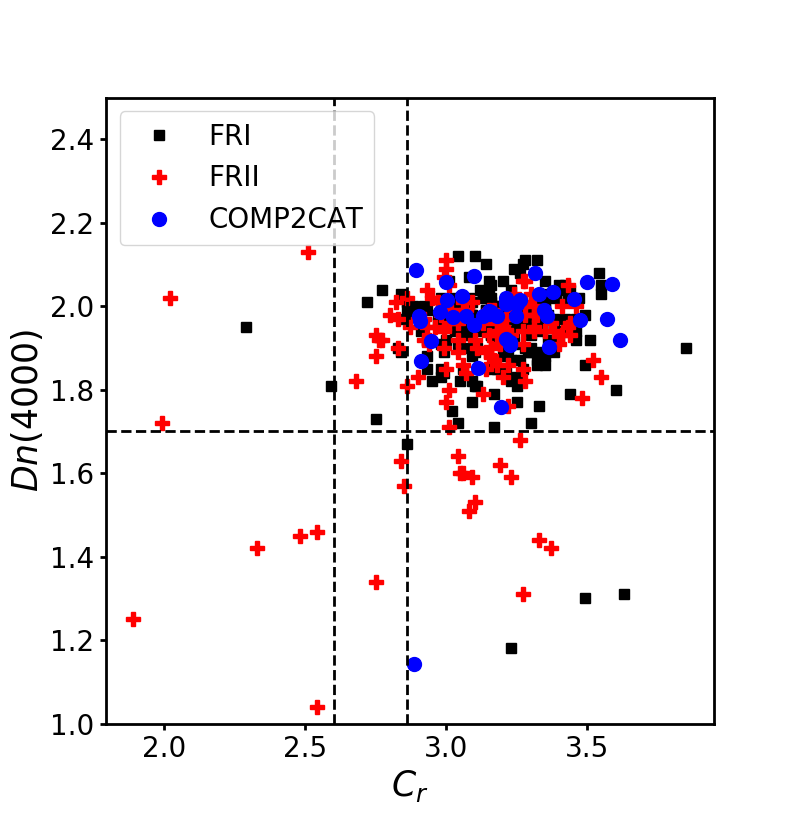}
\includegraphics[width=9.5cm]{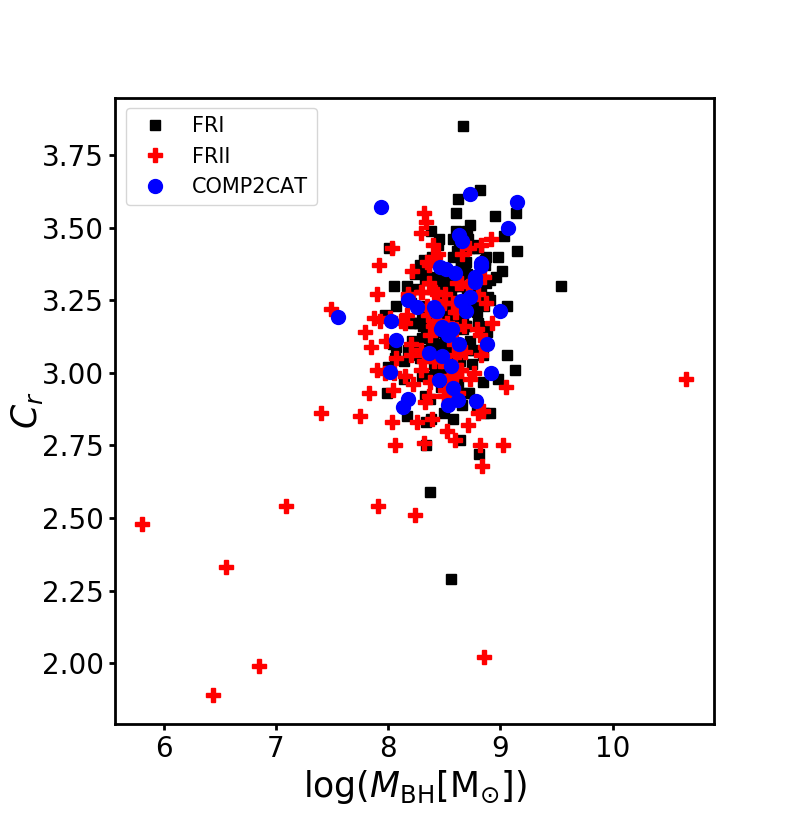}
\caption{Left: Dn(4000) index vs. concentration index $C_r$ for COMP2$CAT$ (blue dots), \FR\ (black squares) and FRII$\sl{CAT}$ (red crosses). The black dashed lines represent the values of $C_r$ and Dn(4000) indices that, according to \citet{nakamura03,shen03} and \citet{strateva01,kauffmann03b,bell03}, and \citet{capetti15} correspond to ETGs. Right: concentration index $C_r$ vs. logarithm of the black hole mass (in solar units) with the same color code that the image on the left.}
\label{crdn}
\end{figure*}

The left panel of Fig. \ref{crdn} (in which the Dn(4000) index versus the $C_{r}$ index are represented for the sources in the three catalogs) shows that most of the sources in COMP2$CAT$ are ETGs since they present high values of both indices.
In the same figure, the right panel shows $C_{r}$ versus the $M_{\rm BH}$. This plot shows no change of $M_{\rm BH}$ with $C_{r}$.

\begin{figure*}
\includegraphics[width=9.5cm]{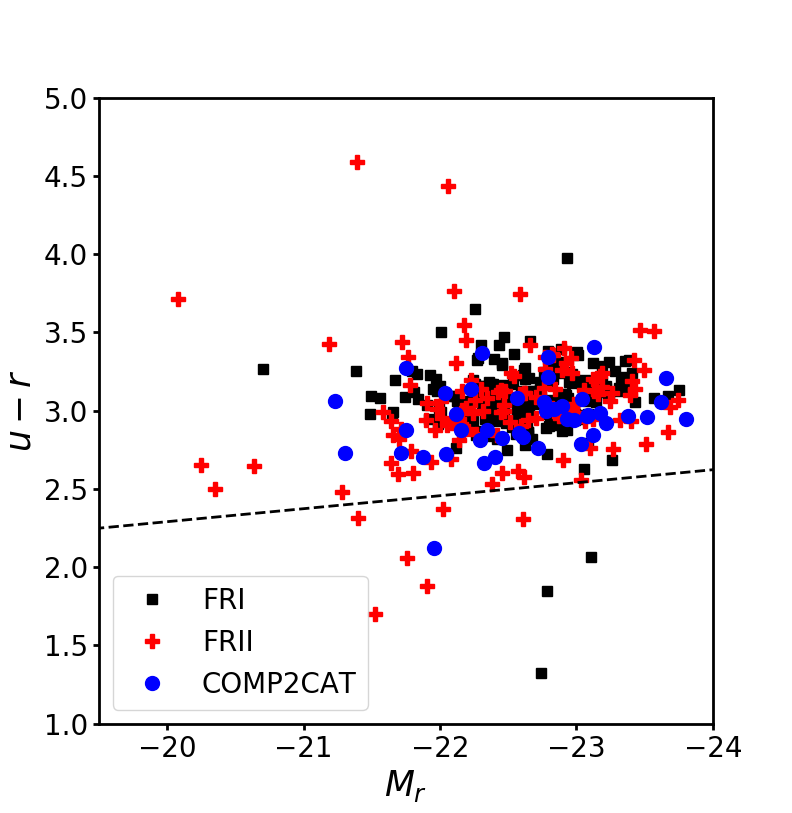} 
\includegraphics[width=9.5cm]{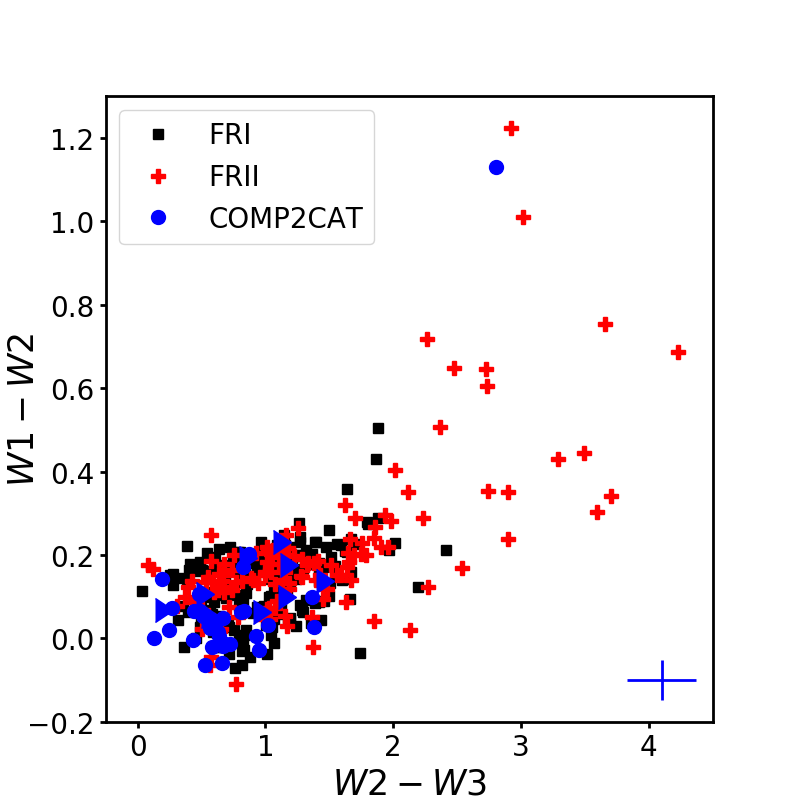}
\caption{Left: $u-r$ color vs. absolute $r$ band magnitude, $M_r$, for COMP2$CAT$, \FR\ and FRII$\sl{CAT}$ hosts (blue dots, black squares and red crosses). The dashed line separates the ``blue'' ETG from the red sequence, following the definition published by \citet{schawinski09}. Right: {\em{WISE}} mid-IR colors of COMP2$CAT$, \FR\ and FRII$\sl{CAT}$ hosts with the same color code as the previous figure. The blue cross in the bottom right corner of the plot represents the average error in the colors.}
\label{mrur}
\end{figure*}

We also show the $u-r$ color of the host galaxy versus its $M_r$ (Fig. \ref{mrur}, left panel) since the $u-r$ color gives information on the properties of the whole source, while the Dn(4000) index only gives information about the region of 3$\arcsec$ in diameter covered by the SDSS spectroscopic aperture. The $u-r$ color was not corrected for galactic extinction, since the correction is $\leq 5\%$. We see that most of the COMP2$CAT$ sources are in the region of red ETGs \citep{schawinski09}. The only source in our sample that is not an ETG according to this diagnostic is the HERG. 

As previously carried out for the other radio galaxy catalogs we also checked the mid-IR colors of selected COMP2$CAT$ sources. To obtain the {\em{WISE}} magnitudes of the sources in COMP2$CAT$, we associated the position of their sources adopting a 3\farcs3 angular separation, which corresponds to the combination of the typical positional uncertainty of the {\em{WISE}} all-sky survey \citep{wright10} and that of the FIRST \citep{dabrusco14, Massaro2014}.

The $WISE$ magnitudes in the [3.4], [4.6], [12], and [22] $\mu$m nominal filters ($W1$, $W2$, $W3$ and $W4$ respectively) are in the Vega system. Their values and those of the colors derived using them have not been corrected for Galactic extinction, due to the fact that it can be considered negligible since it only affects to the magnitude at 3.4 $\mu$m of sources at low Galactic latitudes, and, even in those cases, the correction is less than $\sim3\%$ \citep{dabrusco14}.  

The right panel of Fig. \ref{mrur} is a color-color plot of the COMP2$CAT$, the \FR\ and the FRII{\sl{CAT}} sources. In general, COMP2$CAT$ sources display bluer mid-IR colors than the sources in \FR\ and FRII{\sl{CAT}}. This could be explained by a lower amount of dust in COMP2$CAT$ sources. However, the $W3$ magnitude distributions of COMP2$CAT$, \FR\ and FRII{\sl{CAT}} sources are similar, peaking in all cases at $W3 \sim 12$, while the COMP2$CAT$ $W1$ and $W2$ magnitude distributions seem to peak at lower values ($\sim 12.25$ in both cases) than those for \FR\ and FRII{\sl{CAT}} sources (which peak at $13.25$). Nevertheless, the $W3$ magnitudes of 9 sources (between 11.8 and 12.4 magnitudes) are actually upper limits, so the distribution of $W3$ for COMP2$CAT$ sources could show the same differences (of about $\sim 8\%$) with respect to FRI and FRII sources as the $W1$ and $W2$ magnitudes.

\section{Comparison between optical and radio powers}
\label{sec:discussion}

Here we compare multi-frequency behavior of COMP2$CAT$ sources with that of other radio galaxy catalogs.

\begin{figure}
    \centering
    \includegraphics[width=9.5cm]{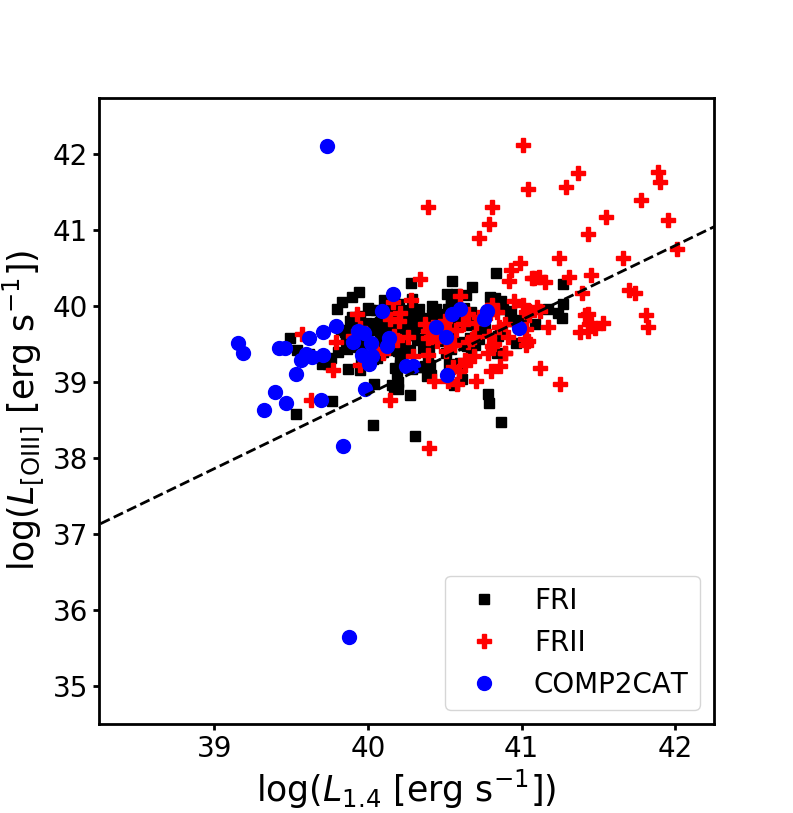}
    \caption{[O III] line luminosity versus radio luminosity at 1.4 GHz for COMP2$CAT$ sources (blue dots), \FR\ sources (black squares) and FRII$CAT$ sources (red crosses). The black dashed line shows the linear correlation between these two quantities derived from the FR~Is of the 3C sample from \citet{buttiglione10}.}
    \label{ropt}
\end{figure}

The comparison of the [O III] line luminosity, $L_{[OIII]}$, to the NVSS radio luminosity at 1.4 GHz, $L_{1.4}$, shown in Fig. \ref{ropt}, is consistent with COMP2$CAT$ sources being the low radio power tail of FR-IIs and highlights the absence of HERGs in it. Having low radio power and showing lower values of $EW_{[OIII]}$ than radio galaxies in FRII$\sl{CAT}$, the line production could be mainly due to stellar processes rather than to the central AGN. This makes the $L_{[OIII]}$ $vs$ $L_{1.4}$ flatter towards the COMP2$CAT$ region. 

\begin{figure}
    \centering
\includegraphics[width=9.5cm]{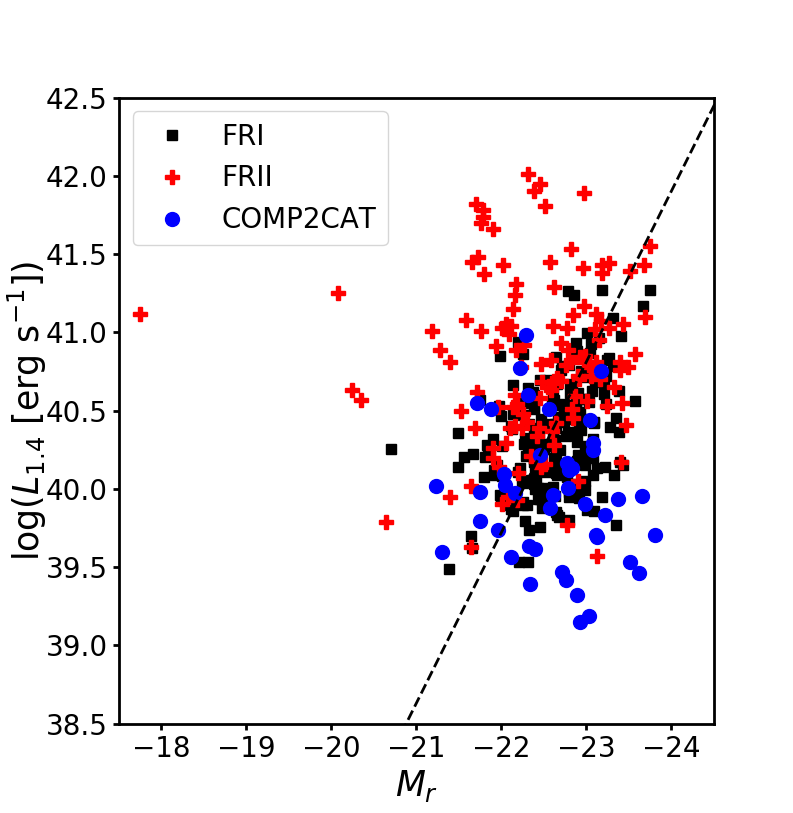}
\caption{Radio luminosity (NVSS) vs. host absolute magnitude , $M_r$, for COMP2$CAT$, \FR\ and  FRII$\sl{CAT}$ (blue dots, black squares and red crosses respectively). The dashed line shows the separation between FR~I and FR~II reported by \citet{ledlow96} to which we applied a correction of 0.34 mag to account for the different magnitude definition and the color transformation between the SDSS and Cousin systems.}
\label{radm}
\end{figure}

COMP2$CAT$ sources lie in the lower part of the optical-radio luminosity plane (a.k.a. Ledlow-Owen plot, see Fig. \ref{radm}). In particular, only $\sim 33 \%$ of COMP2$CAT$ sources lie above the dashed line in Fig. \ref{radm}, which corresponds to that reported in \citet{ledlow96} and marks the separation between the different FR classes of radio galaxies, while for FRIs and FRIIs the number of sources above this separation is $\sim 42 \%$ and $\sim 74 \%$, respectively. The higher fraction of COMP2$CAT$ sources in the FRI region of the Ledlow-Owen plane is consistent with them being LERGs, like FRI$CAT$ sources, and in contrast with the FRII$CAT$ population, which is composed by a $\sim 10 \%$ of HERGs.

\begin{figure}
    \centering
\includegraphics[width=9.5cm]{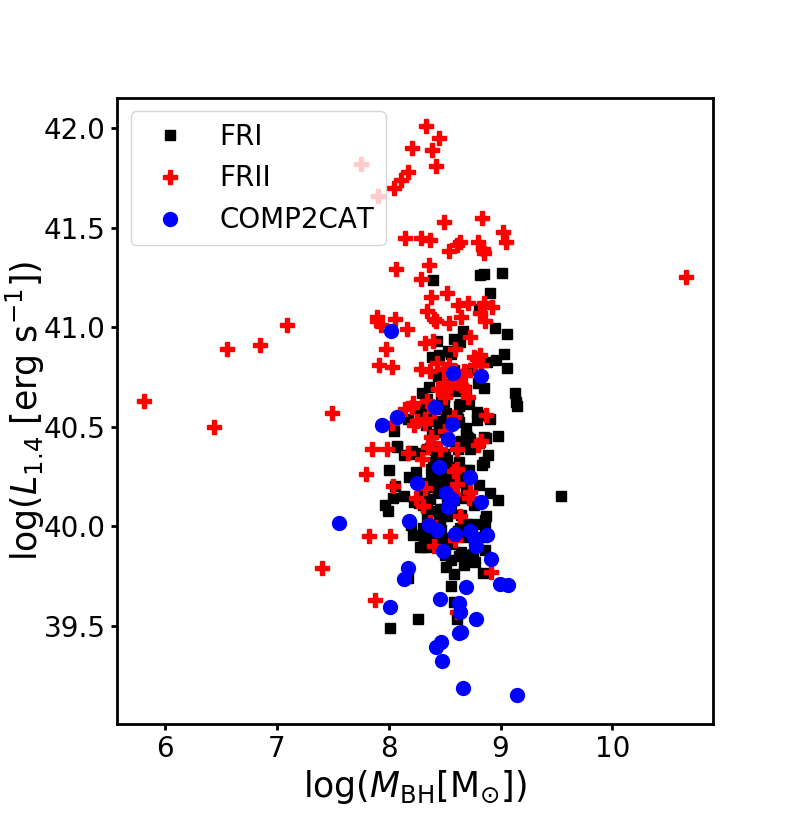}
    \caption{Radio luminosity vs. black hole mass for COMP2$CAT$ (blue dots), \FR\ (black squares) and  FRII$\sl{CAT}$ (red crosses).}
    \label{fig:radmbh}
\end{figure}

Although COMP2$CAT$ sources appear to be the low radio luminosity tail of FR IIs, the three populations (i.e., FRI$CAT$, FRII$CAT$ and COMP2$CAT$ sources) present the same ranges of $M_{BH}$, as shown in Fig. \ref{fig:radmbh}. Thus, the differences in the radio luminosity of these populations could be arise from differences in their accretion rates and/or accretion mechanisms, having COMP2$CAT$ sources less efficient accretion mechanisms than FR IIs. While COMP2$CAT$ sources are almost exclusively LERGs, there are HERGs among the FR IIs in FRII$CAT$, so the hypothesis of the two populations having different accretion rates is consistent with HERGs having more efficient accretion mechanisms than LERGs, as proposed by several authors, such as (\citealt{Hardcastle2007}, \citealt{Balmaverde2008} and \citealt{best12}).

\section{Discussion and conclusions}
\label{sec:conclusions}

We built a catalog of 43 compact double sources selected from the Best \& Heckman (2012) sample restricted to the AGN with redshift $z < 0.15$. Sources were selected if they fulfilled the following criteria: (i) their radio emission does not extend beyond a 30 kpc radius from the position of the optical host galaxy, and (ii) they show FRII-like morphologies, like the sources selected in FRII$\sl{CAT}$ \citep[see][]{Capetti2017II}. This selection was carried out by visually inspecting the FIRST radio images of the sources. Only those identified as compact doubles by, at least, three out of five authors were included in the sample. 

In order to improve the selection, we defined the asymmetric index, $A$ (eq. \ref{eq:asym}), in such a way that very asymmetric sources, i.e., sources with one of the lobes much closer to the SDSS optical position than the other one, have $A \sim 1$, whereas symmetric ones have $A \sim 0$. We excluded from our selection sources with $A \geq 0.5$.

Lastly, we dropped from the selection sources with large-scale NVSS and TGSS radio emission and with FRI morphologies revealed by high-resolution VLA radio maps. The resulting sample of 43 sources constitutes COMP2$CAT$. Since VLA radio maps are not available for all COMP2$CAT$ sources, our sample may still be contaminated by sFRIs; the degree of contamination could be determined with future observations from the VLA Sky Survey (VLASS\footnote{Available at: \newline {\tt https://science.nrao.edu/science/surveys/vlass/vlass}.}; \citealt{Murphy2014}).

Although our aim was to build a complete catalog of compact doubles, COMP2$CAT$ is only $\sim 47 \%$ complete. $\sim 53\%$ of sources are lost due to their low radio luminosities, their small projected linear sizes and the incompleteness of the SDSS. However, we are not missing those with higher luminosities unless they have a projected linear size $\lesssim 10$ kpc; so the fact remains that COMP2$CAT$ sources are consistent with being the lower radio power tail of both FRIs and FRIIs.

Given the estimated incompleteness of the catalog, we would expect to find $\sim 91$ sources up to $z = 0.15$. This number is comparable with the number of FRI (219), FRII (122) and FR0 (108) sources found in this redshift range. Therefore, COMP2$CAT$ sources constitute a significant fraction of the radio sources up to $z = 0.15$.

All but one of COMP2$CAT$ sources are LERGs. This implies that either most compact doubles are LERGs or that HERGs mainly lay at $z > 0.1$, where our catalog is the most incomplete. However, HERGs tend to have higher radio luminosities than LERGs and, therefore, we would expect to find them if they existed at $0.1 < z < 0.15$. This lack of HERGs is consistent with COMP2$CAT$ sources being the lower radio luminosity tail of the FRII sources.

Based on the purely morphological classification presented by \citet{Readhead1995}, COMP2$CAT$ sources can be considered LSOs ($> 15$ kpc) and MSOs ($1 - 15$ kpc). On the other hand, following \citet{Orienti2014} criteria, only one of the COMP2$CAT$ sources could be considered a CSS source; therefore, we conclude that COMP2$CAT$ sources constitute a different population than GPS/CSSs. This lack of GPS/CSSs in the sample could be due to the fact that we miss those high radio luminosity sources with projected linear sizes smaller than 10 kpc (CSOs/MSOs), which could be classified as CSSs.

The differences in the position of COMP2$CAT$ sources with respect to FRIs and FRIIs in the $L_{\rm{[OIII]}}$ $vs$ $L_{1.4}$, $L_{1.4}$ $vs$ $M_{\rm{BH}}$ planes and in the Ledlow-Owen plot are consistent with COMP2$CAT$ sources having lower radio luminosities than FR Is and FR IIs and with COMP2$CAT$ sources being mostly LERGs, like FRI$CAT$ sources and in contrast with FRII$CAT$ sources, which include a $\sim 10 \%$ of HERGs. Thus, these discrepancies between COMP2$CAT$ sources and FR IIs could stem from differences in the accretion between LERGs and HERGs. The accretion of LERGs is indeed thought to be less efficient than that of HERGs. Were this hypothesis correct, COMP2$CAT$ sources would be a population of radio galaxies with the lowest accretion rates.

An additional step to understand COMP2$CAT$ sources would be to carry out a complete multi-frequency study of the catalog, including observations at low radio-frequencies with the Low Frequency Array (LOFAR) and X-ray observations. Currently, only two COMP2$CAT$ sources were observed with $Chandra$ (SDSS J081023.27+421625.8 and SDSS J113305.52+592013.7), both of them in galaxy clusters, identified using the 7th and 4th SDSS releases by \citet{Yang2005, Yang2007} and \citet{Koester2007}, respectively. Another two sources were observed with $XMM-NEWTON$ (SDSS J095341.37+014202.3 and SDSS J103801.77+414625.8), also in galaxy clusters identified using the SDSS by \citet{Shen2008} and \citet{Tempel2012}. Lastly, one source more was observed with $SWIFT$ (SDSS J101944.27-003817.8). This source is also part of a galaxy cluster and was identified using the SDSS by \citet{Tempel2012}. More $Chandra$ observations are needed in order to characterize completely these sources. 

On the other hand, LOFAR and VLASS observations would enable us to characterize source radio spectra as well as to study their morphology at higher resolution, to eventually quantify the degree of contamination of our sample by sFRIs.

Lastly, this catalog can be used in the future to better understand the role of compact double sources in the general evolutionary scheme of radio sources. In particular, using LOFAR data we could compare our catalog to those selected at low radio frequencies as recently done by \citet{Hardcastle2018}. Their sample lists 23244 radio-loud AGN, with 150 MHz luminosities ranging from 10$^{36}$ to 10$^{45}$ erg s$^{-1}$ and projected linear sizes between 1 pc and 1 Mpc, obtained from the LOFAR Two Metre Sky Survey (LoTSS).

\begin{acknowledgements}
We are grateful to the anonymous referee for her/his constructive and valuable comments that improved the presentation of our manuscript.
A. J. acknowledges the financial support (MASF\_CONTR\_FIN\_18\_01) from the Italian National Institute of Astrophysics under the agreement with the Instituto de Astrofisica de Canarias for the ``Becas Internacionales para Licenciados y/o Graduados Convocatoria de 2017".
This work is supported by the ``Departments of Excellence 2018 - 2022" Grant awarded by the Italian Ministry of Education, University and Research (MIUR) (L. 232/2016).
This research has made use of resources provided by the Compagnia di San Paolo for the grant awarded on the BLENV project (S1618\_L1\_MASF\_01) and by the Ministry of Education, Universities and Research for the grant MASF\_FFABR\_17\_01.
F.M. acknowledges financial contribution from the agreement ASI-INAF n.2017-14-H.0
A.P. acknowledges financial support from the Consorzio Interuniversitario per la Fisica Spaziale (CIFS) under the agreement related to the grant MASF\_CONTR\_FIN\_18\_02.
L.O. acknowledges partial support from the INFN Grant InDark.
A.S. was supported by the NASA contract  contract NAS8-03060 (Chandra X-ray Center)
\end{acknowledgements}

\appendix
\clearpage
\newpage
\onecolumn
\section{Excluded sources based on their VLA emission.}
\label{ap:VLA}
\begin{figure}[h]
\includegraphics[width=6.cm,height=6.cm]{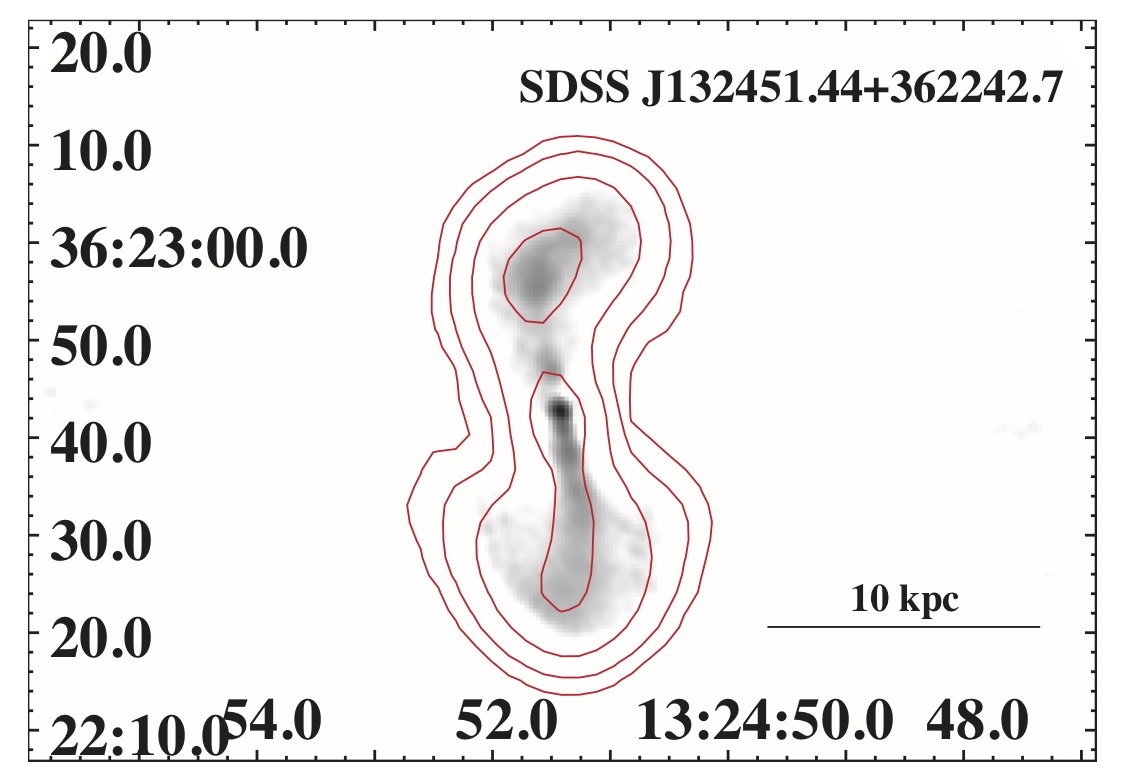}
\includegraphics[width=6.cm,height=6.cm]{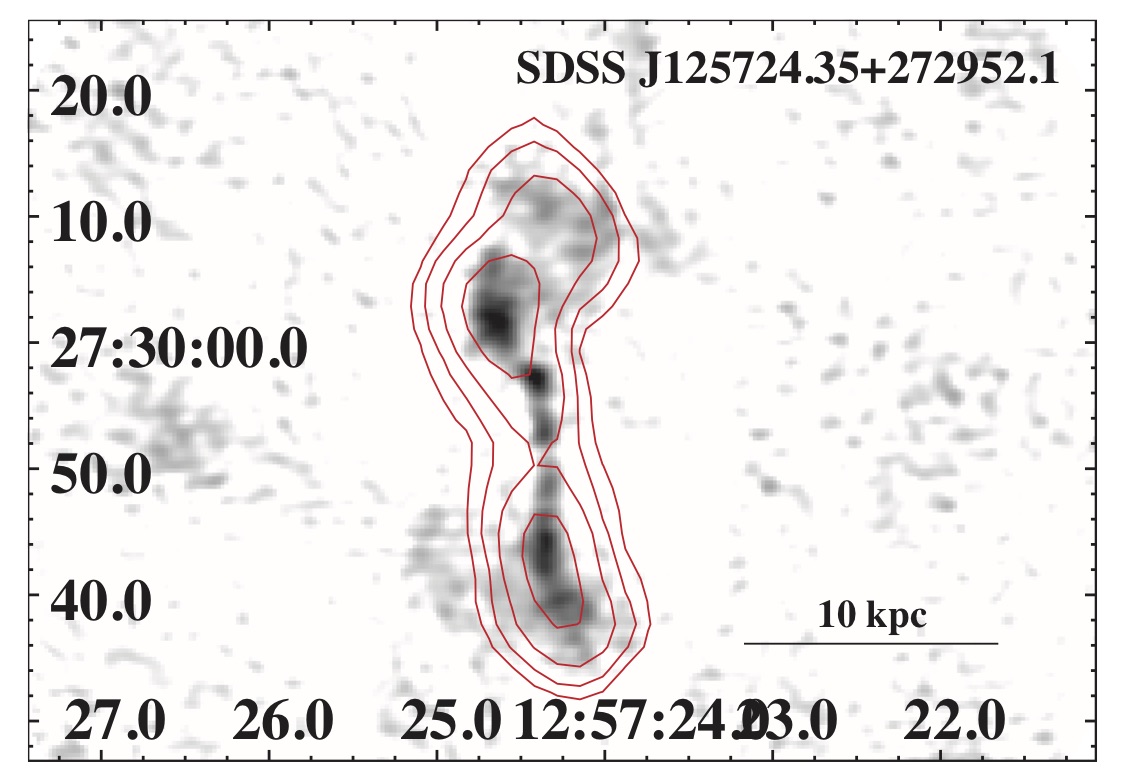}
\includegraphics[width=6.cm,height=6.cm]{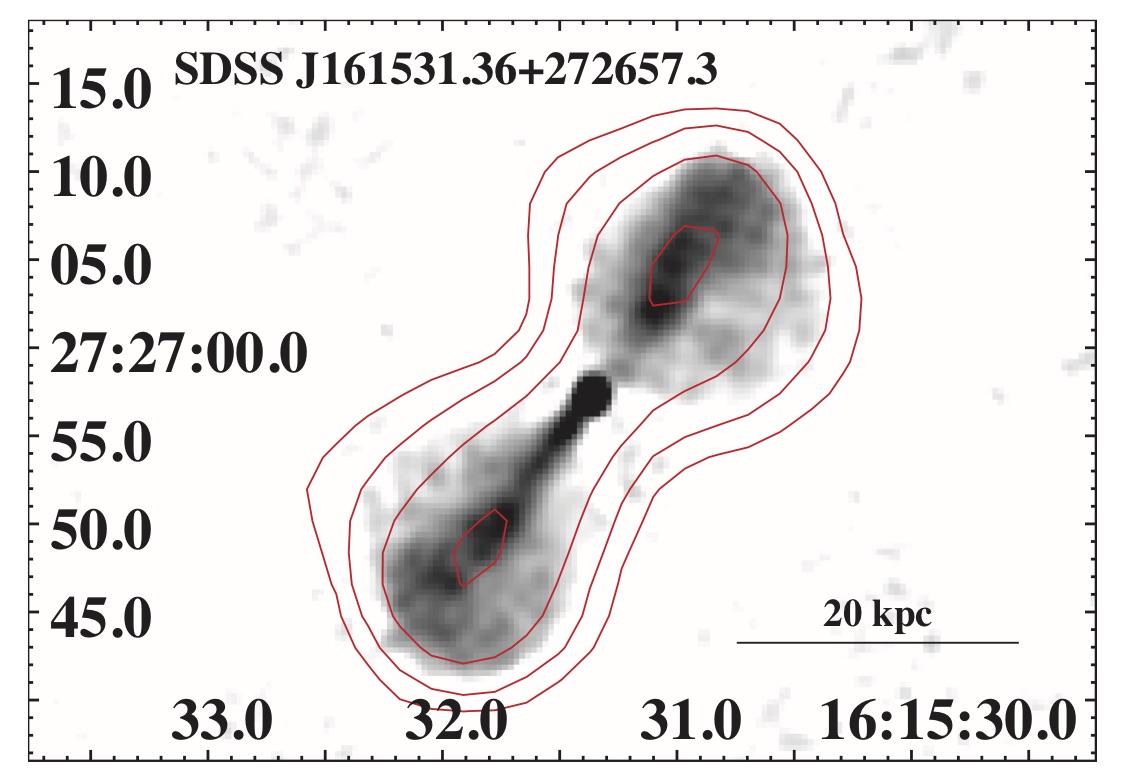}
\includegraphics[width=6.cm,height=6.cm]{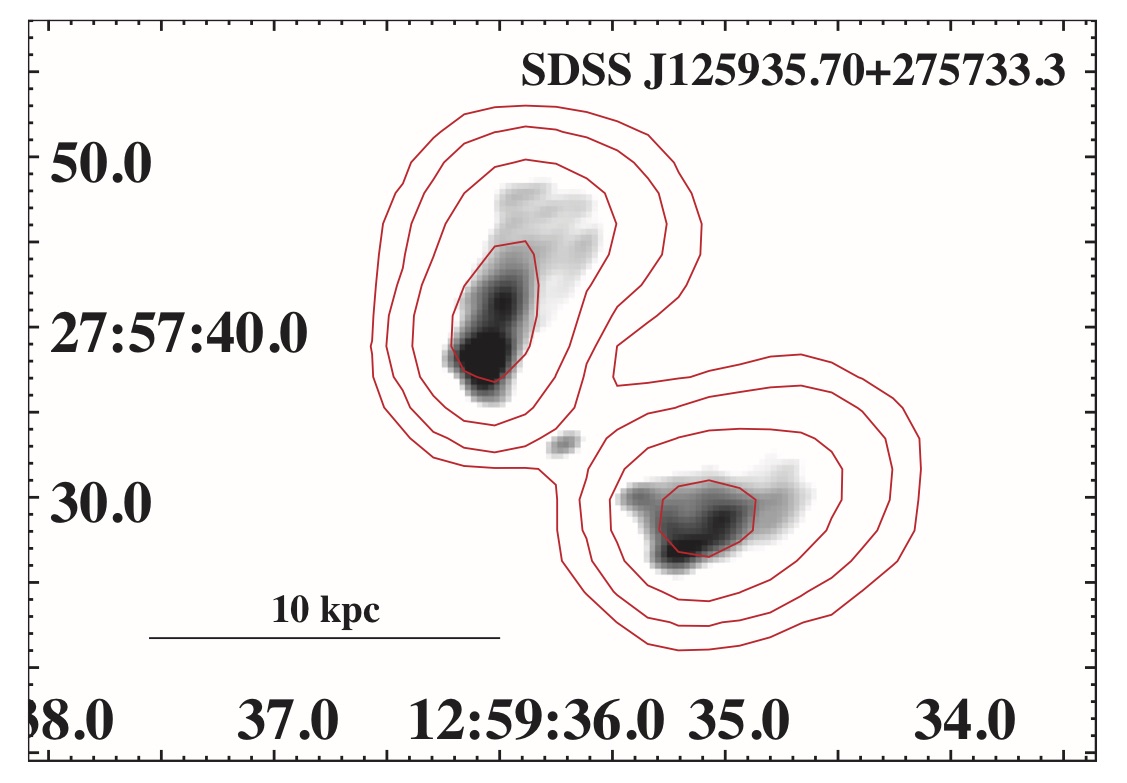}
\includegraphics[width=6.cm,height=6.cm]{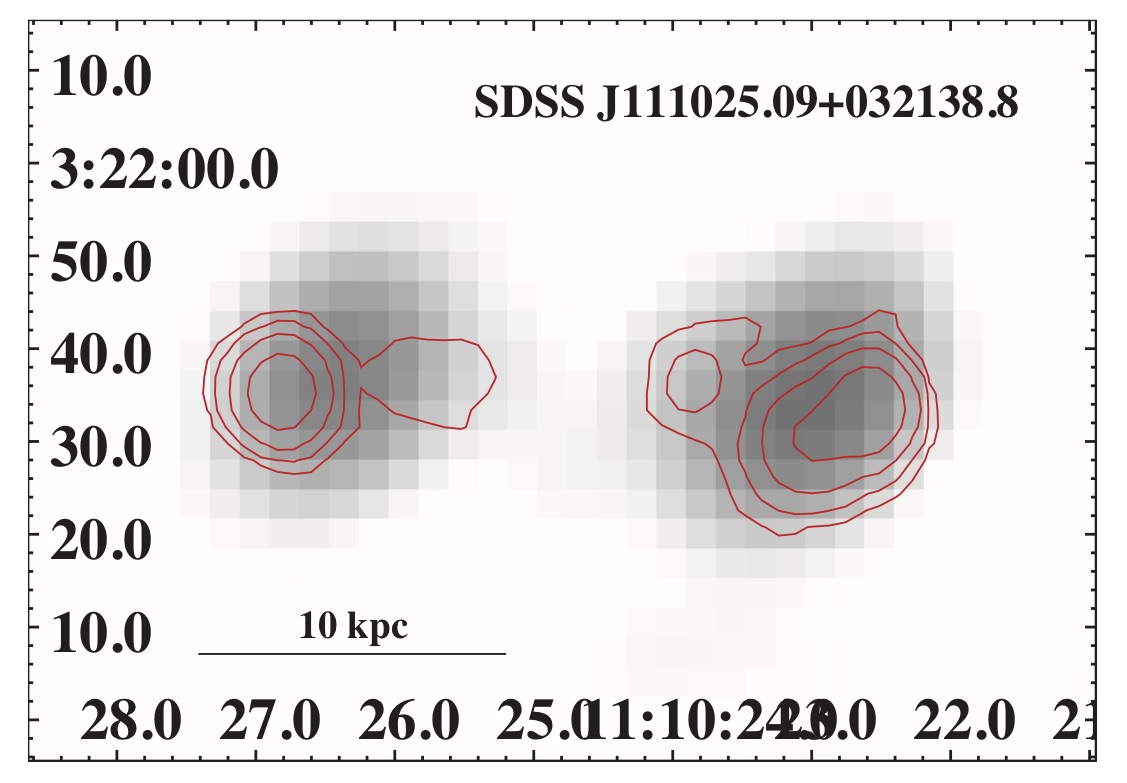}
\caption{Images of sources with different radio morphologies seen in VLA. Contours are drawn using the parameters shown in App. \ref{ap:contours}, in Table \ref{tab:contoursexc}. The first three images correspond to small FRI sources like the ones selected in \citealt{Capetti2017I} (the first one was excluded from our sample because of its FIRST morphology). The fourth image corresponds to a WAT source (see \citealt{owen76}) and the last image shows a star-forming galaxy excluded from the sample.}
\label{excluded}
\end{figure}

\newpage
\section{Sources with extended emisson in NVSS and TGSS.}
\label{ap:NVSS}

\begin{figure}[h]
\includegraphics[width=6.cm,height=6.cm]{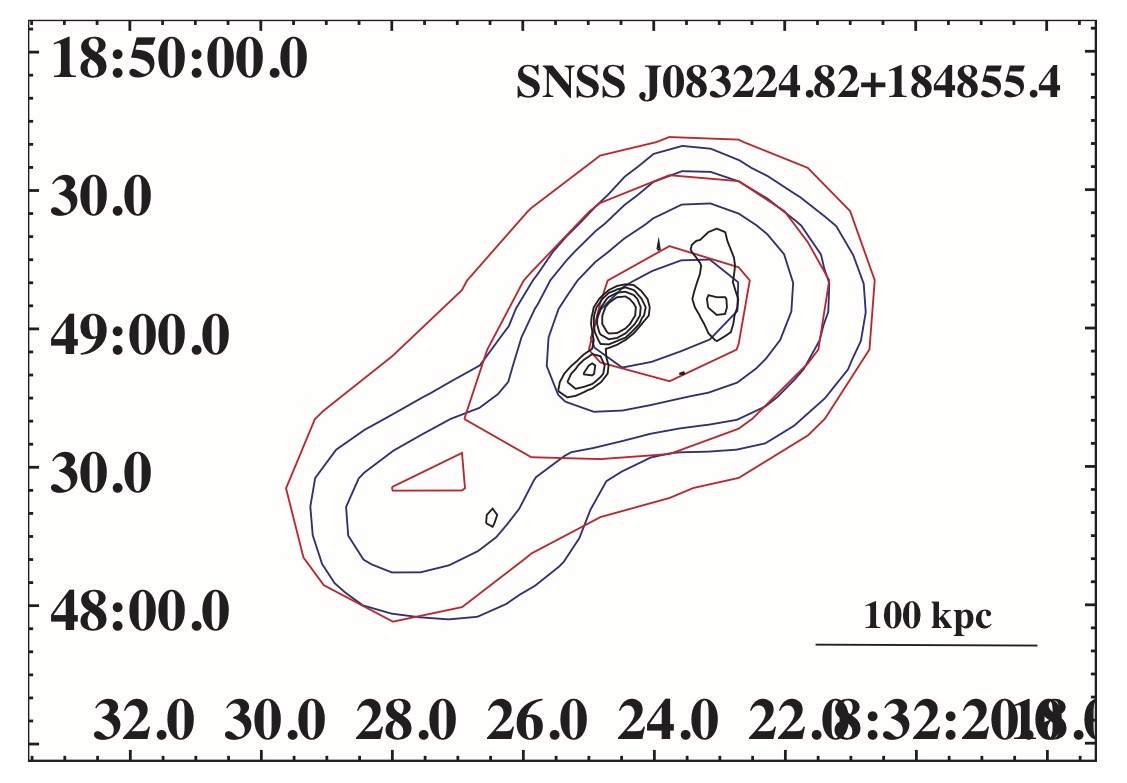}
\includegraphics[width=6.cm,height=6.cm]{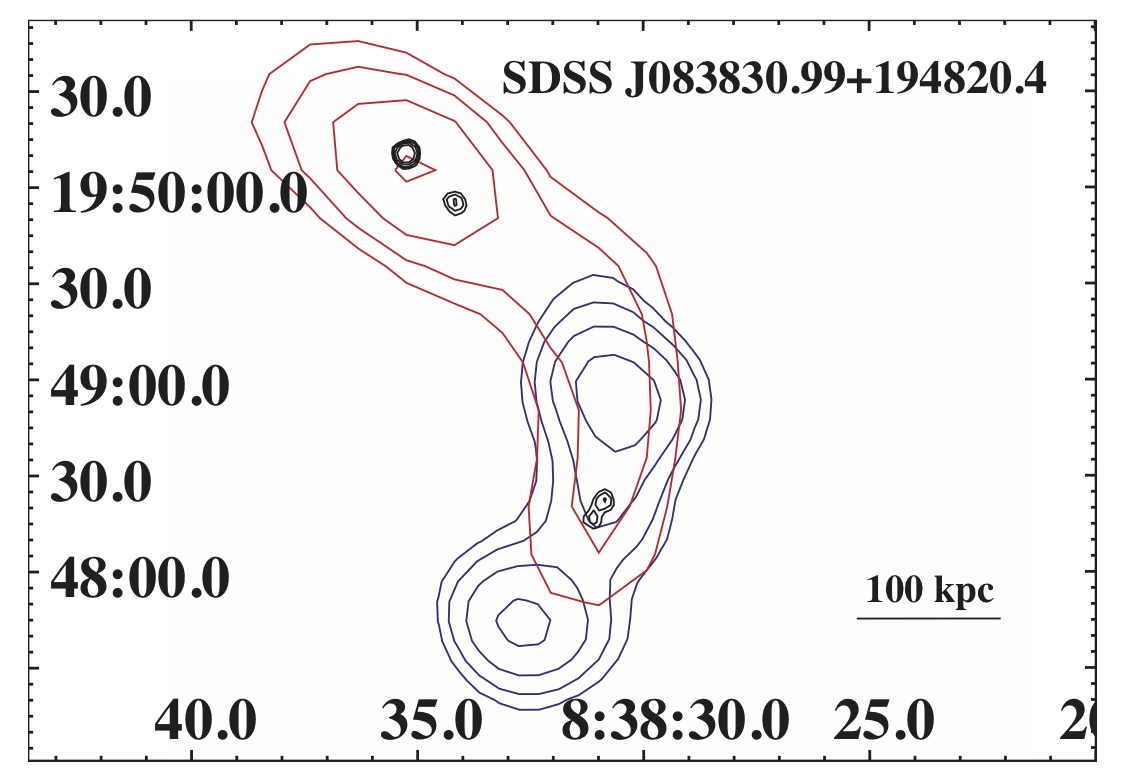}
\includegraphics[width=6.cm,height=6.cm]{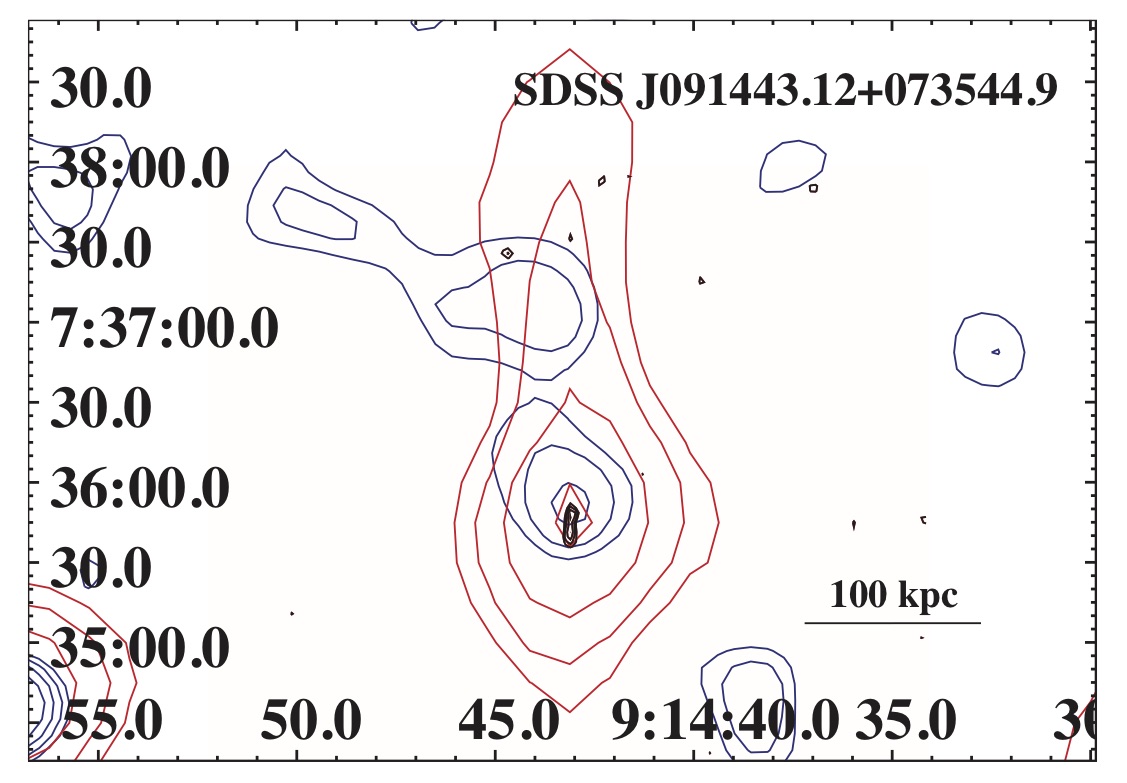}
\includegraphics[width=6.cm,height=6.cm]{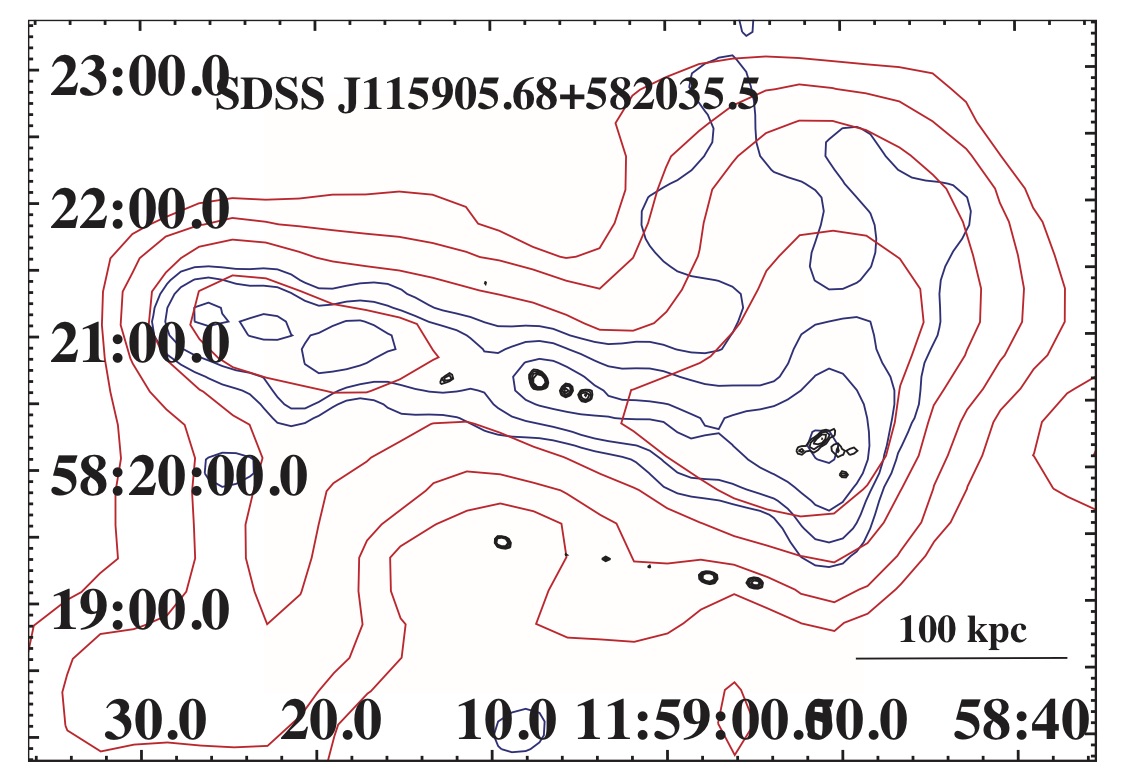}
\includegraphics[width=6.cm,height=6.cm]{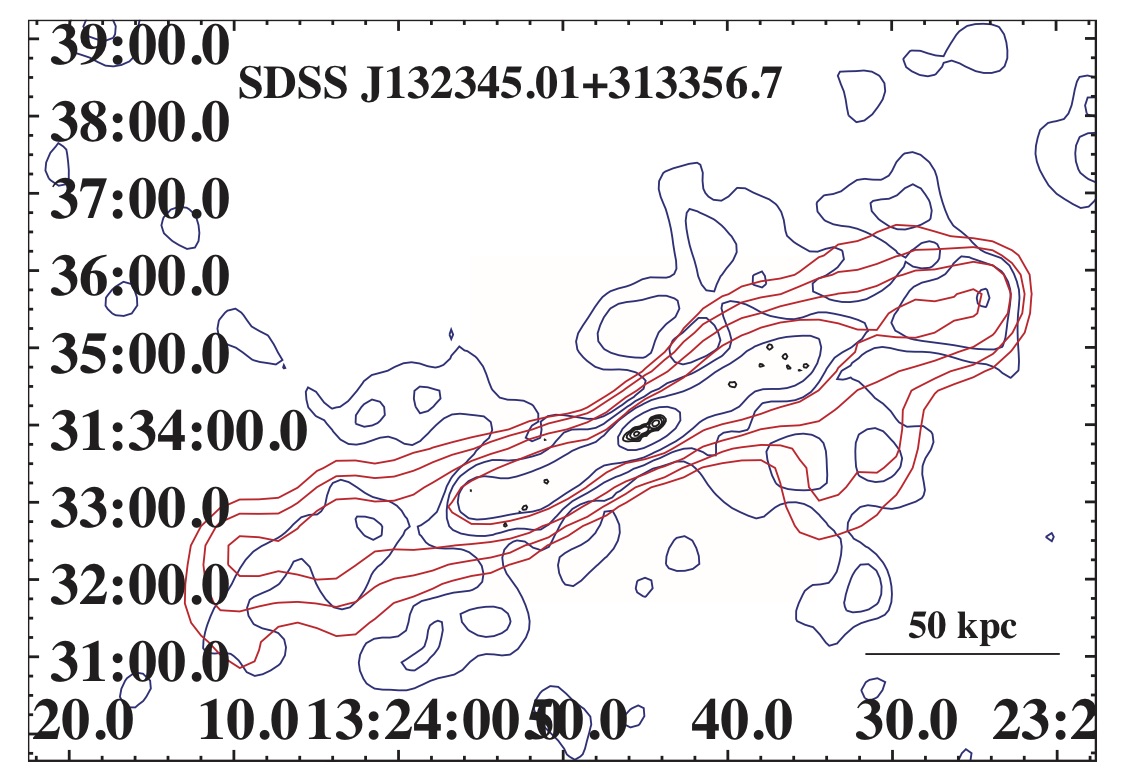}
\includegraphics[width=6.cm,height=6.cm]{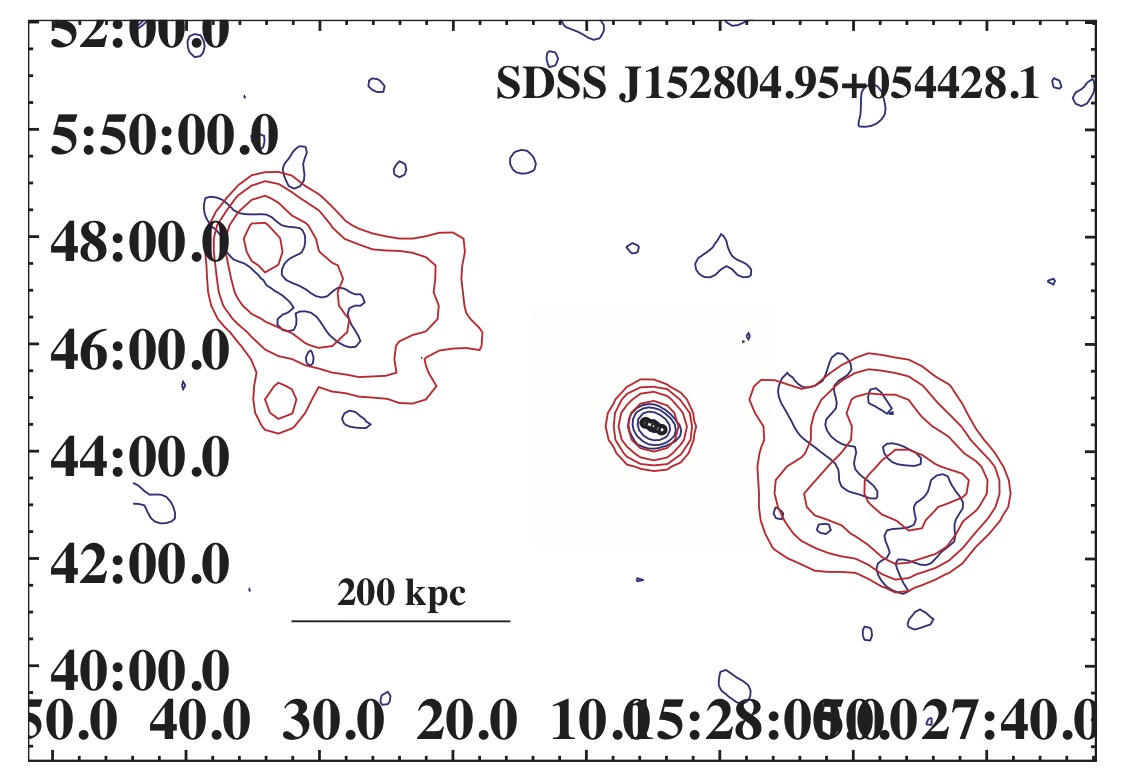}

\includegraphics[width=6.3cm,height=6.3cm]{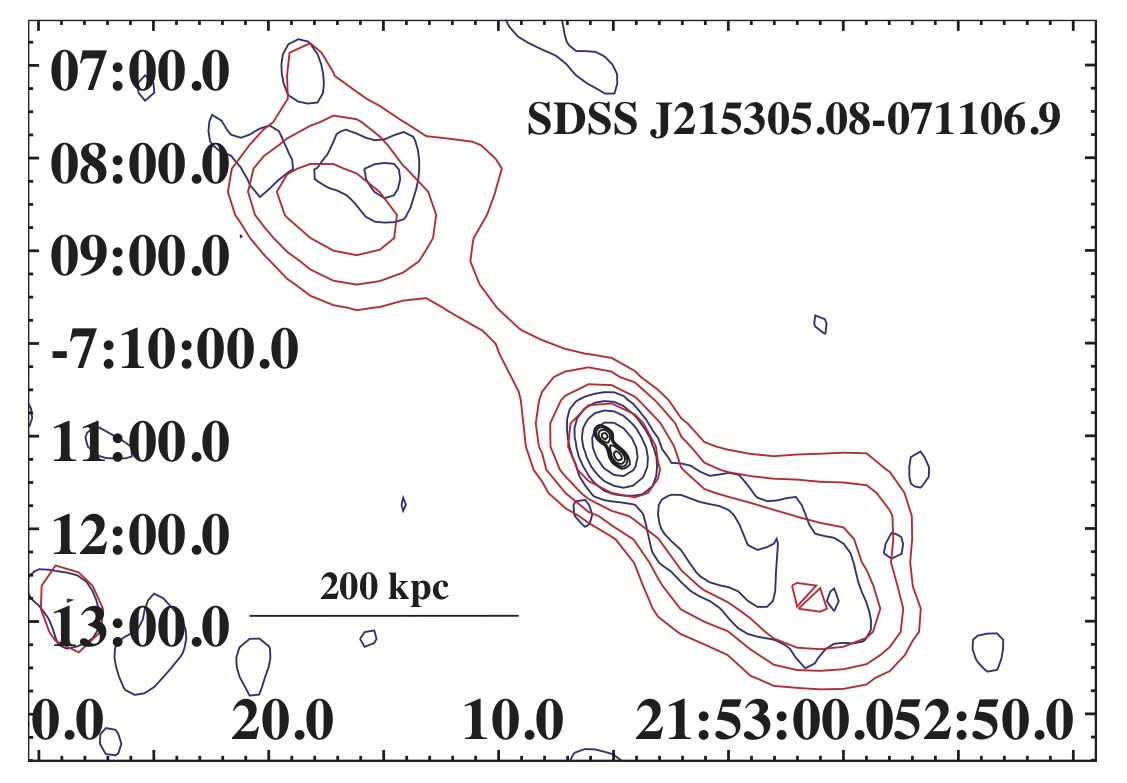}
\caption{Images of sources with extended emission detected in NVSS. Black, red and blue contours correspond to the emission seen in FIRST, NVSS and TGSS. Contours are drawn using the parameters shown in App. \ref{ap:contours}, in Table \ref{tab:contoursext}.}
\label{extended}
\end{figure}

\newpage
\section{Porperties of COMP2$CAT$ sources.}
\label{ap:properties}
\begin{center}
\begin{longtable}{lrrrrrrrrrrrrr}
\caption[Properties of the COMP2$CAT$ sources.]{Properties of COMP2$CAT$ sources.} 
\label{tab:compcat} \\
\hline
  \multicolumn{1}{c}{SDSS Name} &
  \multicolumn{1}{c}{$z$} &
  \multicolumn{1}{c}{$S_{1.4}$}&
  \multicolumn{1}{c}{$S_{\rm{[OIII]}}$}& 
  \multicolumn{1}{c}{$M_{r}$} &
  \multicolumn{1}{c}{Dn(4000)} &
  \multicolumn{1}{c}{$\sigma_{*}$} &
  \multicolumn{1}{c}{$C_r$} &
  \multicolumn{1}{c}{$L_{1.4}$} &
  \multicolumn{1}{c}{$L_{\rm{[OIII]}}$} &
  \multicolumn{1}{c}{$M_{\rm{BH}}$} &  
  \multicolumn{1}{c}{$LS$} &
  \multicolumn{1}{c}{$A$} &
  \multicolumn{1}{c }{$\alpha$} \\
\hline
  J073600.87+273926.0 & 0.079 & 12.0 & 18.2 & -22.766 & 1.99 & 3.15 & 243 & 39.42 & 39.46 & 8.5 & 14.6 & 0.34 & 0.70\\
  J074132.98+475215.6 & 0.127 & 12.4 & 0.0 & -22.579 & 2.02 & 3.06 & 245 & 39.88 & 35.64 & 8.5 & 29.2 & 0.06\\
  J074641.45+184405.4 & 0.051 & 17.4 & 38.4 & -23.037 & 2.02 & 3.45 & 271 & 39.19 & 39.39 & 8.7 & 12.1 & 0.21 & 0.66\\
  J081023.27+421625.8 & 0.064 & 10.3 & 33.0 & -22.93 & 2.05 & 3.59 & 358 & 39.15 & 39.51 & 9.1 & 14.3 & 0.13\\
  J081104.30+355908.3 & 0.082 & 58.6 & 23.0 & -22.835 & 1.98 & 3.02 & 256 & 40.14 & 39.58 & 8.6 & 11.8 & 0.03\\
  J082033.79+395142.4 & 0.102 & 25.4 & 3.0 & -21.751 & 2.02 & 3.21 & 238 & 39.98 & 38.91 & 8.4 & 30.7 & 0.05 & 0.53\\
  J083053.58+231035.7 & 0.145 & 13.0 & 5.7 & -21.231 & 1.76 & 3.19 & 144 & 40.02 & 39.51 & 7.6 & 25.8 & 0.47 & 0.67\\
  J084517.83+303027.4 & 0.106 & 6.0 & 2.5 & -22.342 & 1.92 & 3.21 & 237 & 39.39 & 38.87 & 8.4 & 44.2 & 0.16\\
  J090311.14+540351.6 & 0.083 & 73.0 & 9.5 & -23.084 & 2.01 & 3.26 & 281 & 40.25 & 39.22 & 8.7 & 11.8 & 0.01 & 0.68\\
  J091134.75+125538.1 & 0.05 & 394.6 & 67.1 & -21.878 & 1.97 & 3.57 & 179 & 40.51 & 39.6 & 7.9 & 40.0 & 0.02 & 0.69\\
  J095341.37+014202.3 & 0.098 & 9.8 & 5.1 & -23.519 & 1.98 & 2.9 & 290 & 39.53 & 39.1 & 8.8 & 40.3 & 0.26 & 1.21\\
  J100622.41+301332.9 & 0.114 & 30.7 & 42.7 & -22.775 & 1.98 & 3.36 & 249 & 40.17 & 40.16 & 8.5 & 17.1 & 0.06 & 0.36\\
  J101653.82+002857.0 & 0.116 & 10.9 & 3684.6 & -21.957 & 1.14 & 2.88 & 201 & 39.73 & 42.12 & 8.1 & 34.0 & 0.28 & 0.53\\
  J101944.27-003817.8 & 0.094 & 9.3 & 12.7 & -23.621 & 1.96 & 2.91 & 266 & 39.46 & 39.45 & 8.6 & 16.5 & 0.2 & 0.57\\
  J103801.77+414625.8 & 0.125 & 22.9 & 7.2 & -22.791 & 0.0 & 3.37 & 298 & 40.12 & 39.47 & 8.8 & 22.3 & 0.16 & 0.76\\
  J103842.52+120315.6 & 0.092 & 31.5 & 20.9 & -22.156 & 1.92 & 3.61 & 282 & 39.98 & 39.65 & 8.7 & 18.9 & 0.02 & 0.51\\
  J104254.02+282559.0 & 0.055 & 40.4 & 52.0 & -22.402 & 1.97 & 3.47 & 266 & 39.62 & 39.58 & 8.6 & 27.4 & 0.21 & 0.46\\
  J111109.58+393552.0 & 0.078 & 40.4 & 31.0 & -23.378 & 2.03 & 3.33 & 289 & 39.94 & 39.68 & 8.8 & 8.5 & 0.07 & 0.56\\
  J113305.52+592013.7 & 0.133 & 13.6 & -3.5 & -23.655 & 1.96 & 3.1 & 307 & 39.96 &  & 8.9 & 17.1 & 0.14 & 0.91\\
  J113643.49+545446.8 & 0.055 & 41.8 & 29.1 & -22.321 & 1.9 & 3.37 & 242 & 39.64 & 39.33 & 8.5 & 10.0 & 0.02 & 0.55\\
  J115050.98-031113.0 & 0.129 & 52.6 & 2.8 & -22.567 & 1.98 & 3.15 & 257 & 40.51 & 39.09 & 8.6 & 20.8 & 0.09 & 0.48\\
  J122208.81+073329.6 & 0.137 & 14.3 & 3.4 & -22.788 & 1.98 & 3.07 & 229 & 40.01 & 39.24 & 8.4 & 25.7 & 0.23 & 0.81\\
  J125319.21+475335.2 & 0.139 & 16.9 & 16.4 & -22.037 & 1.98 & 3.13 & 252 & 40.09 & 39.93 & 8.5 & 16.1 & 0.04 & 0.39\\
  J130107.54-032652.5 & 0.083 & 112.6 & 30.1 & -23.041 & 2.09 & 2.89 & 252 & 40.44 & 39.72 & 8.5 & 16.9 & 0.07 & 0.45\\
  J131705.93+435713.2 & 0.052 & 43.7 & 36.6 & -21.305 & 2.02 & 3.0 & 187 & 39.59 & 39.37 & 8.0 & 23.4 & 0.19 & 0.5\\
  J131945.31+603043.0 & 0.07 & 208.4 & 65.9 & -21.714 & 1.85 & 3.11 & 194 & 40.55 & 39.9 & 8.1 & 26.7 & 0.17 & 0.75\\
  J132031.47-012718.5 & 0.083 & 15.4 & 11.5 & -22.119 & 2.07 & 3.1 & 267 & 39.57 & 39.29 & 8.6 & 32.5 & 0.04\\
  J132602.39+364759.3 & 0.054 & 957.0 & 71.5 & -22.292 & 1.98 & 3.18 & 188 & 40.98 & 39.71 & 8.0 & 21.4 & 0.11 & 0.62\\
  J132649.30+164948.0 & 0.08 & 41.0 & 14.5 & -22.61 & 1.99 & 3.35 & 261 & 39.96 & 39.36 & 8.6 & 25.2 & 0.01 & 0.37\\
  J133917.34-015048.7 & 0.089 & 58.2 & -0.0 & -22.456 & 1.91 & 3.23 & 215 & 40.22 &  & 8.3 & 12.8 & 0.34 & 0.45\\
  J135338.43+360802.4 & 0.027 & 128.9 & 32.6 & -22.716 & 1.98 & 3.25 & 269 & 39.47 & 38.72 & 8.7 & 39.0 & 0.07 & 0.5\\
  J135347.34+515734.3 & 0.132 & 90.4 & 18.7 & -22.225 & 1.92 & 2.95 & 258 & 40.77 & 39.94 & 8.6 & 17.0 & 0.2 & 0.62\\
  J144647.43+032527.1 & 0.125 & 18.3 & 5.4 & -22.047 & 1.87 & 2.91 & 206 & 40.03 & 39.35 & 8.2 & 12.2 & 0.23\\
  J144731.24+330606.2 & 0.088 & 73.0 & 8.4 & -23.076 & 1.99 & 2.98 & 240 & 40.3 & 39.21 & 8.4 & 10.7 & 0.2 & 0.52\\
  J145604.88+472712.4 & 0.087 & 211.4 & 35.9 & -23.175 & 2.03 & 3.38 & 298 & 40.75 & 39.84 & 8.8 & 23.8 & 0.04 & 0.61\\
  J145858.83+130145.9 & 0.112 & 10.7 & 1.8 & -23.13 & 2.01 & 3.21 & 276 & 39.69 & 38.76 & 8.7 & 18.9 & 0.17\\
  J151135.87+191228.0 & 0.08 & 22.6 & 29.0 & -23.805 & 2.06 & 3.5 & 342 & 39.71 & 39.67 & 9.1 & 22.1 & 0.16 & 0.74\\
  J155749.61+161836.6 & 0.037 & 113.4 & 70.4 & -23.118 & 2.0 & 3.21 & 328 & 39.71 & 39.35 & 9.0 & 19.4 & 0.19 & 0.43\\
  J160818.19+374335.3 & 0.102 & 5.6 & 1.6 & -22.894 & 1.99 & 3.16 & 244 & 39.32 & 38.63 & 8.5 & 15.2 & 0.06\\
  J162401.10+204018.4 & 0.1 & 17.2 & 21.2 & -21.749 & 2.01 & 3.25 & 205 & 39.79 & 39.74 & 8.2 & 21.1 & 0.26 & 0.33\\
  J164452.86+341251.3 & 0.085 & 31.2 & 18.6 & -22.982 & 2.08 & 3.31 & 290 & 39.9 & 39.53 & 8.8 & 14.8 & 0.1 & 0.54\\
  J165644.31+324321.8 & 0.147 & 48.4 & 15.9 & -22.309 & 1.91 & 3.23 & 235 & 40.6 & 39.97 & 8.4 & 20.1 & 0.07 & 0.45\\
  J171659.25+321445.0 & 0.111 & 15.3 & 0.5 & -23.219 & 2.06 & 3.0 & 313 & 39.83 & 38.16 & 8.9 & 23.4 & 0.23 & 0.64\\
  \hline
  \caption{Column description: (1) source name; (2) redshift; (3) NVSS 1.4 GHz flux density [mJy]; (4) [O III] flux density [10$^{-17}$ erg cm$^{-2}$ s$^{-1}$]; (5) SDSS DR7 $r$-band $AB$ magnitude; (6) Dn(4000) index; (7) stellar velocity dispersion [km s$^{-1}$]; (8) concentration index; (9) logarithm of the radio luminosity [erg s$^{-1}$]; (10) logarithm of the [O III] line luminosity [erg s$^{-1}$]; (11) logarithm of the black hole mass [M$_{\odot}$]; (12) radio linear size [kpc]; (13) asymmetric index; (14) spectral index between 1.4 GHz and 150 MHz.}
\end{longtable}

\end{center}

 \newpage
\section{FIRST, NVSS and TGSS images of the 43 COMP2$CAT$ sources.}
\label{ap:compcat}
\begin{figure}[h]
\includegraphics[width=6.cm,height=6.cm]{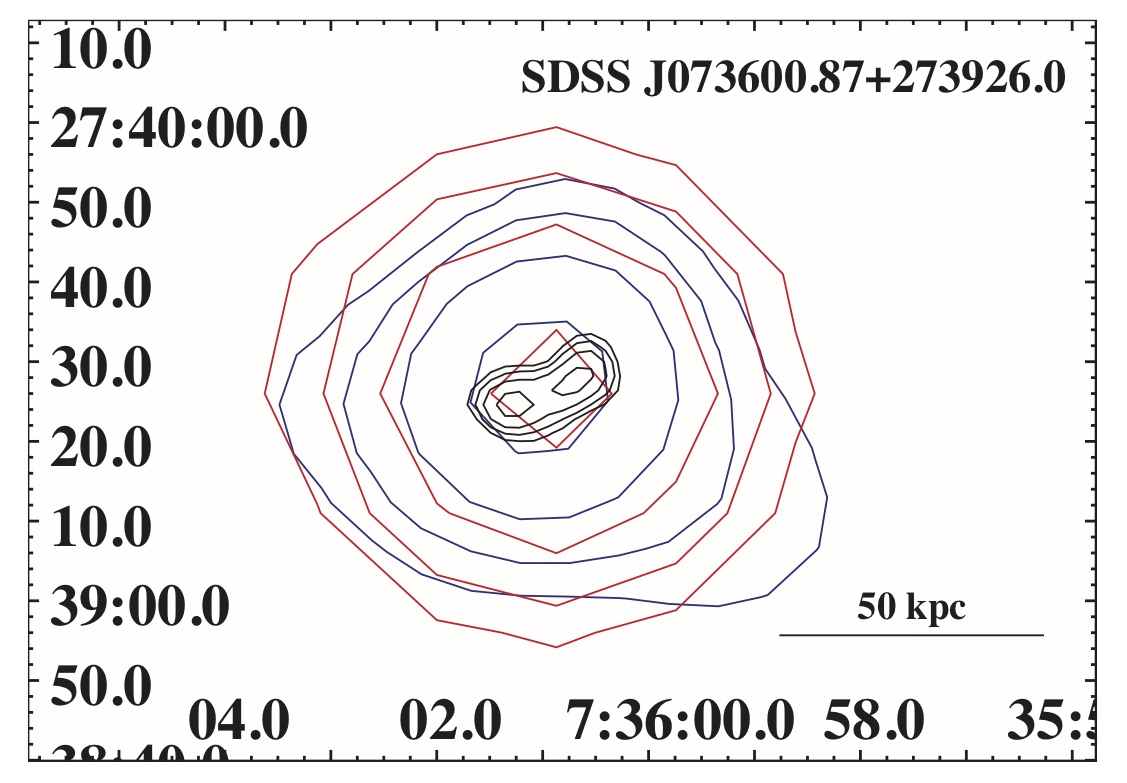}
\includegraphics[width=6.cm,height=6.cm]{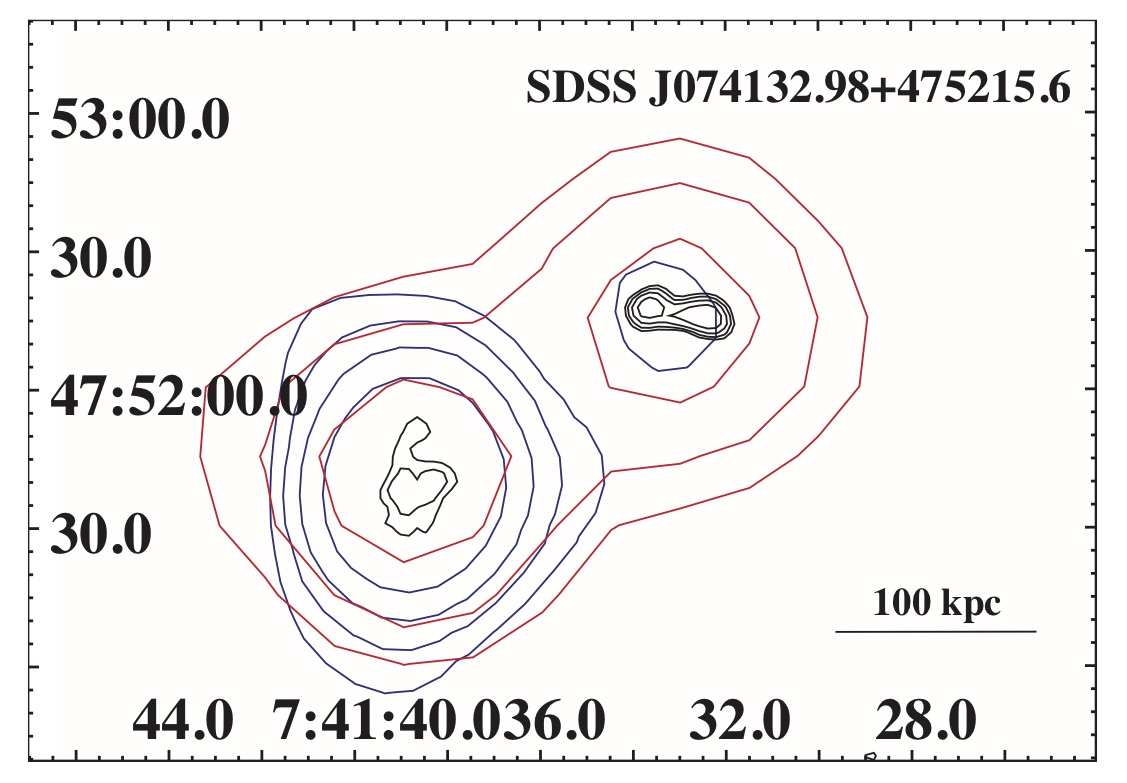} 
\includegraphics[width=6.cm,height=6.cm]{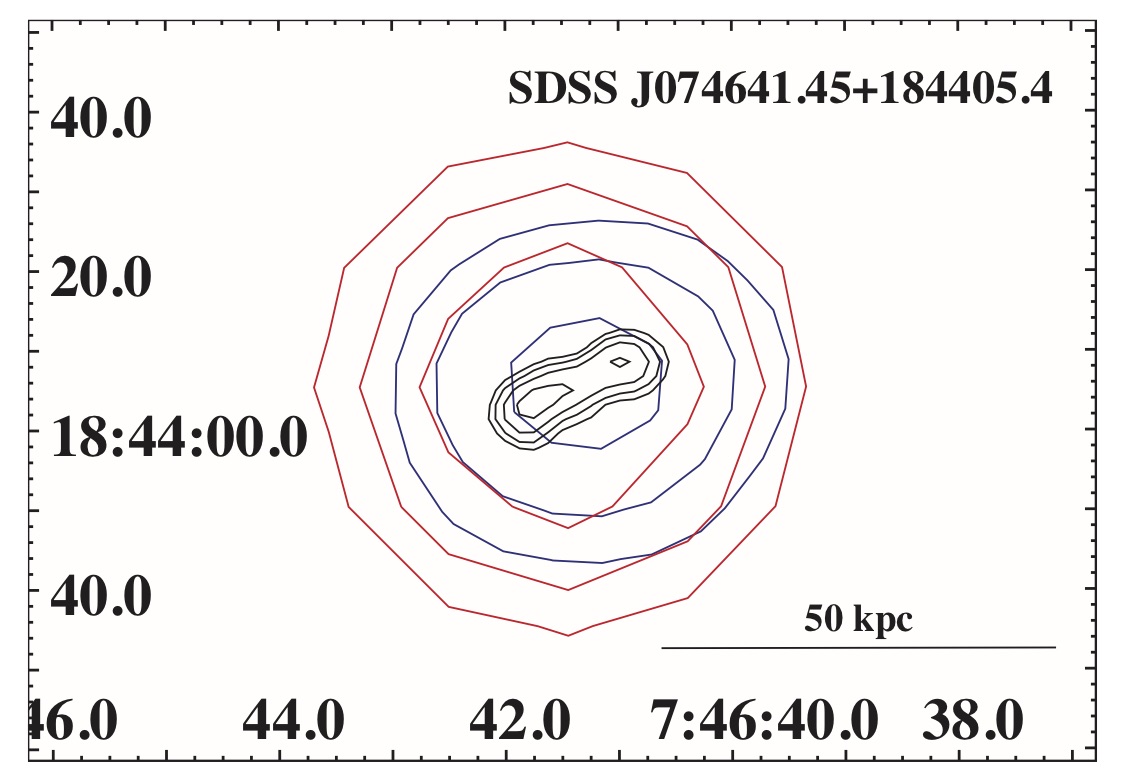}

\includegraphics[width=6.cm,height=6.cm]{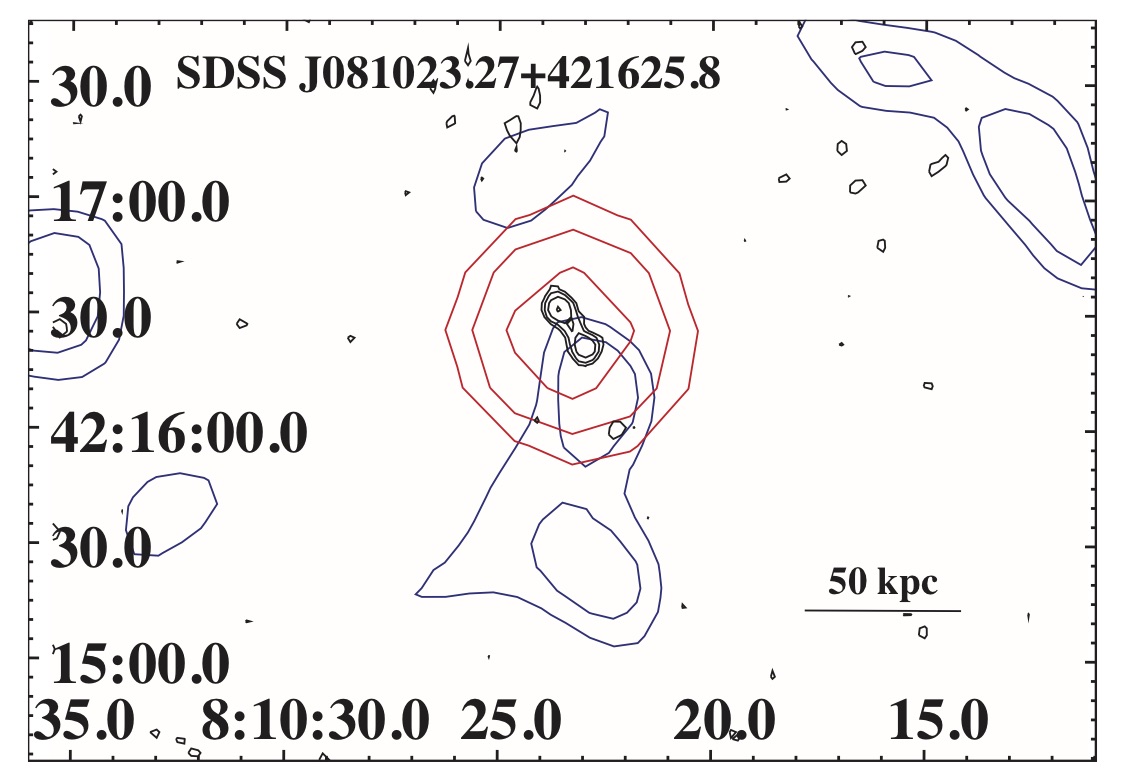} 
\includegraphics[width=6.cm,height=6.cm]{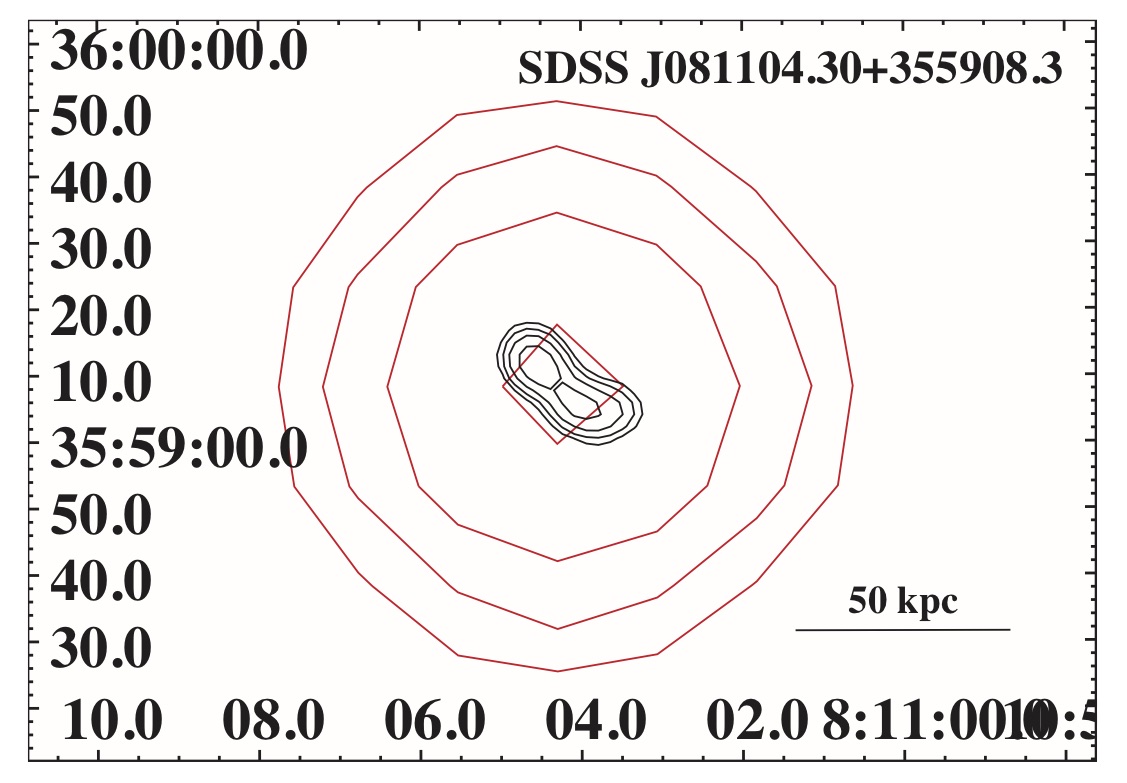}
\includegraphics[width=6.cm,height=6.cm]{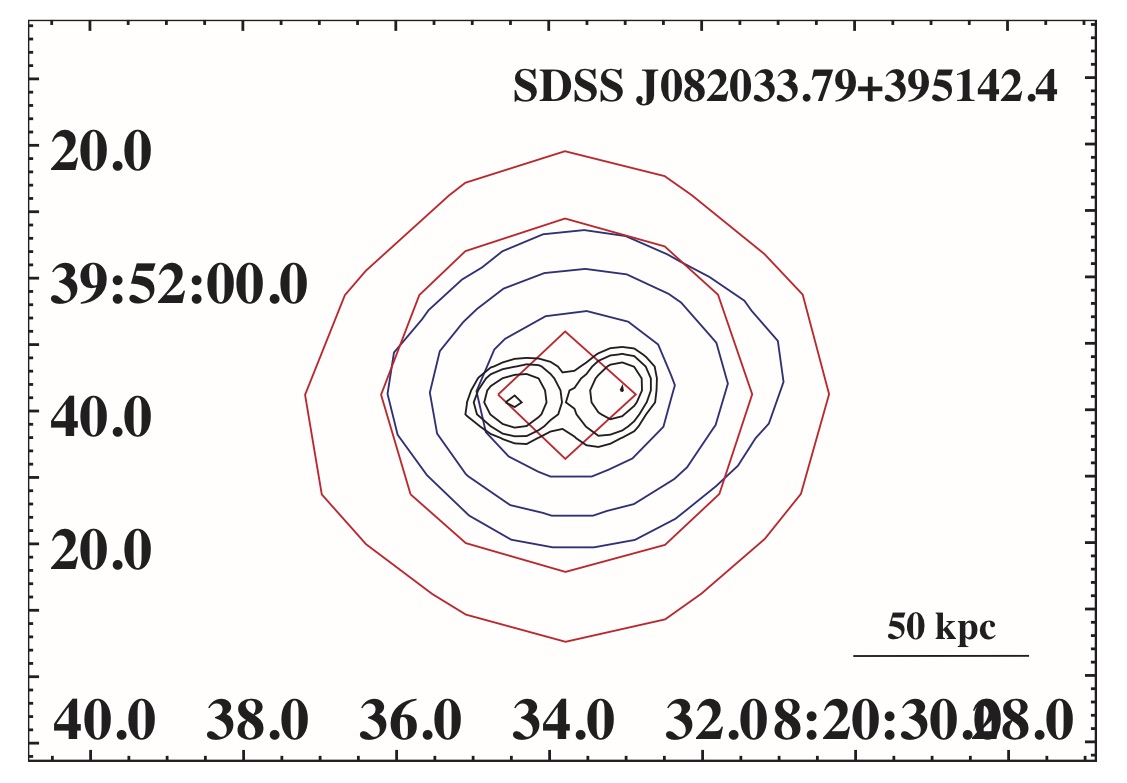}

\includegraphics[width=6.cm,height=6.cm]{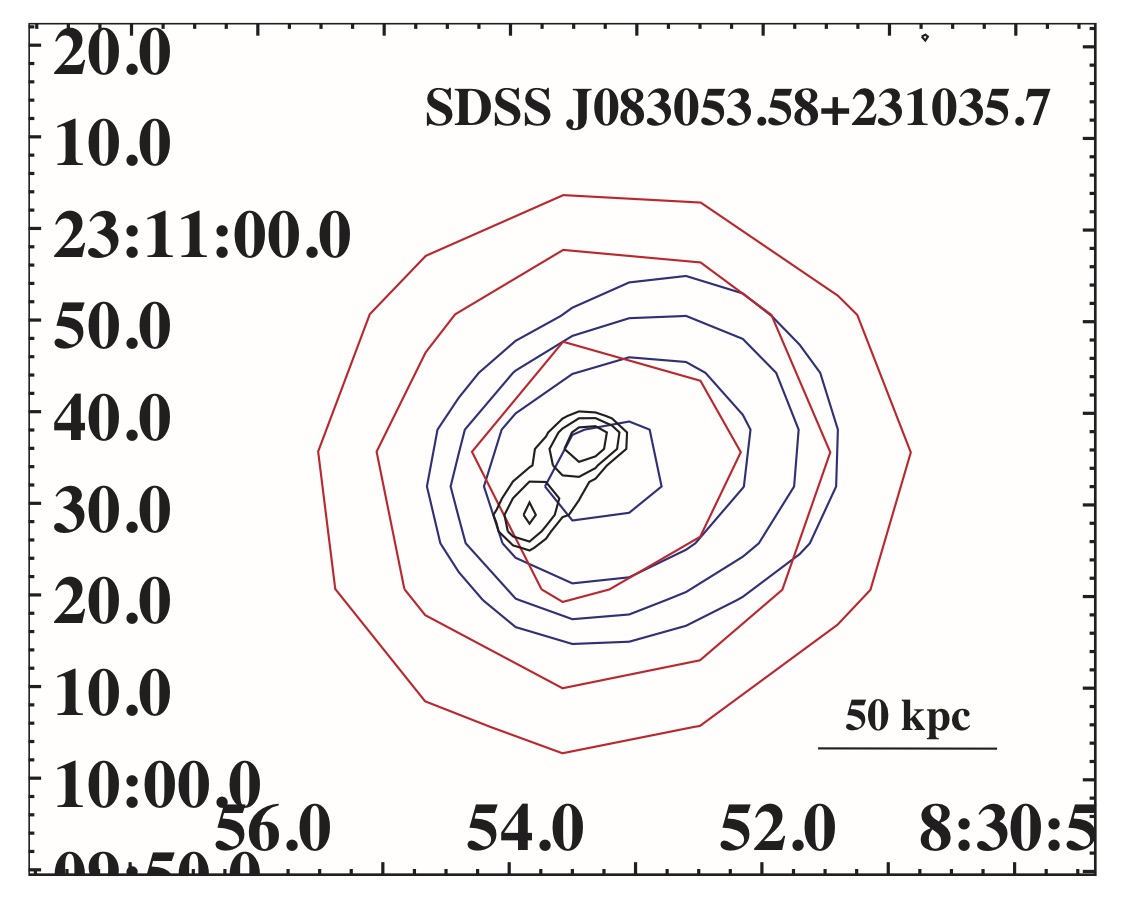}
\includegraphics[width=6.cm,height=6.cm]{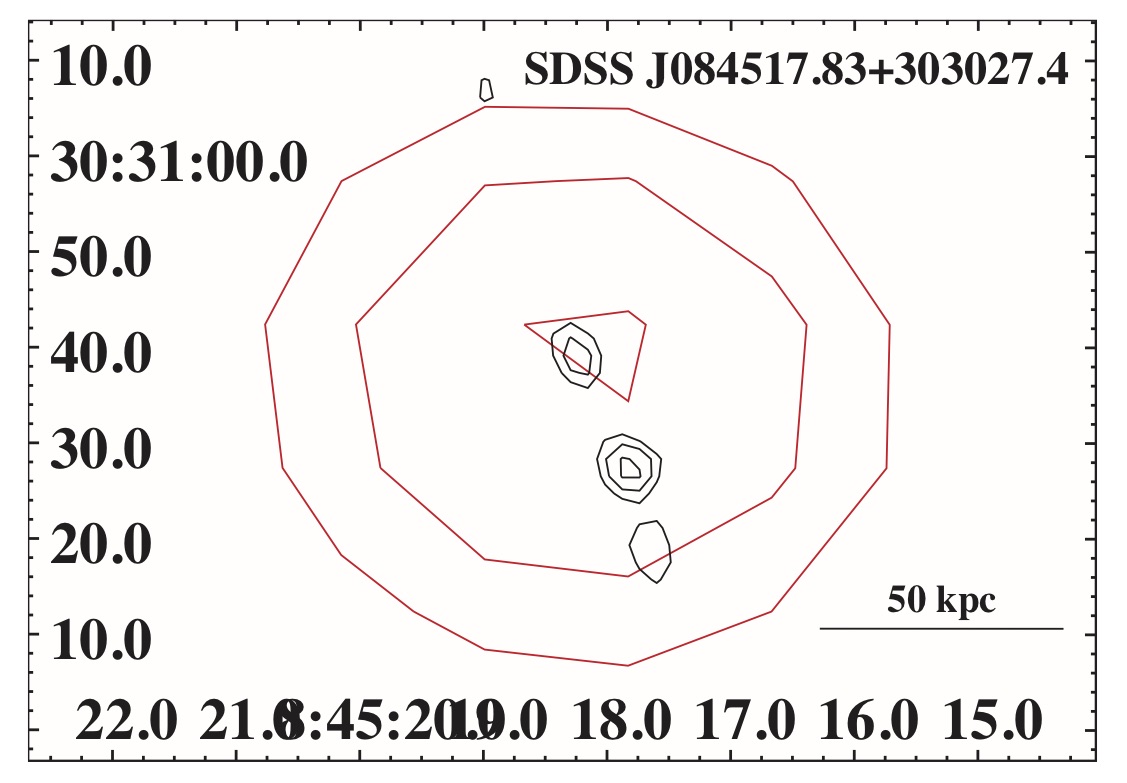}
\includegraphics[width=6.cm,height=6.cm]{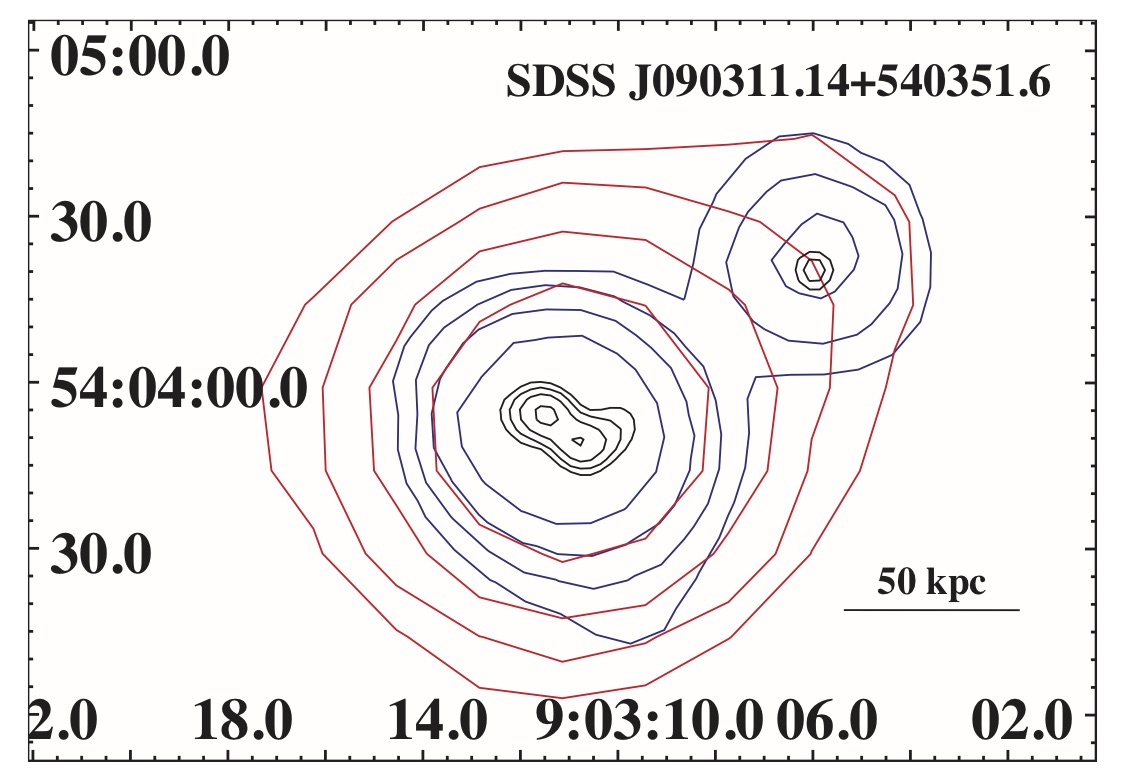}

\caption{Images of COMP2$CAT$ sources. Black, red and blue contours correspond to the emission seen in FIRST, NVSS and TGSS. Contours are drawn using the parameters shown in App. \ref{ap:contours}, in Table \ref{tab:contourscomp}.}
\label{images1}
\end{figure}

\addtocounter{figure}{-1}
\begin{figure}
\includegraphics[width=6.cm,height=6.cm]{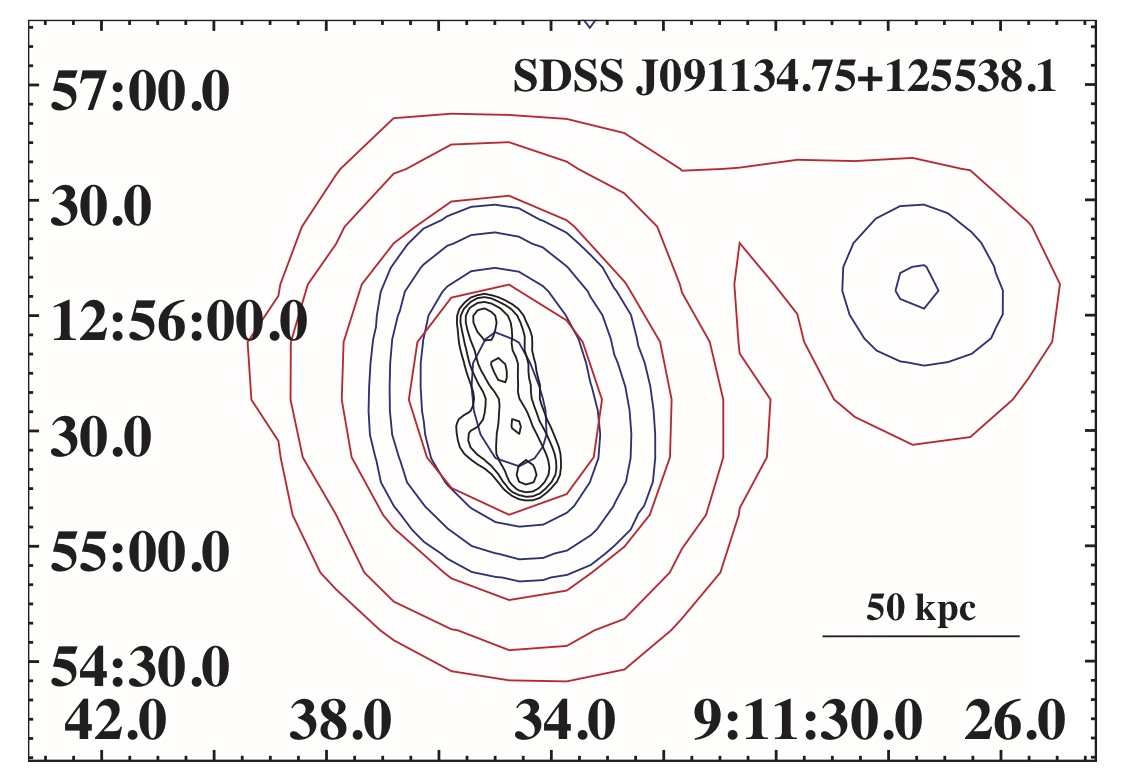}
\includegraphics[width=6.cm,height=6.cm]{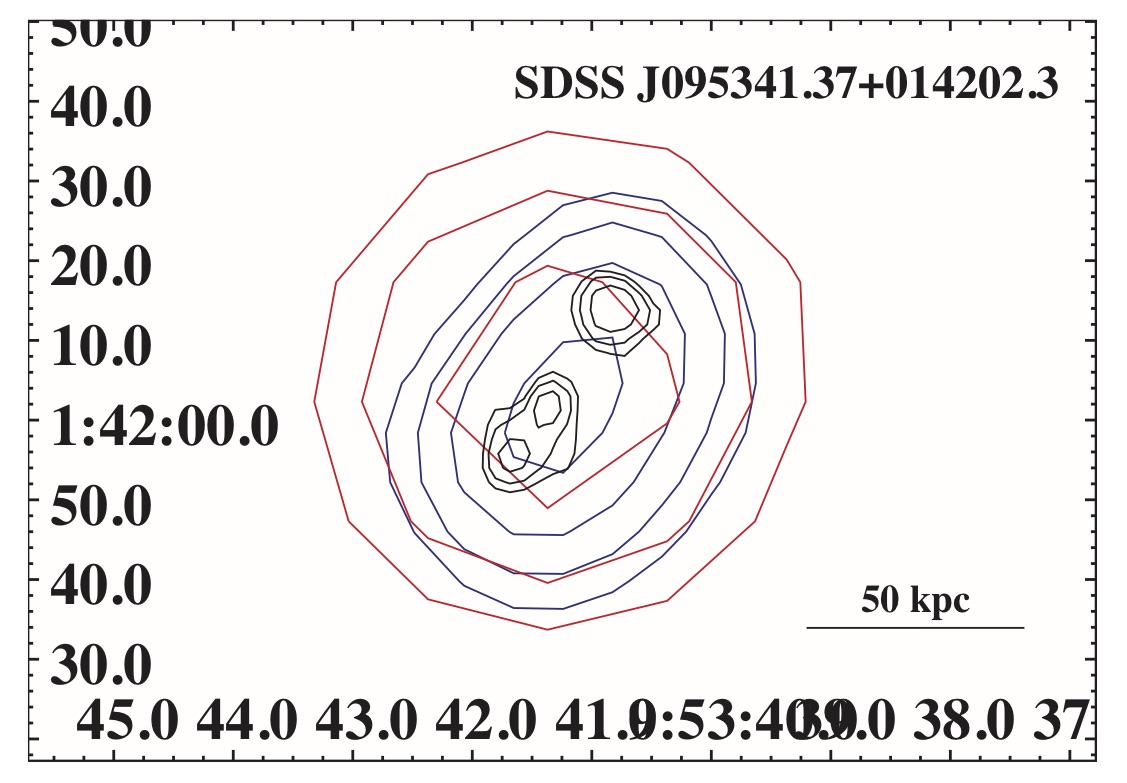}
\includegraphics[width=6.cm,height=6.cm]{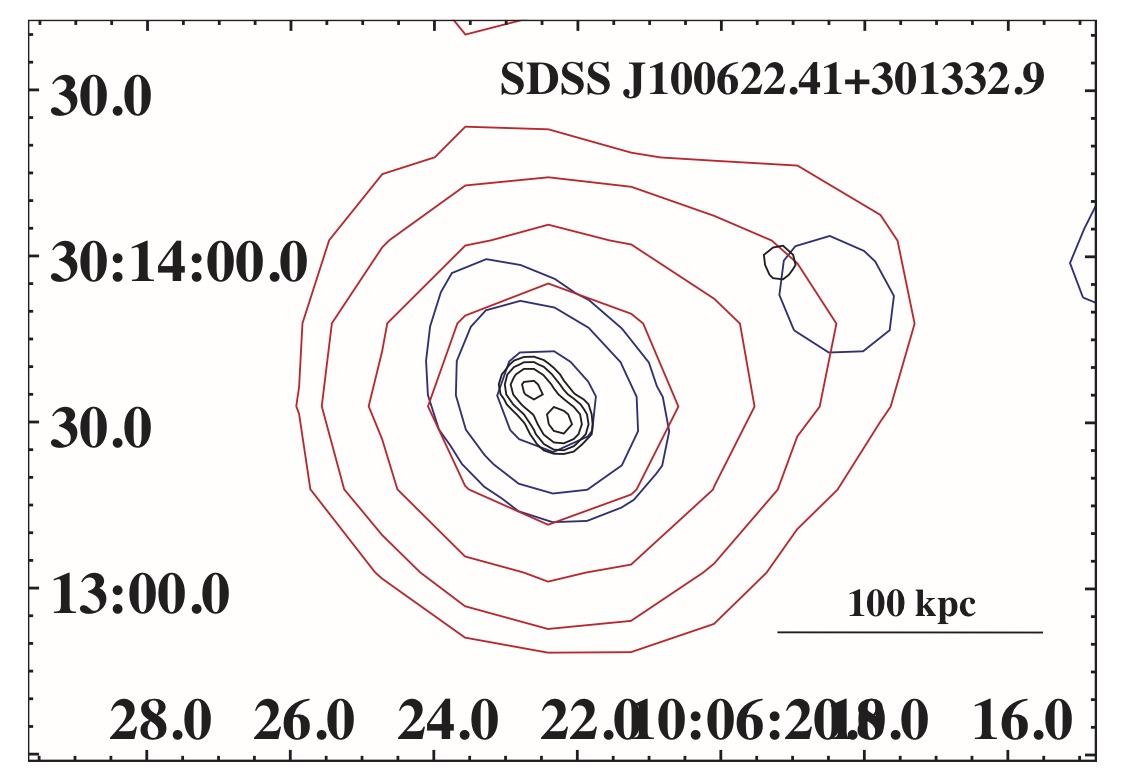}

\includegraphics[width=6.cm,height=6.cm]{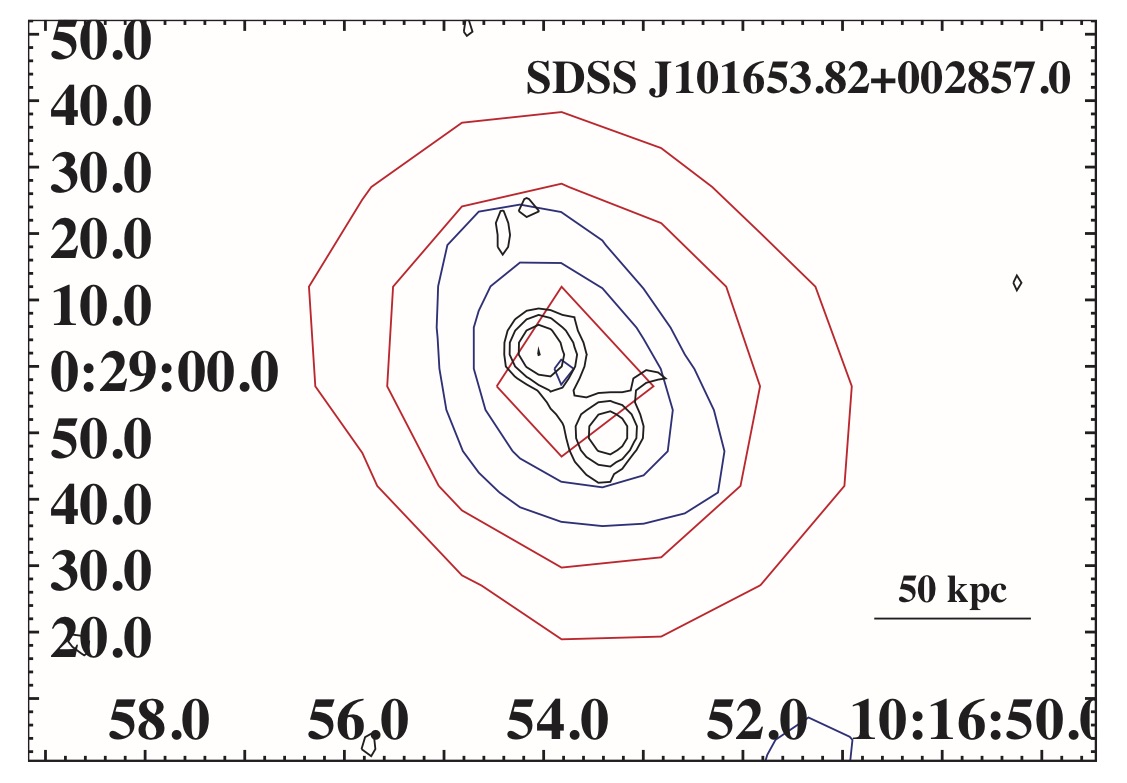}
\includegraphics[width=6.cm,height=6.cm]{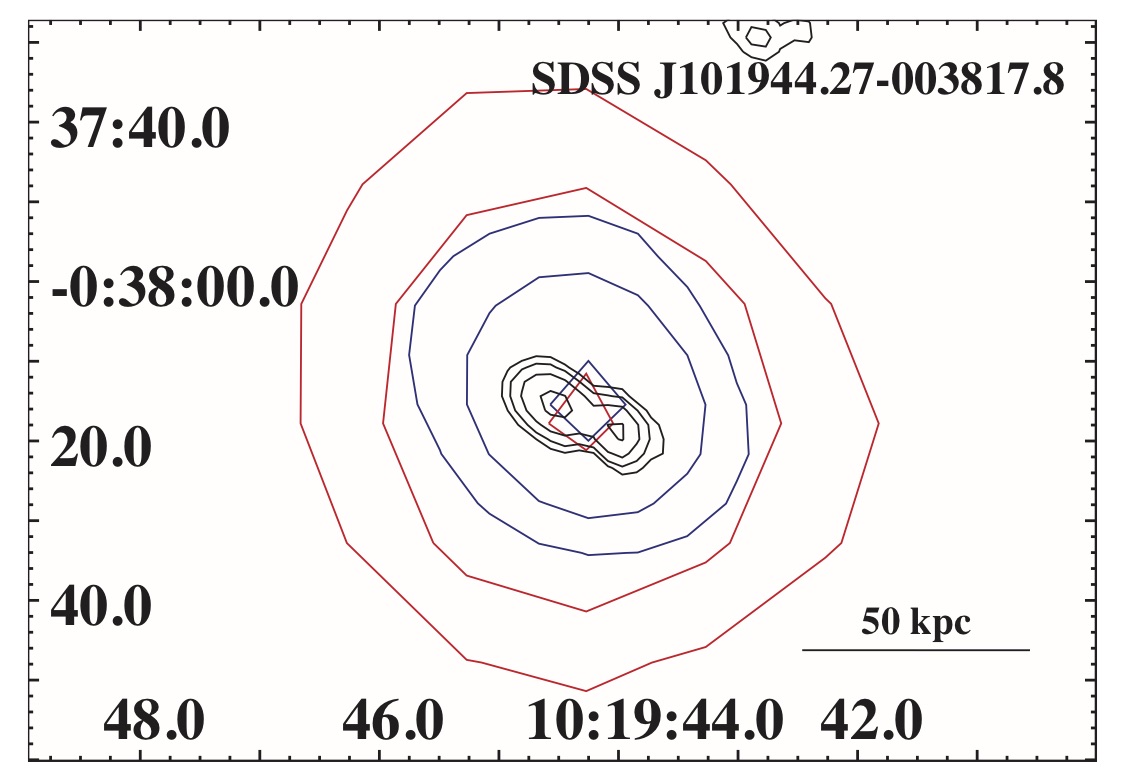} 
\includegraphics[width=6.cm,height=6.cm]{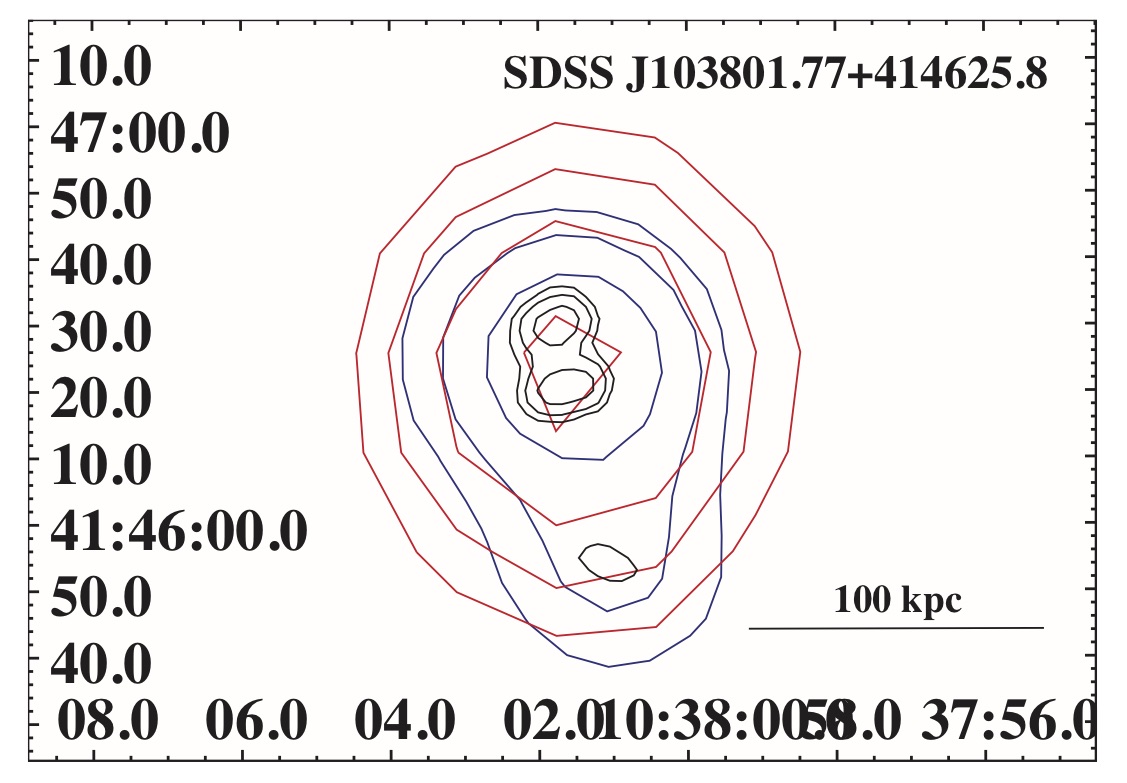} 

\includegraphics[width=6.cm,height=6.cm]{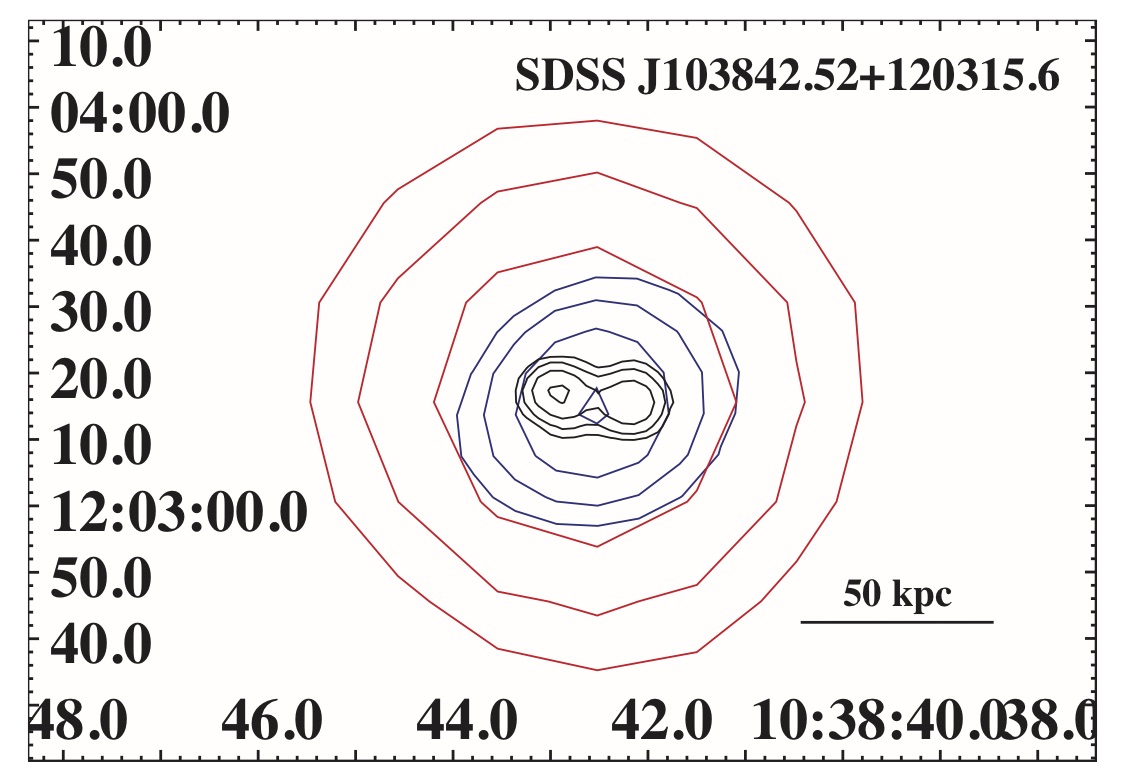} 
\includegraphics[width=6.cm,height=6.cm]{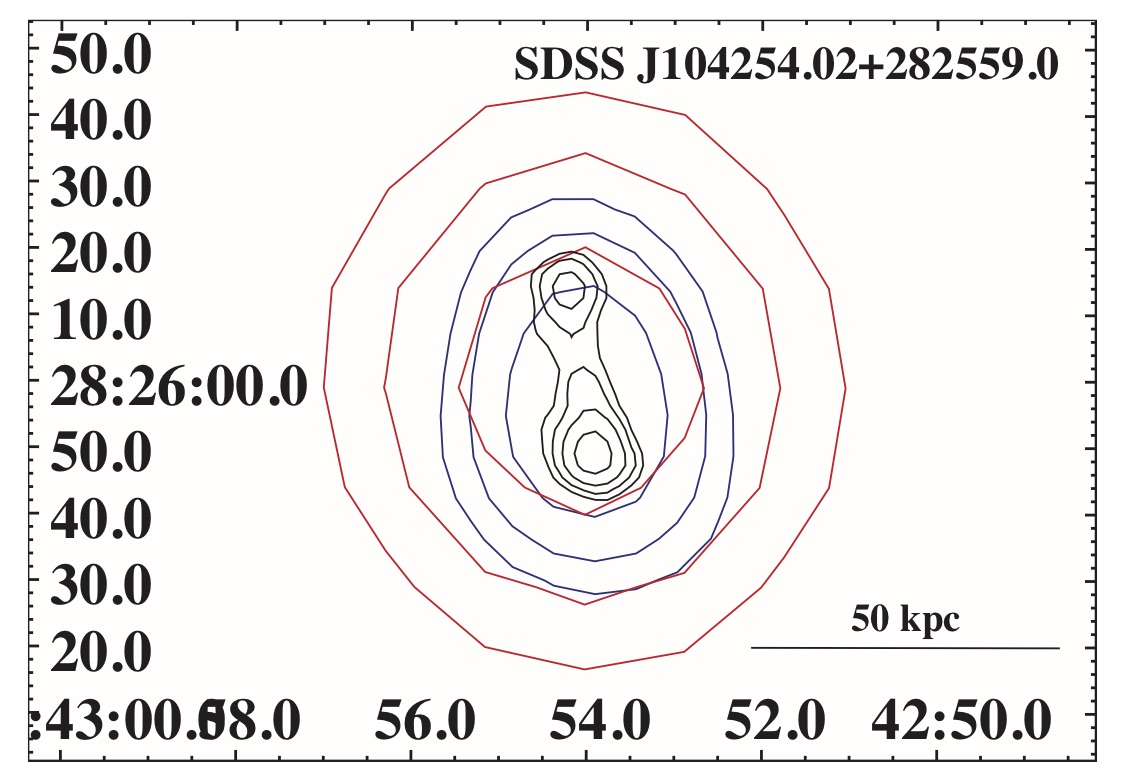} 
\includegraphics[width=6.cm,height=6.cm]{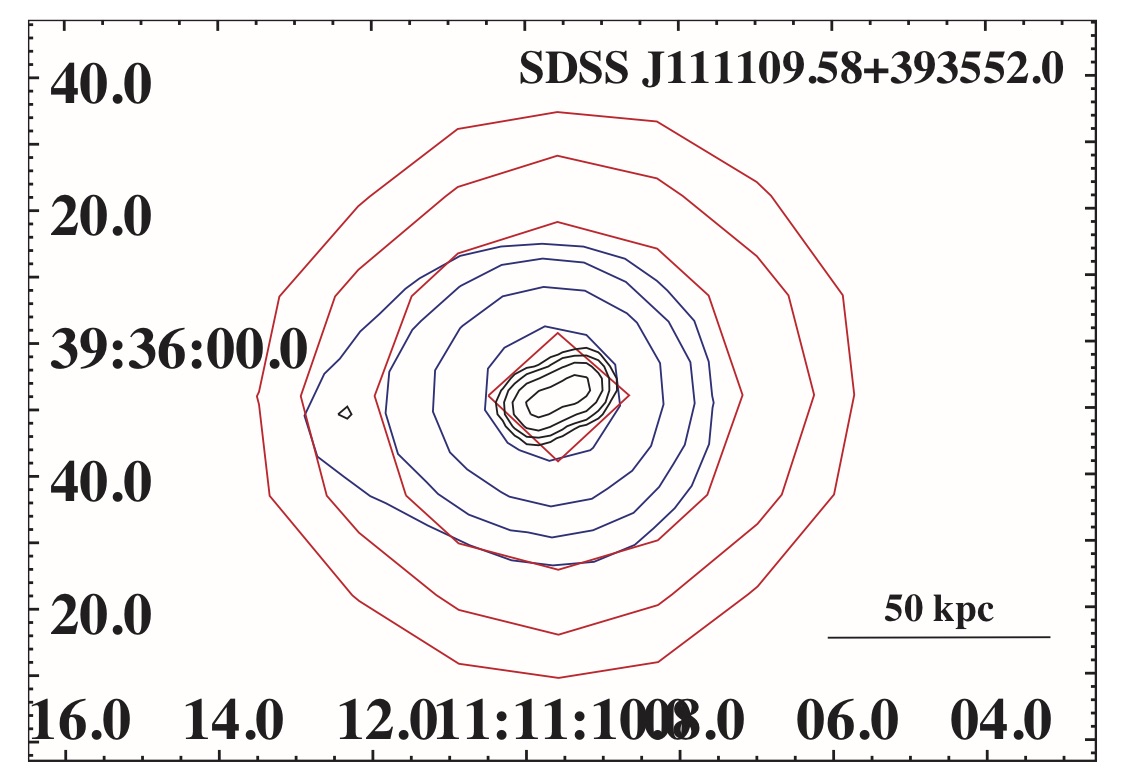} 

\includegraphics[width=6.cm,height=6.cm]{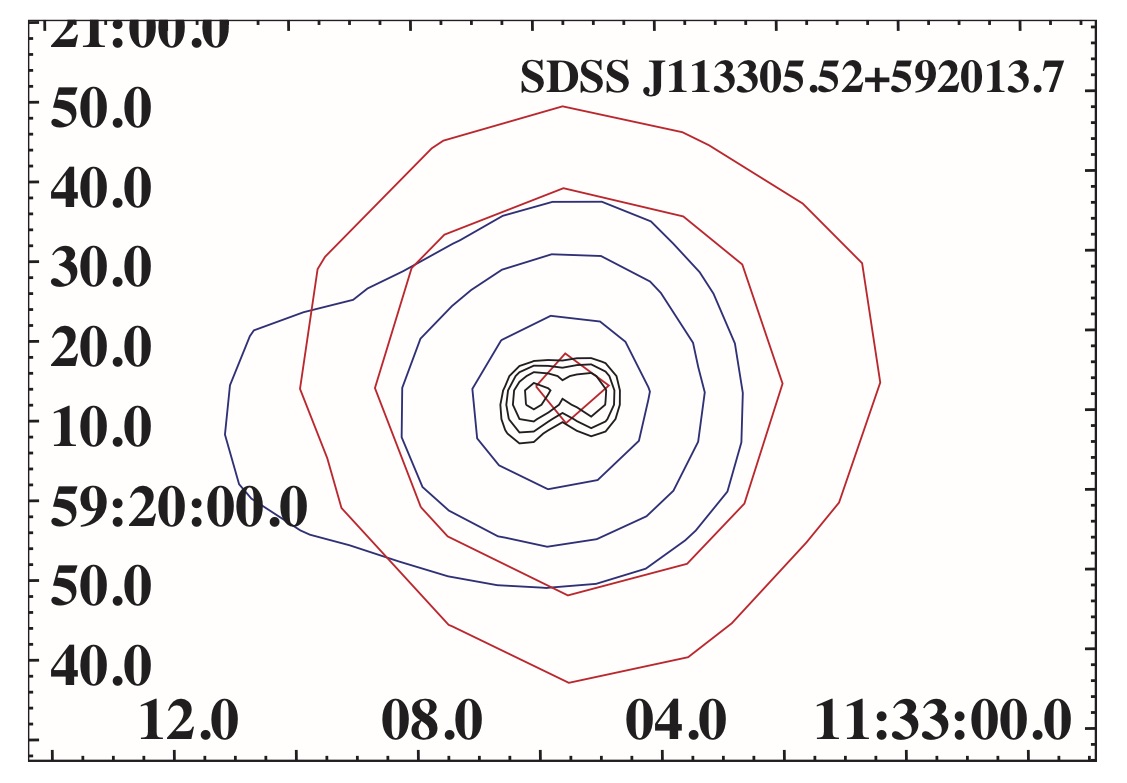} 
\includegraphics[width=6.cm,height=6.cm]{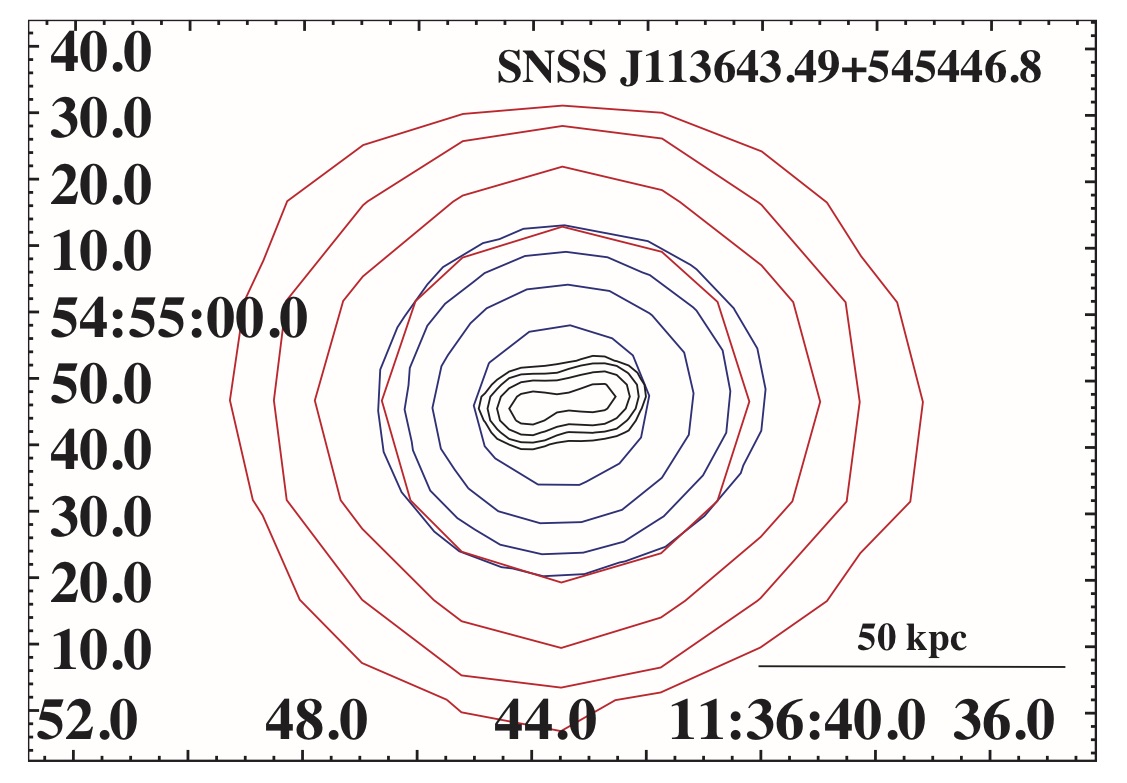} 
\includegraphics[width=6.cm,height=6.cm]{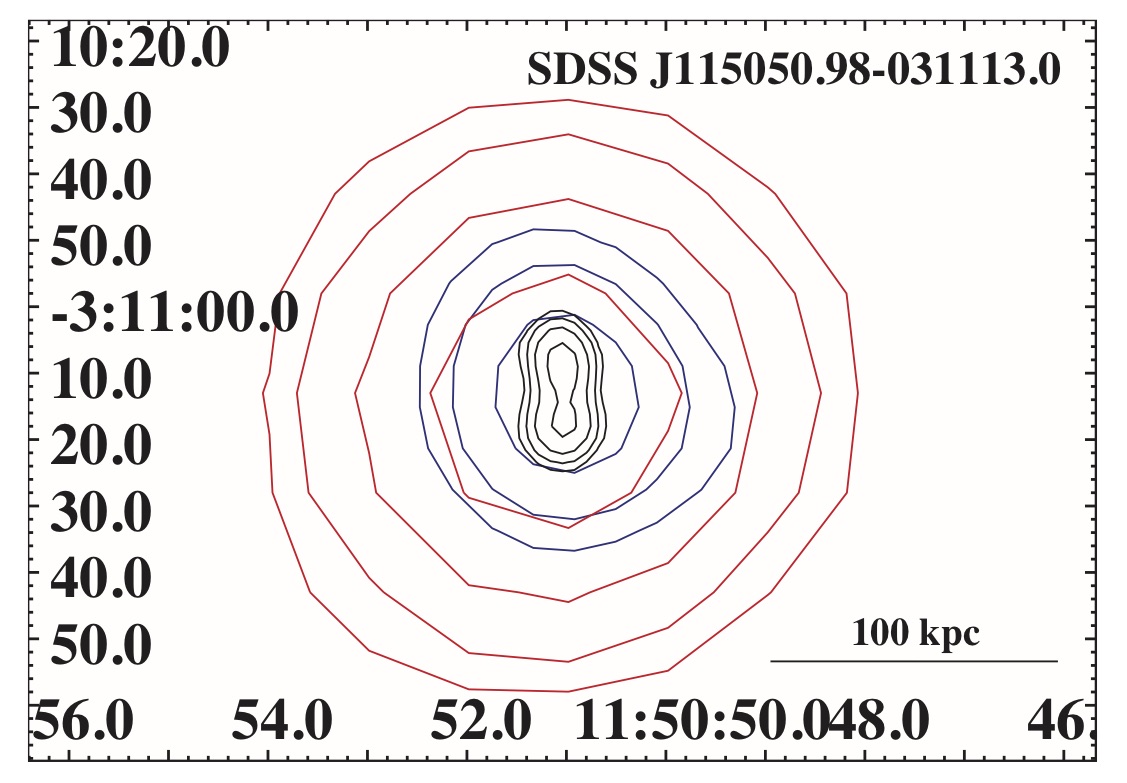}

\caption{(continued)}
\end{figure}

\addtocounter{figure}{-1}
\begin{figure}

\includegraphics[width=6.cm,height=6.cm]{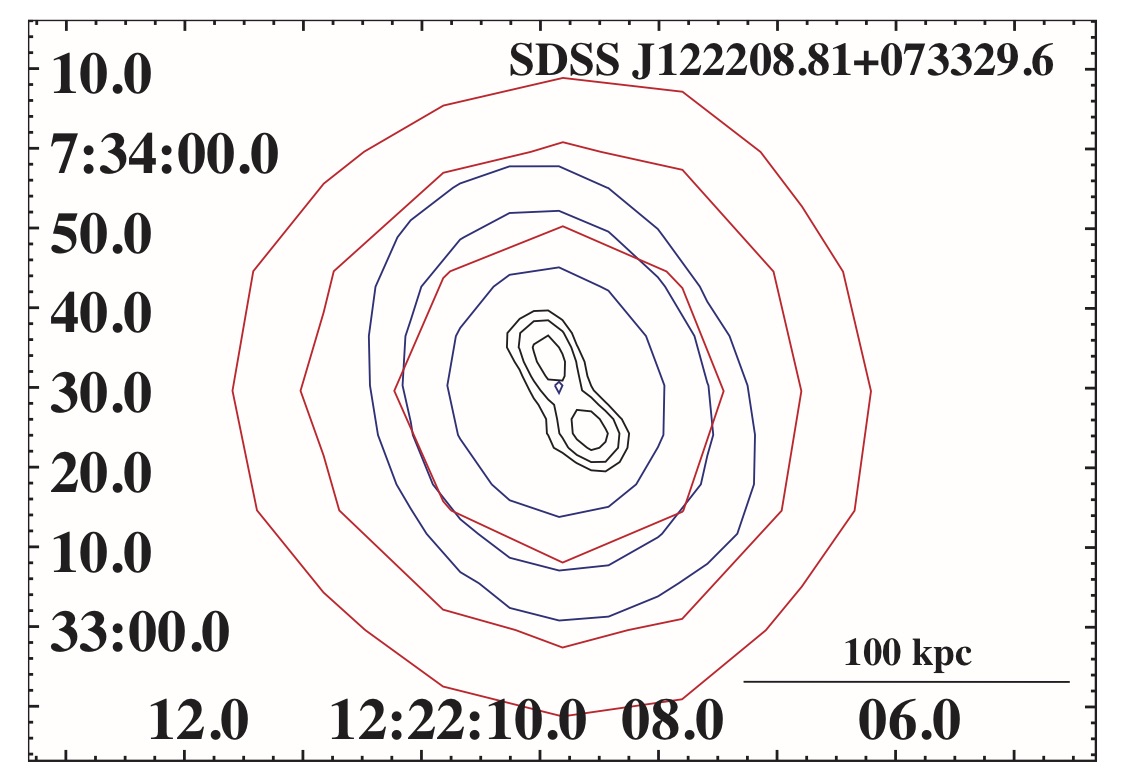} 
\includegraphics[width=6.cm,height=6.cm]{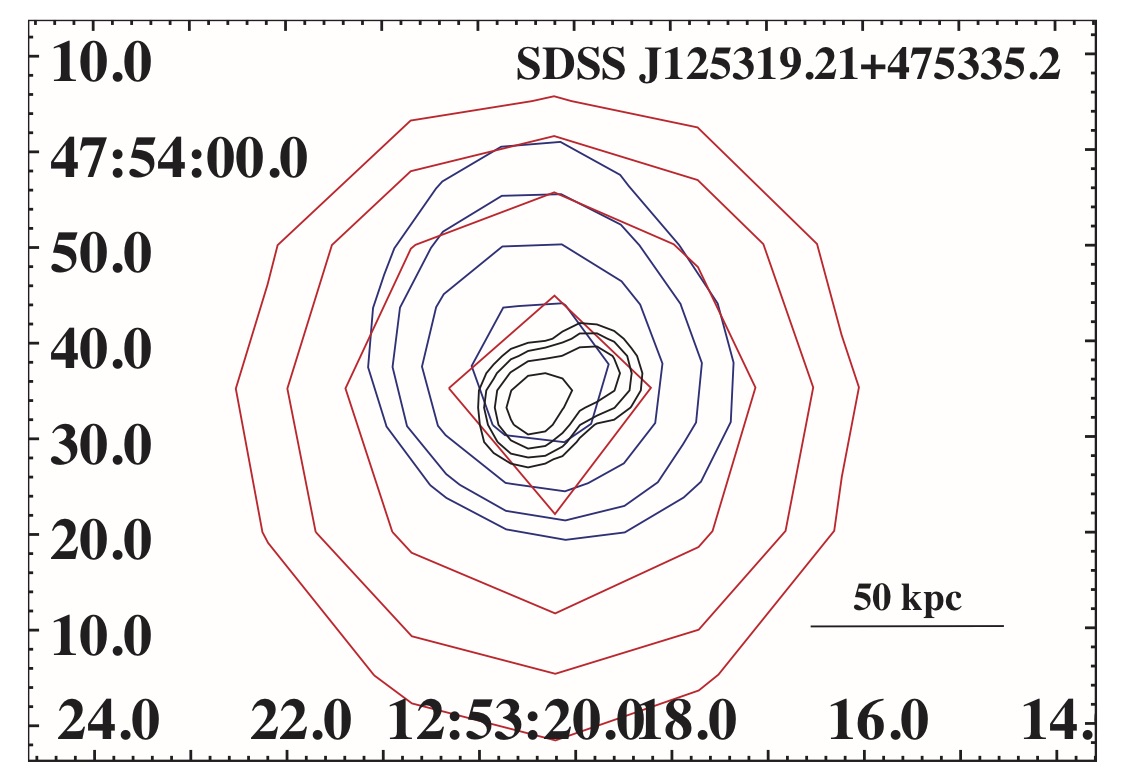}
\includegraphics[width=6.cm,height=6.cm]{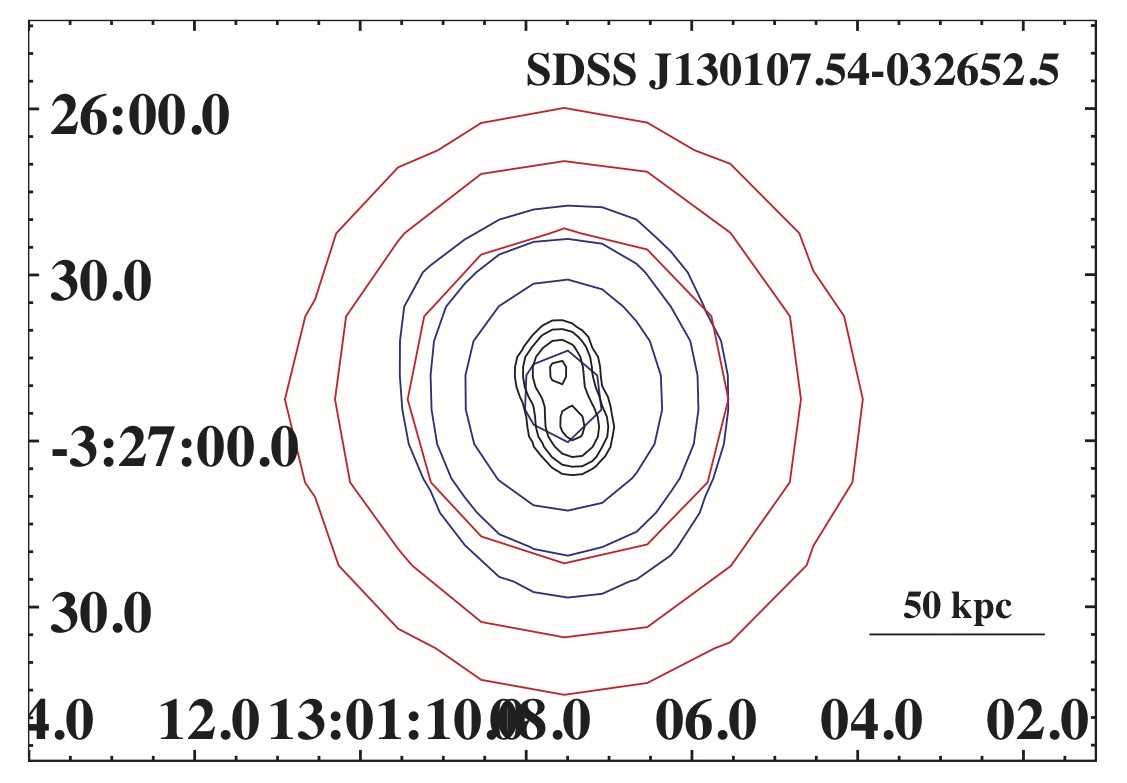}

\includegraphics[width=6.cm,height=6.cm]{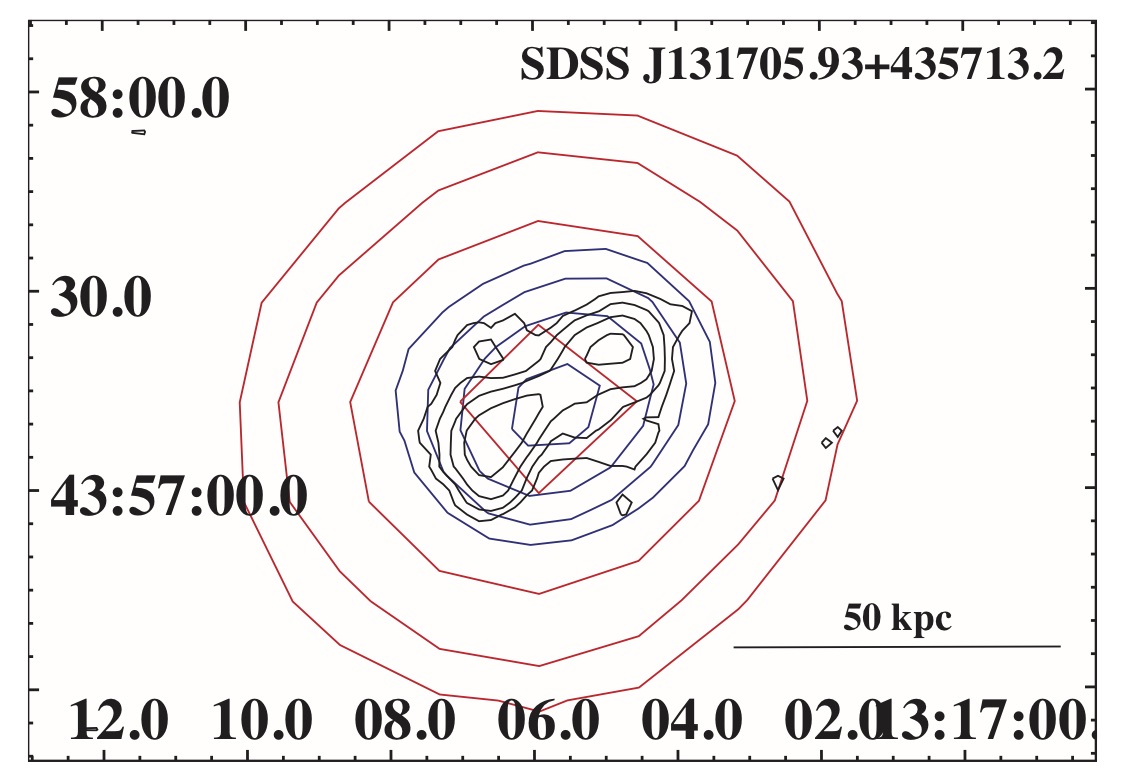} 
\includegraphics[width=6.cm,height=6.cm]{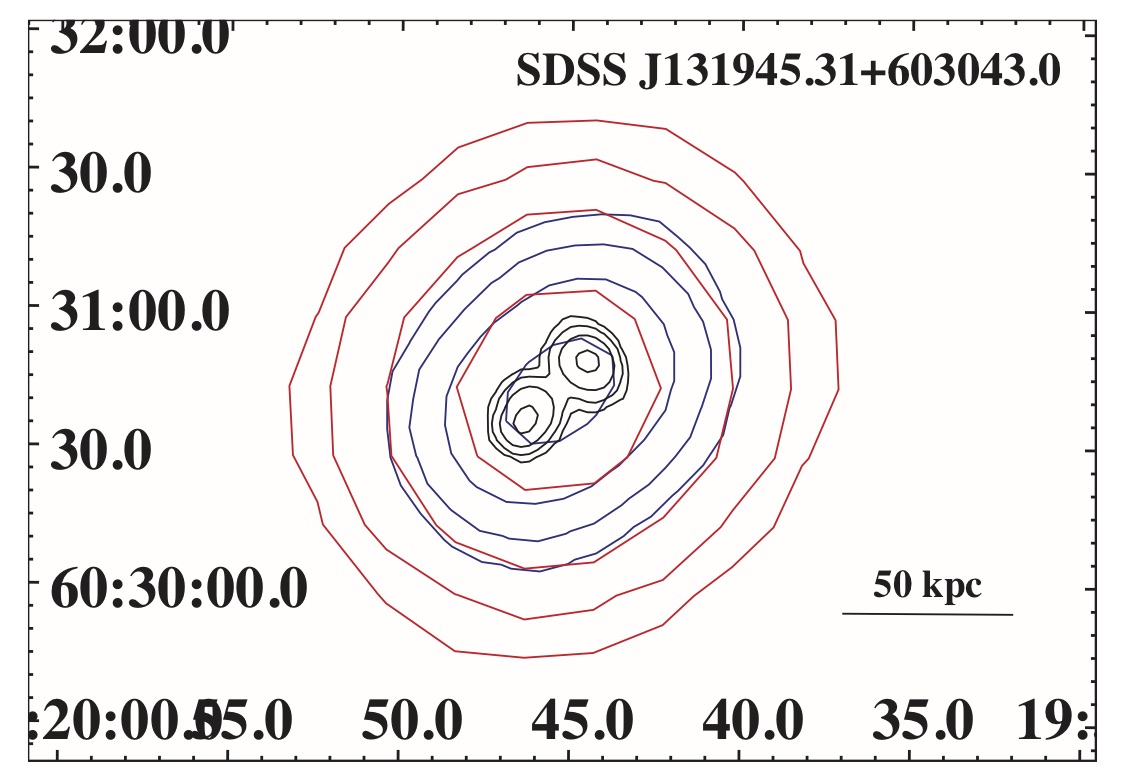} 
\includegraphics[width=6.cm,height=6.cm]{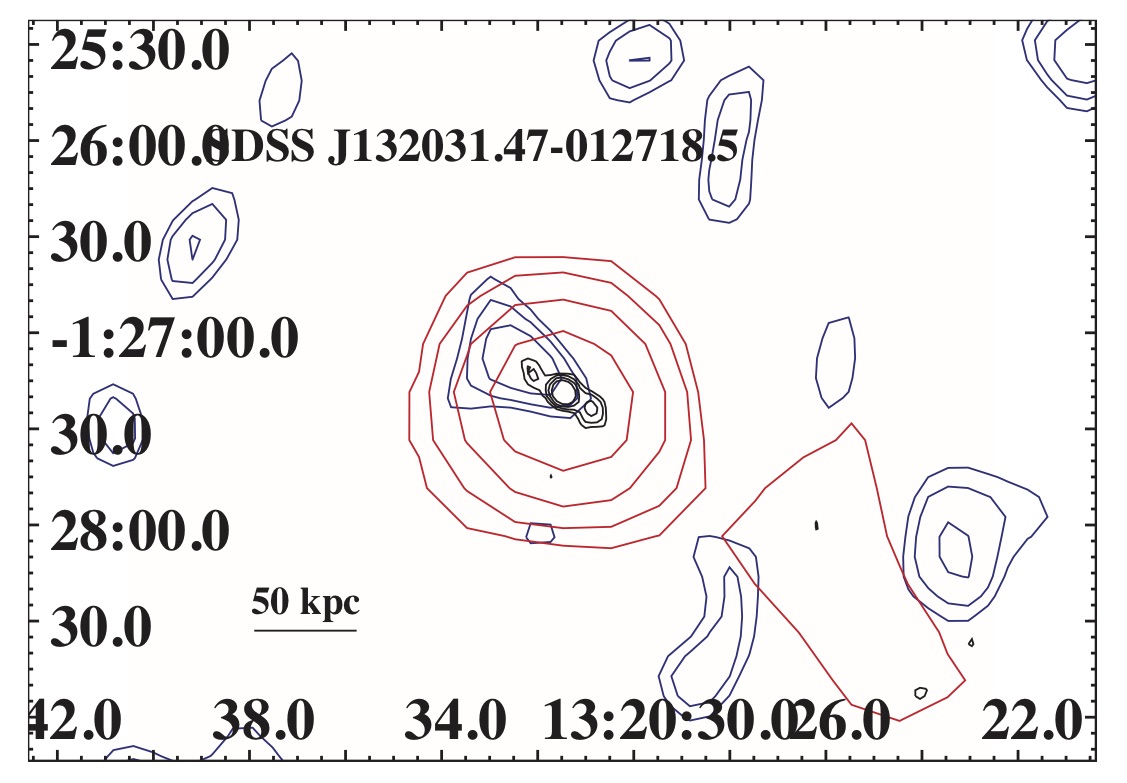}

\includegraphics[width=6.cm,height=6.cm]{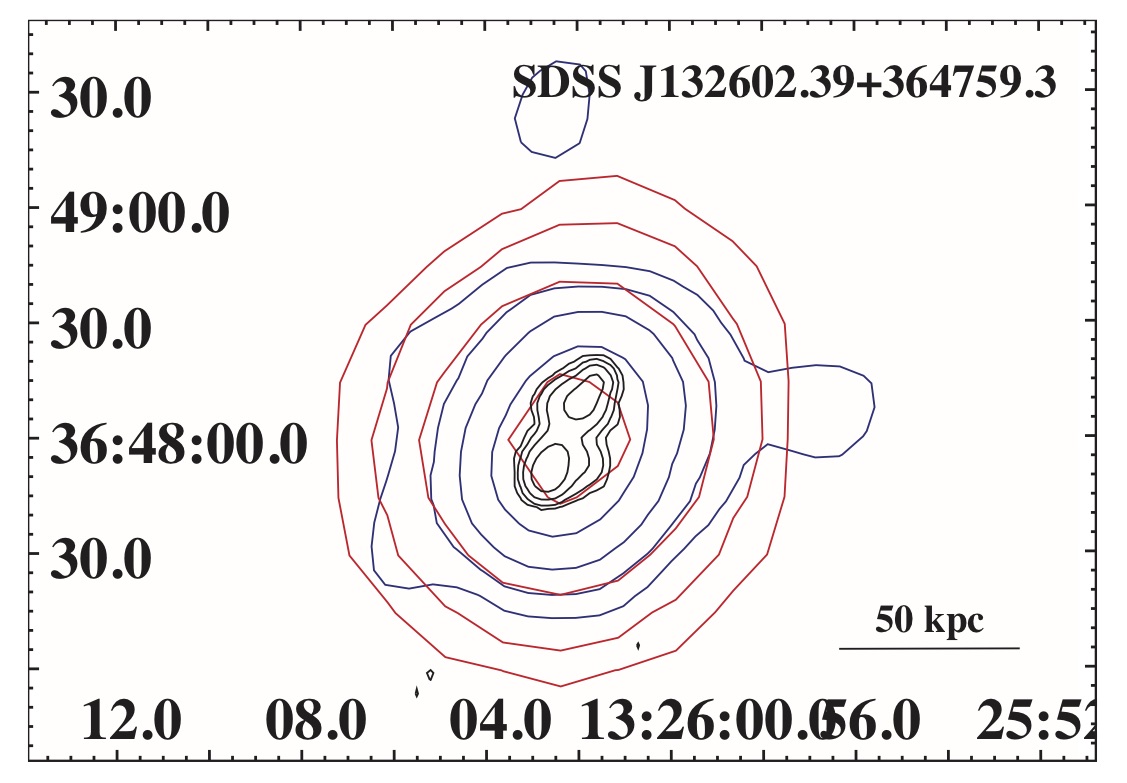} 
\includegraphics[width=6.cm,height=6.cm]{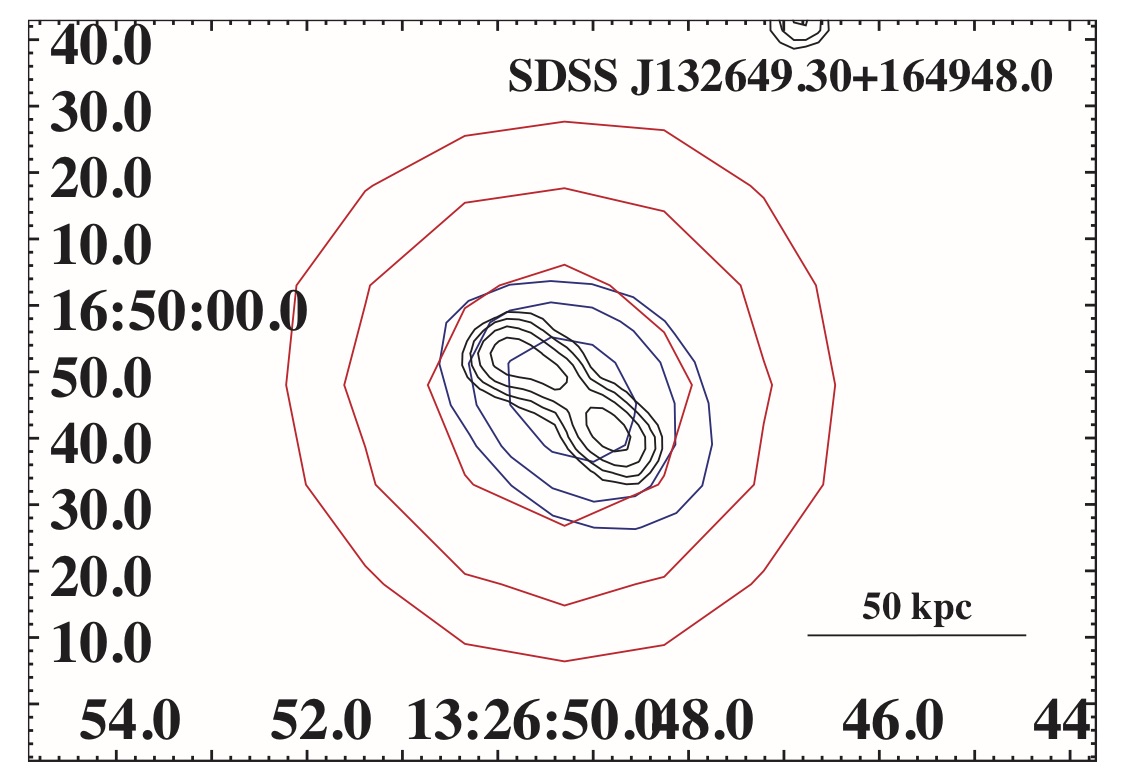} 
\includegraphics[width=6.cm,height=6.cm]{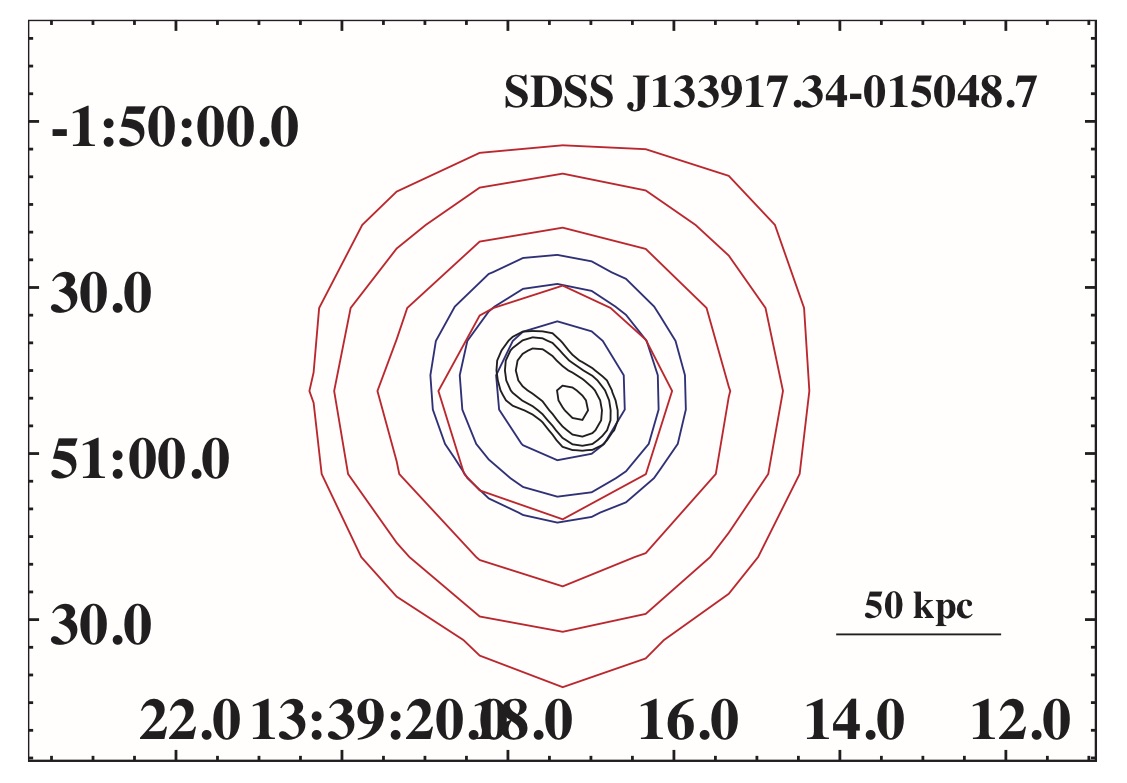}

\includegraphics[width=6.cm,height=6.cm]{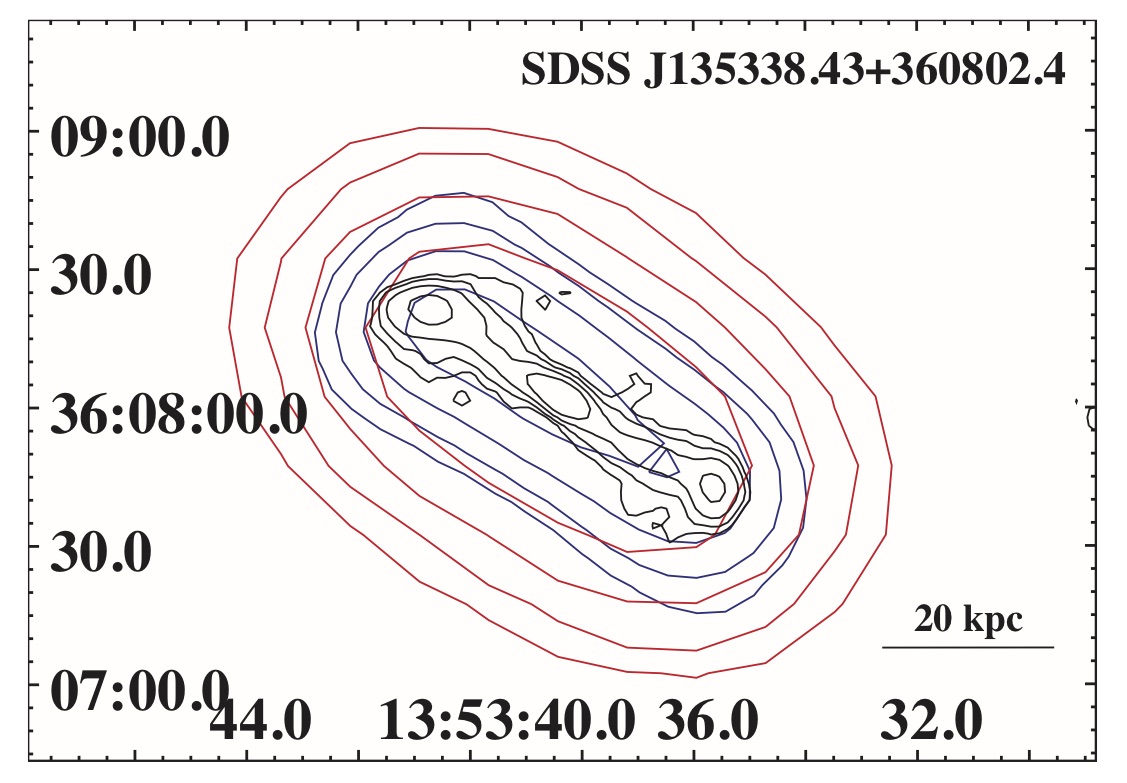}
\includegraphics[width=6.cm,height=6.cm]{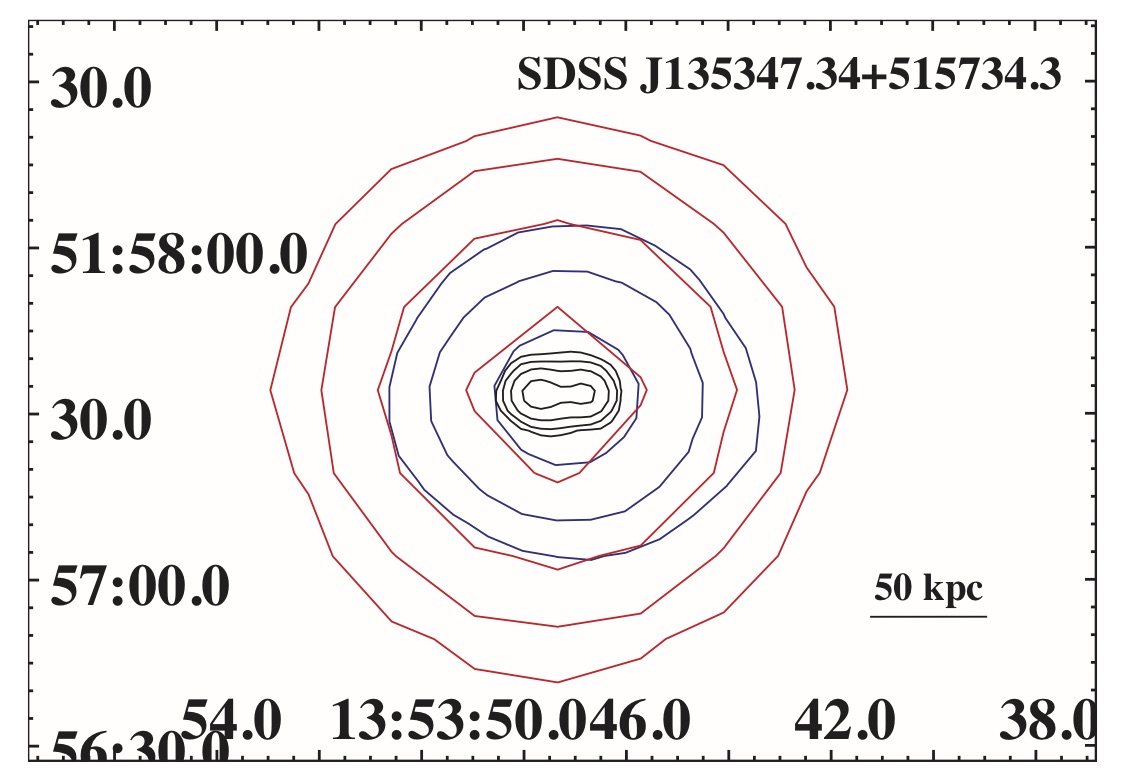} 
\includegraphics[width=6.cm,height=6.cm]{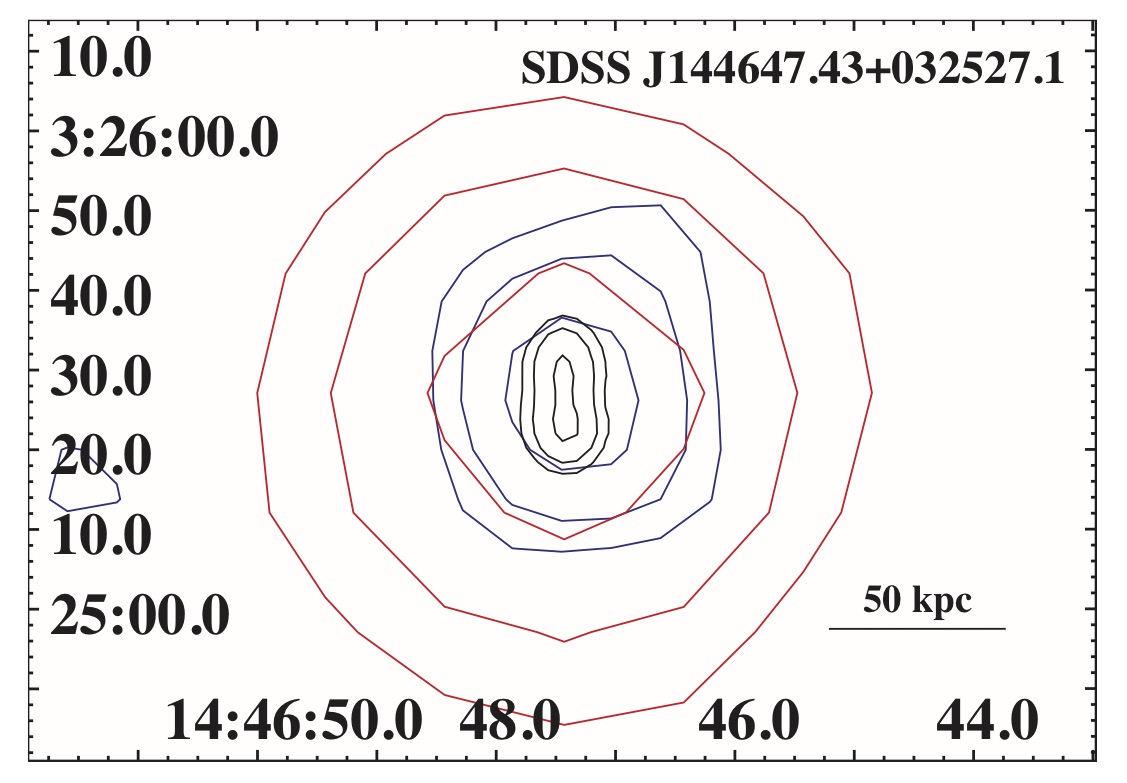} 
\caption{(continued)}
\end{figure}

 \addtocounter{figure}{-1}
 \begin{figure}
 \includegraphics[width=6.cm,height=6.cm]{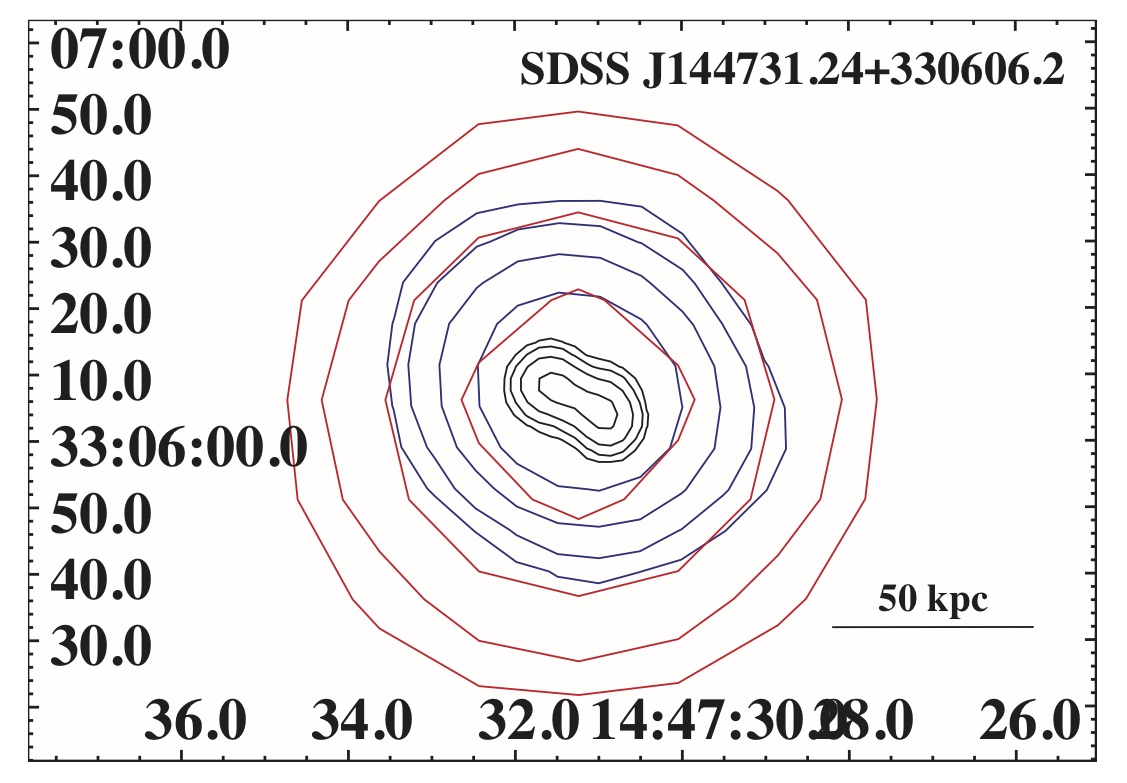}
\includegraphics[width=6.cm,height=6.cm]{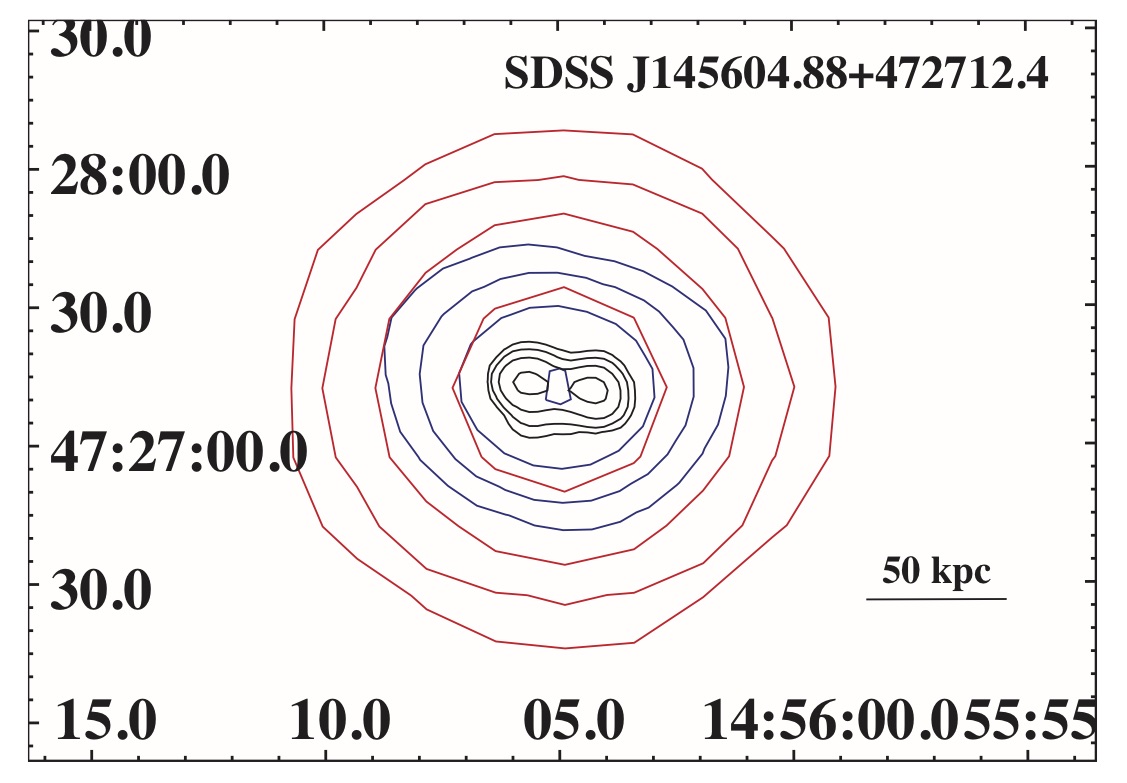} 
\includegraphics[width=6.cm,height=6.cm]{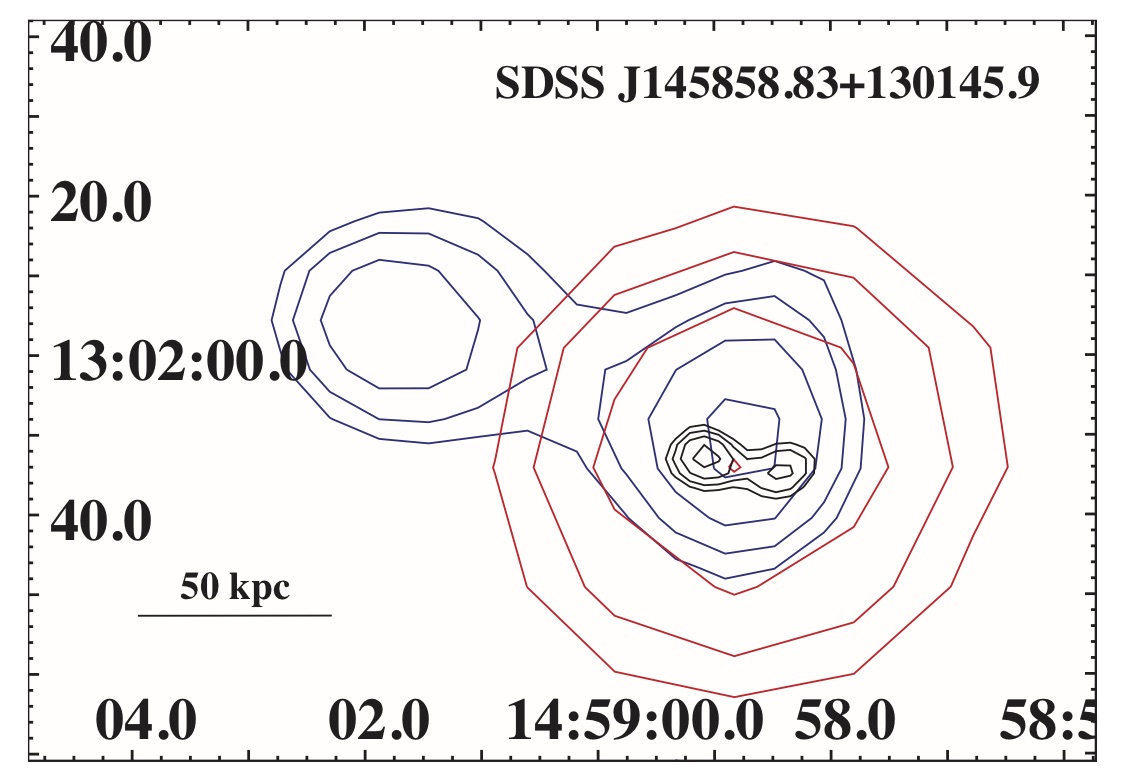}

 \includegraphics[width=6.cm,height=6.cm]{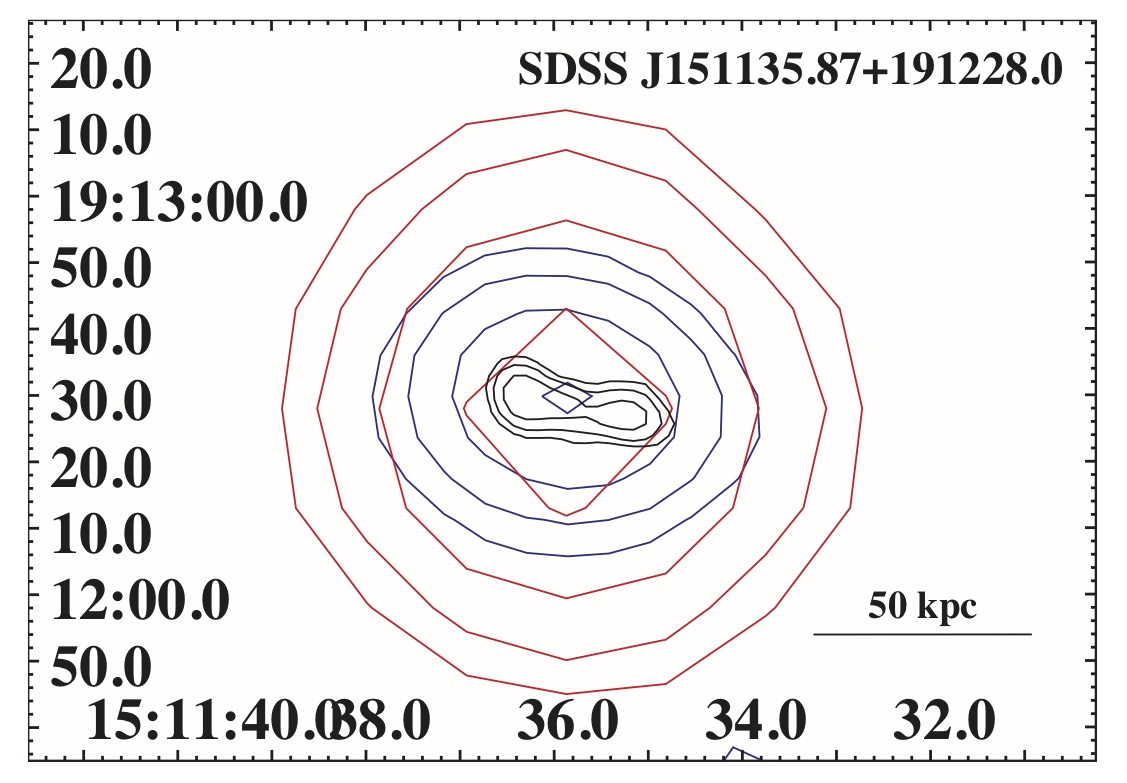} 
 \includegraphics[width=6.cm,height=6.cm]{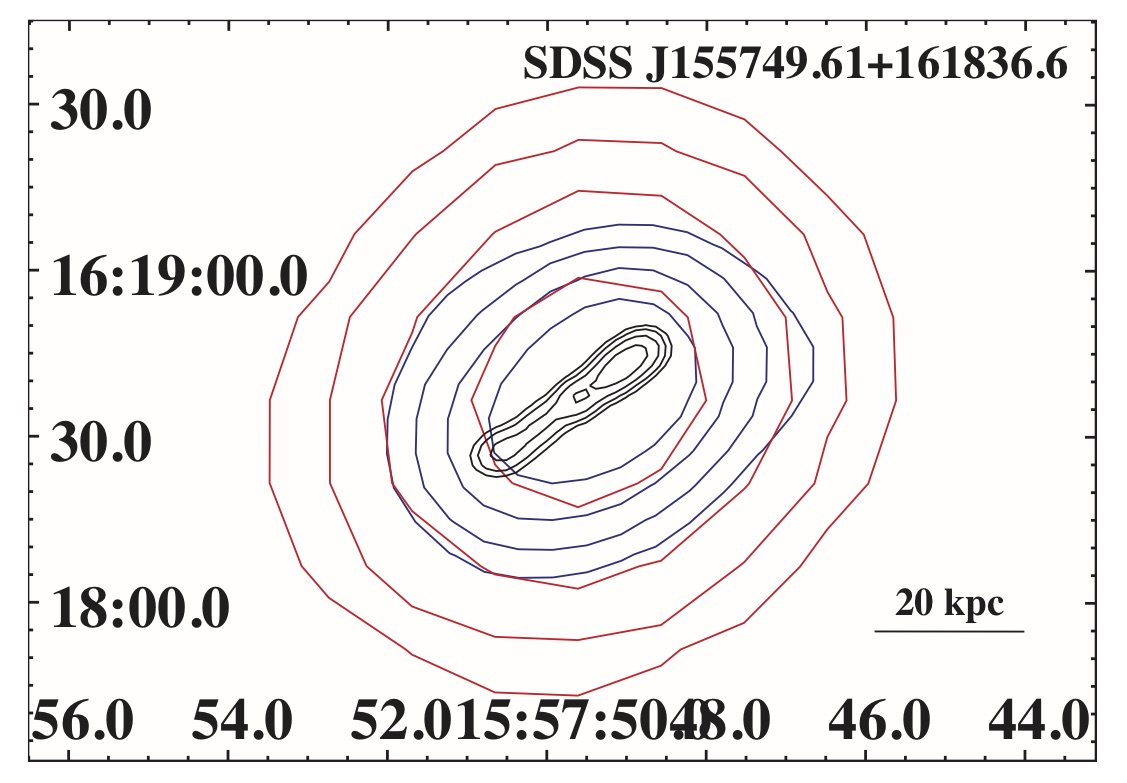}
 \includegraphics[width=6.cm,height=6.cm]{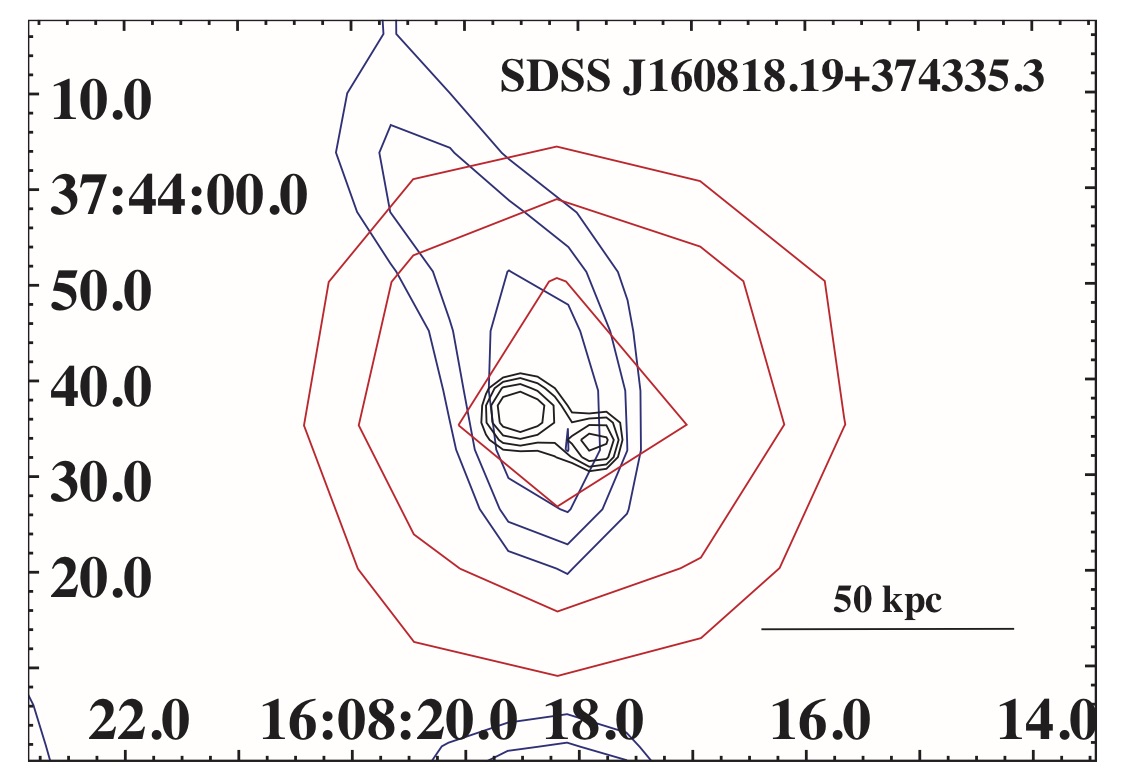} 
 
 \includegraphics[width=6.cm,height=6.cm]{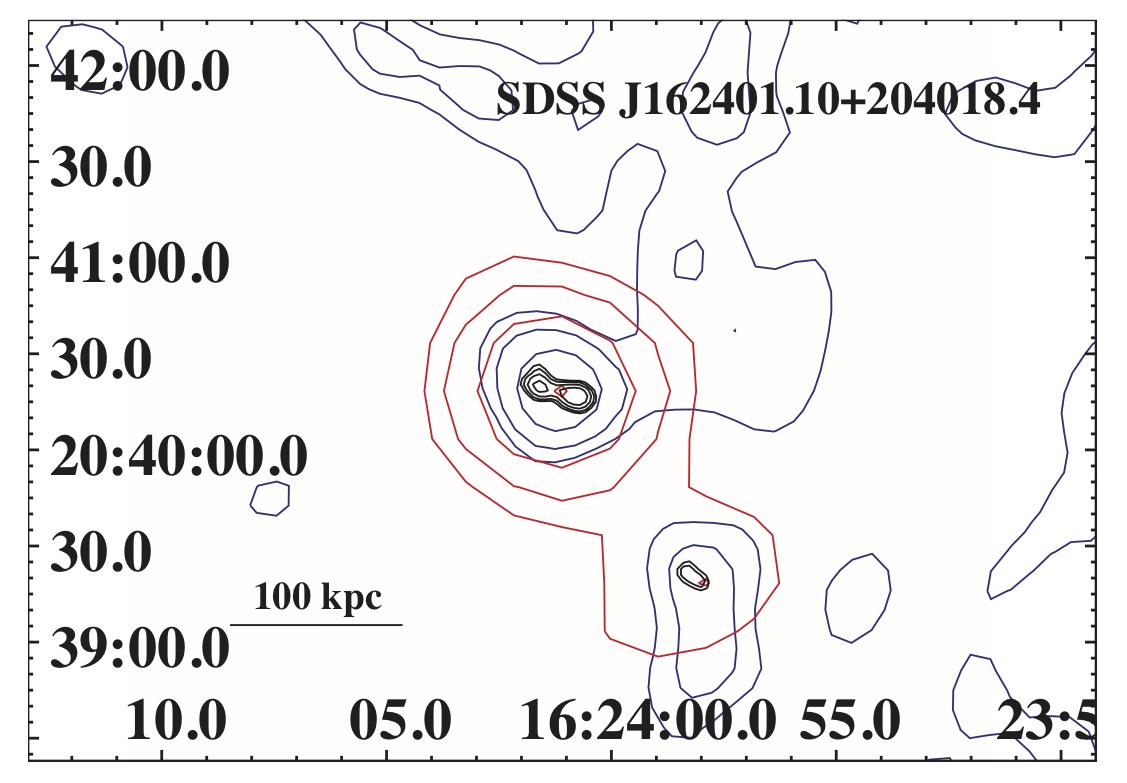} 
 \includegraphics[width=6.cm,height=6.cm]{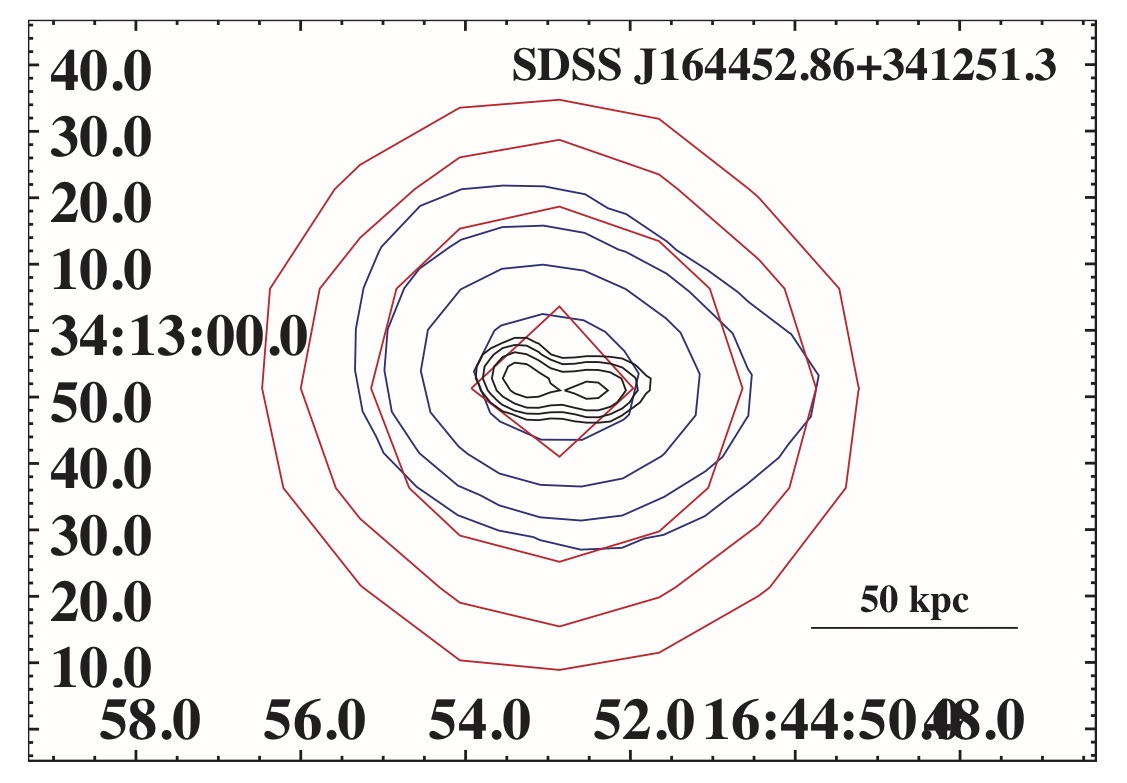}
 \includegraphics[width=6.cm,height=6.cm]{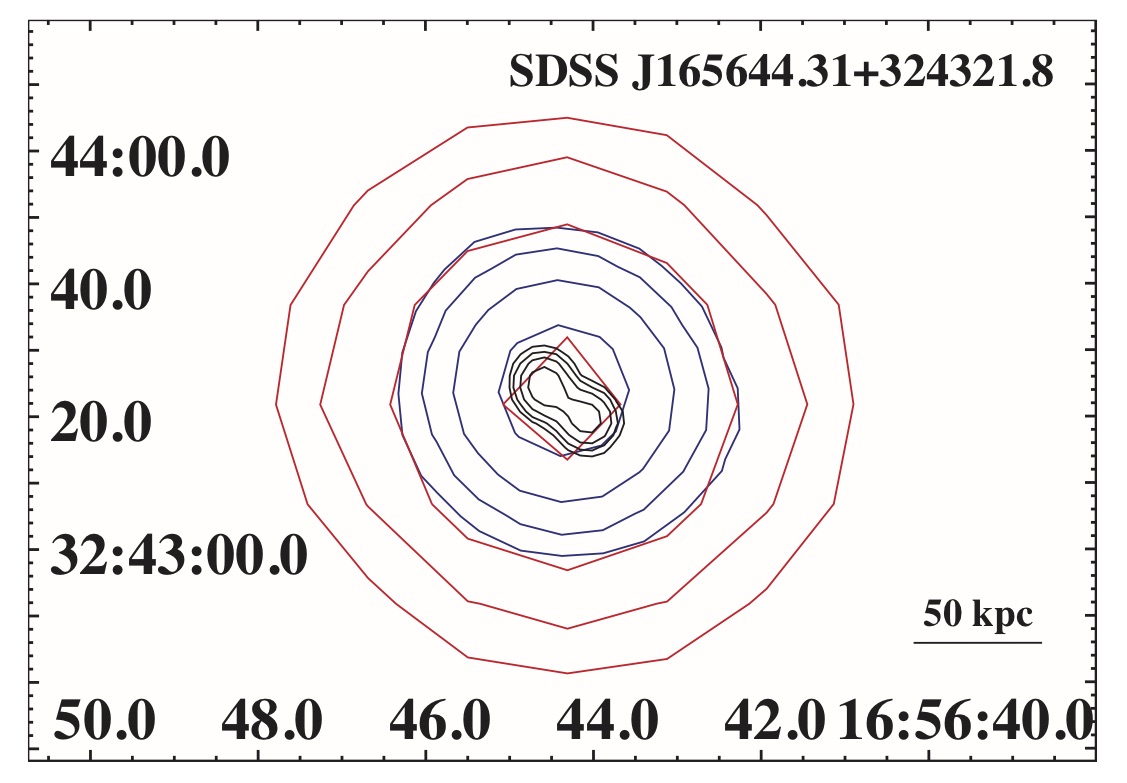} 
 \includegraphics[width=6.3cm,height=6.3cm]{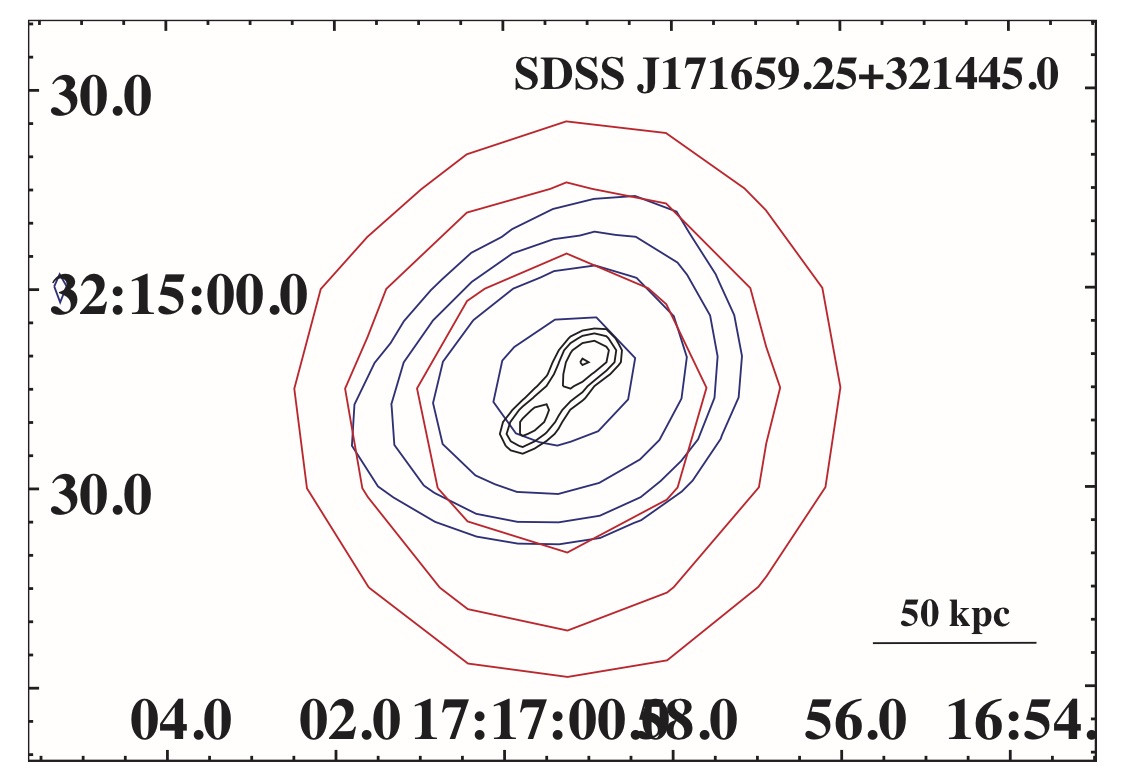} 
 \caption{(continued)}
 \end{figure}
 
\newpage
\section{Images parameters: sources excluded and COMP2$CAT$ sources.}
\label{ap:contours}

\begin{center}
\begin{longtable}{cccccc}
\caption[Parameters of the images of the sources excluded because of their high-resolution radio morphologies.]{Parameters of the images of the sources excluded because of their high-resolution radio morphologies.} 
\label{tab} \\
\hline
  \multicolumn{1}{c}{SDSS} &
  \multicolumn{2}{c}{FIRST} &
  \multicolumn{1}{c}{VLA}\\
  \multicolumn{1}{c}{Name} &
  \multicolumn{1}{c}{$l$ [mJy/beam]} &
  \multicolumn{1}{c}{$f$}&
  \multicolumn{1}{c}{band [GHz]}\\
  \hline
  \hline
    J111025.09+032138.8&1&4&1.4\\
    J125724.35+272952.1&0.6&2&1.4\\
    J125935.70+275733.3&1&3&4.8\\
    J132451.44+362242.7&0.8&4&1.4\\
    J161531.36+272657.3&2&2.25&4.8\\
    \hline
    	\caption{Col. (1): SDSS name of the sources. Col. (2): $l$, value of the starting contour level of the FIRST radio map and $f$, factor increase of the FIRST radio contours. Col. (3): same parameters as in Col. (2) for the NVSS radio map. Col. (4): same parameters as in Col. (2) for the TGSS radio map.}
\label{tab:contoursexc}
\end{longtable}

\end{center}

\begin{center}
\begin{longtable}{ccccccc}
\caption[Parameters of the images of the sources with large-scale extended emission.]{Parameters of the images of the sources with large-scale extended emission.} 
\label{tab} \\
\hline
	\multicolumn{1}{c}{SDSS} &
	\multicolumn{2}{c}{FIRST} &
	\multicolumn{2}{c}{NVSS}&
	\multicolumn{2}{c}{TGSS}\\
	\multicolumn{1}{c}{Name} &
	\multicolumn{1}{c}{$l$ [mJy/beam]} &
	\multicolumn{1}{c}{$f$} &
	\multicolumn{1}{c}{$l$ [mJy/beam]} &
	\multicolumn{1}{c}{$f$} &
	\multicolumn{1}{c}{$l$ [mJy/beam]} &
	\multicolumn{1}{c}{$f$} \\
	\hline
	\hline
		J083224.82+184855.4&0.6&1.75&10&2&10&2\\
		J083830.99+194820.4&0.6&1.5&3&1.5&10&2\\
		J091443.12+073544.9&0.5&1.25&3&1.5&4&1.5\\
		J115905.68+582035.5&0.8&1.25&1&2.5&8&2\\
		J132345.01+313356.7&0.6&2&4&2&6&2\\
		J152804.95+054428.1&1&2.25&2&2&10&2\\
		J215305.08-071106.9&1&1.75&2&2&6&2\\
	\hline
	\caption{Col. (1): SDSS name of the sources. Col. (2): $l$, value of the starting contour level of the FIRST radio map and $f$, factor increase of the FIRST radio contours. Col. (3): same parameters as in Col. (2) for the NVSS radio map. Col. (4): same parameters as in Col. (2) for the TGSS radio map.}
	\label{tab:contoursext}
\end{longtable}
\end{center}

\begin{center}
\begin{longtable}{ccccccc}
\caption[Parameters of the images of COMP2$CAT$ sources.]{Parameters of the images of COMP2$CAT$ sources.} 
\label{tab} \\
\hline
  \multicolumn{1}{c}{SDSS} &
  \multicolumn{2}{c}{FIRST} &
  \multicolumn{2}{c}{NVSS}&
  \multicolumn{2}{c}{TGSS} \\
  \multicolumn{1}{c}{Name} &
  \multicolumn{1}{c}{$l$ [mJy/beam]} &
  \multicolumn{1}{c}{$f$} &
  \multicolumn{1}{c}{$l$ [mJy/beam]} &
  \multicolumn{1}{c}{$f$} &
  \multicolumn{1}{c}{$l$ [mJy/beam]} &
  \multicolumn{1}{c}{$f$} \\
  \hline
  \hline
  J073600.87+273926.0&0.8&1.5&3&1.5&4&2\\
  J074132.98+475215.6&0.5&1.5&2&2&10&2\\
  J074641.45+184405.4&1&1.5&5&1.5&10&2\\
  J081023.27+421625.8&0.4&2&3&1.5&15&1.5\\
  J081104.30+355908.3&2&1.75&6&2&-&-\\
  J082033.79+395142.4&0.6&2&5&2&8&2\\
  J083053.58+231035.7&0.6&1.5&4&1.5&10&1.5\\
  J084517.83+303027.4&0.6&1.5&2&1.5&-&-\\
  J090311.14+540351.6&2&2&4&2&10&2\\
  J091134.75+125538.1&4&1.75&2.5&4&10&4\\
  J095341.37+014202.3&0.5&1.5&3&1.5&20&1.5\\
  J100622.41+301332.9&1&2&2&2&10&2\\
  J101653.82+002857.0&0.4&2&2&2&6&2\\
  J101944.27-003817.8&0.4&1.75&2&2&6&2\\
  J103801.77+414625.8&0.8&2&6&1.5&10&2\\
  J103842.52+120315.6&1&2&4&2&2&1.5\\
  J104254.02+282559.0&1&1.75&6&2&10&2\\
  J111109.58+393552.0&1&2.25&4&2&10&2\\
  J113305.52+592013.7&1&1.5&3&2&8&2\\
  J113643.49+545446.8&1&2&2&2&8&2\\
  J115050.98-031113.0&1.5&2&4&2&20&2\\
  J122208.81+073329.6&0.8&1.75&2&2&8&2\\
  J125319.21+475335.2&0.6&2&4&1.5&8&1.5\\
  J130107.54-032652.5&1&3&4&3&8&3\\
  J131705.93+435713.2&0.4&2&4&2&20&1.5\\
  J131945.31+603043.0&0.6&4&4&3&8&2\\
  J132031.47-012718.5&0.6&1.5&1&2&4.5&1.25\\
  J132602.39+364759.3&1&4&10&4&10&4\\
  J132649.30+164948.0&0.6&2&6&2&20&1.5\\
  J133917.34-015048.7&1&2.5&4&2&20&2\\
  J135338.43+360802.4&0.45&2&4&2&10&2\\
  J135347.34+515734.3&1&2.75&4&2.5&10&4\\
  J144647.43+032527.1&1&2.25&3&2&10&1.5\\
  J144731.24+330606.2&1&2.5&6&2&10&2\\
  J145604.88+472712.4&1&3&4&3&20&3\\
  J145858.83+130145.9&1&1.5&3&1.5&10&1.25\\
  J151135.87+191228.0&0.8&2&2&2&10&2\\
  J155749.61+161836.6&3&1.75&4&2.5&10&2\\
  J160818.19+374335.3&0.5&1.25&2&1.5&6&1.15\\
  J162401.10+204018.4&0.8&1.5&2&2&4&2\\
  J164452.86+341251.3&1&1.75&3&2&6&2\\
  J165644.31+324321.8&2&1.75&5&2&10&2\\
  J171659.25+321445.0&1&1.5&2&2&4&2\\
\hline
\caption{Col. (1): SDSS name of the sources. Col. (2): $l$, value of the starting contour level of the FIRST radio map and $f$, factor increase of the FIRST radio contours. Col. (3): same parameters as in Col. (2) for the NVSS radio map. Col. (4): same parameters as in Col. (2) for the TGSS radio map.}

  \label{tab:contourscomp}

\end{longtable}
\end{center}


\begin{thebibliography}{40}
\expandafter\ifx\csname natexlab\endcsname\relax\def\natexlab#1{#1}\fi

\bibitem[{{Abazajian} {et~al.}(2009){Abazajian}, {Adelman-McCarthy},
  {Ag{\"u}eros}, {Allam}, {Allende Prieto}, {An}, {Anderson}, {Anderson},
  {Annis}, {Bahcall}, \& et~al.}]{abazajian09}
{Abazajian}, K.~N., {Adelman-McCarthy}, J.~K., {Ag{\"u}eros}, M.~A., {et~al.}
  2009, \apjs, 182, 543

\bibitem[{{Baldi} \& {Capetti}(2010)}]{Baldi2010}
{Baldi}, R.~D. \& {Capetti}, A. 2010, \aap, 519, A48

\bibitem[{{Baldi}, {Capetti} \& {Massaro}(2018)}]{Baldi2018}
{Baldi}, R.~D. and {Capetti}, A. and {Massaro}, F. 2018, \aap, 609, A1

\bibitem[{{Baldi} {et~al.}(2015){Baldi}, {Capetti}, \& {Giovannini}}]{baldi15}
{Baldi}, R.~D., {Capetti}, A., \& {Giovannini}, G. 2015, \aap, 576, A38

\bibitem[{{Balmaverde}, {Baldi}, \& {Capetti}(2008){Balmaverde}, {Baldi}, \& {Capetti}}]{Balmaverde2008}
{Balmaverde}, B., {Baldi}, R.~D. \& {Capetti}, A. 2008, \aap, 486, 119

\bibitem[{{Balogh} {et~al.}(1999){Balogh}, {Morris}, {Yee}, {Carlberg}, \&
  {Ellingson}}]{balogh99}
{Balogh}, M.~L., {Morris}, S.~L., {Yee}, H.~K.~C., {Carlberg}, R.~G., \&
  {Ellingson}, E. 1999, \apj, 527, 54

\bibitem[{{Becker} {et~al.}(1995){Becker}, {White}, \& {Helfand}}]{becker95}
{Becker}, R.~H., {White}, R.~L., \& {Helfand}, D.~J. 1995, \apj, 450, 559

\bibitem[{{Bell} {et~al.}(2003){Bell}, {McIntosh}, {Katz}, \&
  {Weinberg}}]{bell03}
{Bell}, E.~F., {McIntosh}, D.~H., {Katz}, N., \& {Weinberg}, M.~D. 2003, \apjs,
  149, 289

\bibitem[{{Bennett} {et~al.}(2014)}]{bennett14}
{Bennett}, C.~L., {Larson}, D., {Weiland}, J.~L. \& {Hinshaw}, G. 2014, \apj, 794, 135

\bibitem[{{Best} (2009)}]{Best2009}
{Best}, P.~N. 2009, Astron. Nachr., 330, 184

\bibitem[{{Best} \& {Heckman}(2012)}]{best12}
{Best}, P.~N. \& {Heckman}, T.~M. 2012, \mnras, 421, 1569

\bibitem[{{Brinchmann} {et~al.}(2004){Brinchmann}, {Charlot}, {White},
  {Tremonti}, {Kauffmann}, {Heckman}, \& {Brinkmann}}]{bri04}
{Brinchmann}, J., {Charlot}, S., {White}, S.~D.~M., {et~al.} 2004, \mnras, 351, 1151

\bibitem[{{Bodo} {et~al.}(2013){Bodo},{Mamatsashvili}, {Rossi}, \& {Mignone}}]{Bodo2013}
{Bodo}, G., {Mamatsashvili}, G., {Rossi}, P. \& {Mignone}, A. 2013, \mnras, 434, 3030

\bibitem[{{Buttiglione} {et~al.}(2010){Buttiglione}, {Capetti}, {Celotti},
  {Axon}, {Chiaberge}, {Macchetto}, \& {Sparks}}]{buttiglione10}
{Buttiglione}, S., {Capetti}, A., {Celotti}, A., {et~al.} 2010, \aap, 509, A6

\bibitem[{{Capetti} \& {Baldi}(2011)}]{Capetti2011}
{Capetti}, A. \& {Baldi}, R.~D. 2011, \aap, 529, A126

\bibitem[{{Capetti} \& {Raiteri}(2015)}]{capetti15}
{Capetti}, A. \& {Raiteri}, C.~M. 2015, \aap, 580, A73

\bibitem[{{Capetti} {et~al.}(2017a)}]{Capetti2017I}
{Capetti}, A., {Massaro}, F. \& {Baldi}, R.~D. 2017, \aap, 598, A49

\bibitem[{{Capetti} {et~al.}(2017b)}]{Capetti2017II}
{Capetti}, A., {Massaro}, F. \& {Baldi}, R.~D. 2017, \aap, 601, A81

\bibitem[{{Condon} {et~al.}(1998){Condon}, {Cotton}, {Greisen}, {Yin},
  {Perley}, {Taylor}, \& {Broderick}}]{condon98}
{Condon}, J.~J., {Cotton}, W.~D., {Greisen}, E.~W., {et~al.} 1998, \aj, 115,
  1693

\bibitem[{{D'Abrusco} {et~al.}(2014){D'Abrusco}, {Massaro}, {Paggi}, {Smith},
  {Masetti}, {Landoni}, \& {Tosti}}]{dabrusco14}
{D'Abrusco}, R., {Massaro}, F., {Paggi}, A., {et~al.} 2014, \apjs, 215, 14

\bibitem[{{Fanaroff} \& {Riley}(1974)}]{fanaroff74}
{Fanaroff}, B.~L. \& {Riley}, J.~M. 1974, MNRAS, 167, 31P

\bibitem[{{Fanti} {et~al.}(1995)}]{Fanti1995}
{Fanti}, C., {Fanti}, R., {Dallacasa}, D., {Schilizzi}, R.~T., {Spencer}, R.~E., \& {Stanghellini}, C. 1995, \aap, 302, 317

\bibitem[{{Ghisellini}(2011)}]{Ghisellini2011}
{Ghisellini}, G. 2011, AIP Conference Proceedings, 1381, 180

\bibitem[{{Hardcastle}, {Evans}, \& {Croston}(2007)}]{Hardcastle2007}
{Hardcastle}, M.~J., {Evans}, D.~A. \& {Croston}, J.~H. 2007, \mnras, 376, 1849
\bibitem[{{Hardcastle} {et~al.}(2019){Hardcastle}, M.~J., {Williams}, W.~L., {Best}, P.~N., {Croston}, J.~H., {Duncan}, K.~J., {Rottgering}, H.~J.~A., {Sabater}, J., {Shimwell}, T.~W., {Tasse}, C., {Callingham}, J.~R., {Cochrane}, R.~K., {de Gasperin}, F., {Gurkan}, G., {Jarvis}, M.~J., {Mahatma}, V., {Miley}, G.~K., {Mingo}, B., {Mooney}, S., {Morabito}, L.~K., {O'Sullivan}, S.~P., {Prandoni}, I., {Shulevski}, A. \& {Smith}, D.~J.~B.}]{Hardcastle2018}
{Hardcastle}, M.~J., {Williams}, W.~L., {Best}, P.~N. {et~al.} 2019, \aap, 622, A12

\bibitem[{{Kauffmann} {et~al.}(2003){Kauffmann}, {Heckman}, {White}, {Charlot},
  {Tremonti}, {Brinchmann}, {Bruzual}, {Peng}, {Seibert}, {Bernardi},
  {Blanton}, {Brinkmann}, {Castander}, {Cs{\'a}bai}, {Fukugita}, {Ivezic},
  {Munn}, {Nichol}, {Padmanabhan}, {Thakar}, {Weinberg}, \&
  {York}}]{kauffmann03b}
{Kauffmann}, G., {Heckman}, T.~M., {White}, S.~D.~M., {et~al.} 2003, \mnras, 341, 33
  
\bibitem[{{Koester} {et~al.}(2007){Koester}, {McKay}, {Annis}, {et~al.}}]{Koester2007}
{Koester}, B.~P., {McKay}, T.~A. \& {Annis}, J., {et~al.} 2007, \apj, 660, 239

\bibitem[{{Kunert-Bajraszewska} {et~al.}(2010)}]{Kunert2010}
{Kunert-Bajraszewska}, M., {Gawro{\'n}ski}, M.~P., {Labiano}, A. \& {Siemiginowska}, A. 2010, \mnras, 408, 2261

\bibitem[{{Labiano} (2006)}]{Labiano2006}
{Labiano}, A. 2006, {Ph.D. Thesis}, {Space Telescope Science Institute Kapteyn Astronomical Institute}

\bibitem[{{Labiano} (2009)}]{Labiano2009}
{Labiano}, A. 2006, {Astronomische Nachrichten}, 330, 241

\bibitem[{{Ledlow} \& {Owen}(1996)}]{ledlow96}
{Ledlow}, M.~J. \& {Owen}, F.~N. 1996, \aj, 112, 9

\bibitem[{{Lin} {et~al.} (2010)}]{Lin2010}
{Lin}, Y.~T., {Shen}, Y., and {Strauss}, M.~A., {Richards}, G.~T. \& {Lunnan}, R. 2010, \apj, 723, 1119

\bibitem[{{Massaro} {et~al.}(2014){Massaro}, {Masetti}, {D'Abrusco}, {Paggi} \& {Funk}}]{Massaro2014}
{Massaro}, F., {Masetti}, N., {D'Abrusco}, R., {Paggi}, A. \& 
	{Funk}, S. 2014, \aj, 148, 66

\bibitem[{{Montero-Dorta} \& {Prada}(2009)}]{montero09}
{Montero-Dorta}, A.~D. \& {Prada}, F. 2009, \mnras, 399, 1106

\bibitem[{{Murphy} \& {Baum}(2014)}]{Murphy2014}
{Murphy}, E.~J. \& {Baum}, S. 2014, NRAO [White paper]

\bibitem[{{Nakamura} {et~al.}(2003){Nakamura}, {Fukugita}, {Yasuda}, {Loveday},
  {Brinkmann}, {Schneider}, {Shimasaku}, \& {SubbaRao}}]{nakamura03}
{Nakamura}, O., {Fukugita}, M., {Yasuda}, N., {et~al.} 2003, \aj, 125, 1682

\bibitem[{{O'Dea}(1998)}]{Odea1998} {O'Dea}, C.~P. 1998, \pasp, 110, 493

\bibitem[{{Orienti} \& {Dallacasa}(2014)}]{Orienti2014}{Orienti}, M. and {Dallacasa}, D. 2014, \mnras, 438, 463

\bibitem[{{Owen} \& {Rudnick}(1976)}]{owen76}
{Owen}, F.~N. \& {Rudnick}, L. 1976, \apjl, 205, L1

\bibitem[{{Phillips} \& {Mutel}(1982)}]{Phillips1982}
{Phillips}, R.~B. \& {Mutel}, R.~L. 1982, \aap, 106, 21

\bibitem[{{Readhead} (1995)}]{Readhead1995}
{Readhead}, A.~C.~S. 1995, Proceedings of the National Academy of Science, 92, 11447

\bibitem[{{Schawinski} {et~al.}(2009){Schawinski}, {Lintott}, {Thomas},
  {Sarzi}, {Andreescu}, {Bamford}, {Kaviraj}, {Khochfar}, {Land}, {Murray},
  {Nichol}, {Raddick}, {Slosar}, {Szalay}, {Vandenberg}, \&
  {Yi}}]{schawinski09}
{Schawinski}, K., {Lintott}, C., {Thomas}, D., {et~al.} 2009, \mnras, 396, 818

\bibitem[{{Schoenmakers} {et~al.}(2000){Schoenmakers}, {Bruyn}, {R$\"{o}$ttgering}, {van de Laan}, \& {Kaiser}}]{Schoenmakers2000}
{Schoenmakers}, A.~P., {Bruyn}, A.~G., {Röttgering}, H.~J.~A., {et~al.} 2000, \mnras, 315, 371

\bibitem[{{Shen} {et~al.}(2008){Shen}, S., {Kauffmann}, G., {von der Linden}, A., {White}, S.~D.~M., {Best}, P.~N.}]{Shen2008}
{Shen}, S., {Kauffmann}, G., {von der Linden}, A., {White}, S.~D.~M., {Best}, P.~N. 2008, \mnras, 389, 1074

\bibitem[{{Shen} {et~al.}(2003){Shen}, {Mo}, {White}, {Blanton}, {Kauffmann},
  {Voges}, {Brinkmann}, \& {Csabai}}]{shen03}
{Shen}, S., {Mo}, H.~J., {White}, S.~D.~M., {et~al.} 2003, \mnras, 343, 978

\bibitem[{{Snellen} {et~al.}(2004)}]{Snellen2004}
{Snellen}, I.~A.~G., {Mack}, K.-H., {Schilizzi}, R.~T. \& 
	{Tschager}, W. 2004, \mnras, 348, 227
	
\bibitem[{{Sobolewska} {et~al.}(2019)}]{Sobolewska2018}
{Sobolewska}, M., {Siemiginowska}, A., {Guainazzi}, M., {Hardcastle}, M., {Migliori}, G., {Ostorero}, L. \& {Stawarz}, L. 2019, \apj\, 871, 71

\bibitem[{{Strateva} {et~al.}(2001){Strateva}, {Ivezi{\'c}}, {Knapp},
  {Narayanan}, {Strauss}, {Gunn}, {Lupton}, {Schlegel}, {Bahcall}, {Brinkmann},
  {Brunner}, {Budav{\'a}ri}, {Csabai}, {Castander}, {Doi}, {Fukugita}, {Gy{\H
  o}ry}, {Hamabe}, {Hennessy}, {Ichikawa}, {Kunszt}, {Lamb}, {McKay},
  {Okamura}, {Racusin}, {Sekiguchi}, {Schneider}, {Shimasaku}, \&
  {York}}]{strateva01}
{Strateva}, I., {Ivezi{\'c}}, {\v Z}., {Knapp}, G.~R., {et~al.} 2001, \aj, 122,
  1861

\bibitem[{{Strauss} {et~al.}(2002){Strauss}, {Weinberg}, {Lupton}, {Narayanan},
  {Annis}, {Bernardi}, {Blanton}, {Burles}, {Connolly}, {Dalcanton}, {Doi},
  {Eisenstein}, {Frieman}, {Fukugita}, {Gunn}, {Ivezi{\'c}}, {Kent}, {Kim},
 {Knapp}, {Kron}, {Munn}, {Newberg}, {Nichol}, {Okamura}, {Quinn}, {Richmond},
  {Schlegel}, {Shimasaku}, {SubbaRao}, {Szalay}, {Vanden Berk}, {Vogeley},
 {Yanny}, {Yasuda}, {York}, \& {Zehavi}}]{strauss02}
{Strauss}, M.~A., {Weinberg}, D.~H., {Lupton}, R.~H., {et~al.} 2002, \aj, 124,
  1810
 
\bibitem[{{Tempel}, {Tago} \& {Liivam{\"a}gi}(2012)  {Tempel}, E., {Tago}, E. \& {Liivam{\"a}gi}, L.~J.}]{Tempel2012}
{Tempel}, E., {Tago}, E. \& {Liivam{\"a}gi}, L.~J. 2012, \aap, 540, A106

\bibitem[{{Tremaine} {et~al.}(2002){Tremaine}, {Gebhardt}, {Bender}, {Bower},
  {Dressler}, {Faber}, {Filippenko}, {Green}, {Grillmair}, {Ho}, {Kormendy},
  {Lauer}, {Magorrian}, {Pinkney}, \& {Richstone}}]{tremaine02}
{Tremaine}, S., {Gebhardt}, K., {Bender}, R., {et~al.} 2002, \apj, 574, 740

\bibitem[{{Tremonti} {et~al.}(2004){Tremonti}, {Heckman}, {Kauffmann},
  {Brinchmann}, {Charlot}, {White}, {Seibert}, {Peng}, {Schlegel}, {Uomoto},
  {Fukugita}, \& {Brinkmann}}]{tre04}
{Tremonti}, C.~A., {Heckman}, T.~M., {Kauffmann}, G., {et~al.} 2004, \apj, 613,
  898
  
\bibitem[{{Wing} \& {Blanton}(2011)}]{Wing2011}
{Wing}, J.~D., \& {Blanton}, E.~L. 2011, \aj, 141, 3

\bibitem[{{Wright} {et~al.}(2010){Wright}, {Eisenhardt}, {Mainzer}, {Ressler},
  {Cutri}, {Jarrett}, {Kirkpatrick}, {Padgett}, {McMillan}, {Skrutskie},
  {Stanford}, {Cohen}, {Walker}, {Mather}, {Leisawitz}, {Gautier}, {McLean},
  {Benford}, {Lonsdale}, {Blain}, {Mendez}, {Irace}, {Duval}, {Liu}, {Royer},
  {Heinrichsen}, {Howard}, {Shannon}, {Kendall}, {Walsh}, {Larsen}, {Cardon},
  {Schick}, {Schwalm}, {Abid}, {Fabinsky}, {Naes}, \& {Tsai}}]{wright10}
{Wright}, E.~L., {Eisenhardt}, P.~R.~M., {Mainzer}, A.~K., {et~al.} 2010, \aj,
  140, 1868
  
\bibitem[{{Yang} {et~al.}(2005)}]{Yang2005}
{Yang}, X., {Mo}, H.~J., {van den Bosch}, F.~C. \& {Jing}, Y.~P. 2005, \mnras, 356, 1293

\bibitem[{{Yang} {et~al.}(2007)}]{Yang2007}
{Yang}, X., {Mo}, H.~J., {van den Bosch}, F.~C., {Pasquali}, A., {Li}, C. \& {Barden}, M. 2007, \apj, 671, 153

\end{thebibliography}
\end{document}